\documentclass{aa}

\usepackage{graphicx}
\usepackage{subcaption}
\usepackage{txfonts}
\begin{document}

   \title{SaNDi-SHoP: Searching for Satellites'N'Disks with \\ a Star-Hopping Program} \subtitle{ I. Analysis of the close surroundings of DI companions\thanks{Based on observations collected at the European Southern Observatory under ESO programs 111.24UH.001, 112.25QU.001, 113.26GK.001 and 114.276W.001}}
\titlerunning{Searching for Satellites'N'Disks with a Star-Hopping Program}
\authorrunning{Lazzoni et al.}
\author{C. Lazzoni\inst{1}, A. Zurlo\inst{2,3}, S. Desidera\inst{1}, A. Bernardi\inst{2,3}, S. P\'erez\inst{3,4,5}, D. Mesa\inst{1}, D. Barbato\inst{1},\\ P. H. Nogueira\inst{6,3} and A. Dasgupta\inst{2,3}}
         
   \institute{\inst{1} INAF-Osservatorio Astronomico di Padova, Vicolo dell'Osservatorio 5, Padova, Italy, 35122-I \\
\email{cecilia.lazzoni@inaf.it}\\
  \inst{2} Instituto de Estudios Astrof\'isicos, Facultad de Ingenier\'ia y Ciencias, Universidad Diego Portales, Av. Ej\'ercito 441, Santiago, Chile \\
\inst{3} Millennium Nucleus on Young Exoplanets and their Moons (YEMS), Santiago, Chile \\
              \inst{4} Departamento de Física, Universidad de Santiago de Chile, Av. Victor Jara 3659, Santiago, Chile \\
              \inst{5} Center for Interdisciplinary Research in Astrophysics and Space Science (CIRAS), Universidad de Santiago de Chile\\
              \inst{6} Department of Physics and Astronomy, Texas A\&M University, College Station, TX 77843-4242, USA}

   \date{Received ; accepted }

 
  \abstract
   {}
   {We aim to search for satellites and circumplanetary or circumsubstellar disks around directly imaged substellar companions, exploring their immediate environment to constrain the conditions for satellites and disk formation.}
   {We conducted a dedicated survey of {twelve} planets and brown dwarfs with VLT/SPHERE using a novel application of the star hopping technique. By building libraries of contemporaneous point spread function (PSF) references from nearby stars, we applied a frame-by-frame subtraction of each companion’s flux using the Negative Fake Companion method (NEGFC). This approach mitigates temporal PSF variability and enhances sensitivity to faint circumplanetary features. We derived contrast curves, translated them into mass detection limits using evolutionary models, and constrained the dynamically stable regions through estimates of Hill radii from orbital fits.}
   {Our analysis yields stringent limits on the presence of massive satellites, generally excluding companions more massive than a few Jupiter masses at separations beyond $\sim$1–5 au, depending on each system’s Hill radius. In most cases, no convincing point-like or extended residuals were found. However, we identify promising signals for three systems: extended residuals consistent with a circumplanetary disk around CT Cha b, tentative repeated residuals near TYC 8047-232-1 B that may trace a bound satellite companion of $3-6$ M\textsubscript{Jup}, and {marginal residual signals at the location of the previously reported candidate around DH Tau b, whose interpretation, however, remains uncertain due to possible contamination by instrumental effects}. These results confirm the power of star hopping in reducing PSF-related artifacts and provide some of the most stringent constraints to date on the mass and location of potential satellites and disks around directly imaged substellar companions.}
   {}


   \maketitle
%

\section{Introduction}
The direct imaging (DI) technique stands at the forefront of exoplanetary exploration, offering a unique opportunity to directly observe and characterize young planets and disks, and to unveil the architecture of planetary systems \citep[see, for a review,][]{Zurlo2024}. Over the last ten years, DI has undergone a revolution, driven by significant advancements in high-contrast imaging instruments such as VLT/SPHERE \citep{Beuzit2019}, Gemini/GPI \citep{Macintosh2014}, and Magellan/MagAO \citep{Males2018,Males2024}. These technical improvements have opened a new window into the study of nearby planetary systems, leading to landmark discoveries such as the two giant planets around PDS 70 \citep{Keppler2018, Mesa2019, Isella2019, Haffert2019} caught in the process of formation within the gap of their circumstellar disk, or the Jupiter analog AF Lep b with a mass of 2–5 M\textsubscript{Jup} around a young solar-like star \citep{Mesa2023, DeRosa2023,Franson2023}. 

More recently, the James Webb Space Telescope (JWST) has further extended these capabilities, allowing us to reach lower-mass companions, previously not detectable from the ground, and extending the search toward older systems. Examples of JWST capabilities include the detection of a candidate sub-Jovian planet around TWA 7 \citep[0.3 M\textsubscript{Jup} at a projected separation of $\sim$50 au,][]{Lagrange2025}, and the characterization of 14 Her c, a cold 6.9 M\textsubscript{Jup} planet in a dynamically hot, multi-planet configuration with 14 Her b, a 9.1 M\textsubscript{Jup} companion \citep{Bardalez2025}.

The success of DI not only depends upon sophisticated instrumentation but also on pivotal post-processing techniques, such as the Angular Differential Imaging \citep[ADI,][]{Marois2010a} and Principal Component Analysis \citep[PCA, e.g.][]{Soummer2012}. These refined methods enable the detection of faint companions that would otherwise remain hidden in the overwhelming glare of their parent stars.

Lately, Reference Differential Imaging \citep[RDI,][]{Lafreniere2009} has emerged as an indispensable tool to disentangle exoplanetary signals from instrumental artifacts and stellar noise, especially for space-based missions \citep[][]{Carter2023}. By employing reference stars with similar spectral characteristics and subtracting them from the science target, RDI enhances the sensitivity of DI systems to faint planetary companions placed in the innermost regions. The observing method of star-hopping \citep{Wahhaj2021} permits us to efficiently apply the RDI and improve the contrast at close separations. This technique involves repeatedly hopping between the science target and a nearby, similar-brightness reference star. 

In parallel with methodological advances in high-contrast post-processing, alternative statistical frameworks are emerging that may extend the discovery space of direct imaging. Algorithms such as the patch covariance \citep[PACO,][]{Flasseur2018} provide a detection-oriented approach that differs from classical PSF-subtraction pipelines and has demonstrated strong performance at small angular separations. As shown by \citet{Chomez2023}, PACO can efficiently recover faint companions in compact configurations, enabling the detection of a $\sim$15 M\textsubscript{Jup} brown dwarf at very small projected separation within a hierarchical multiple system. 

Beyond the detection of massive companions, DI also allows for the {spectroscopic} and astrometric characterization of exoplanets and, when pushed to its limit, for the characterization of their surroundings. By analogy with the {moons and rings} of our Solar System, we can expect that the area close to an exoplanet is populated by a circumplanetary disk (CPD) and satellites. Nevertheless, only a few exomoon candidates have been proposed from the transits technique \citep[e.g., Kepler-1625 b-i and Kepler-1708 b-i,][]{Teachey2018, Kipping2022}. A candidate Jupiter-like satellite has also been presented using the high-contrast imaging technique around the $\sim10$ M\textsubscript{Jup} substellar companion DH Tau b \citep{Lazzoni2020a}, potentially representing the first binary planets, with a mass ratio of 0.1. Finally, the most compelling detection of a circumplanetary disk around a planet was presented by \citet{Benisty2021}, which has been detected around PDS 70 c in mm dust emission.

Imaging these features is still a challenge, especially when rocky moons and faint rings are involved. However, thanks to the exquisite performances of state-of-the-art DI instruments, attempts to detect massive satellites, in the form of binary planets, and bright circumplanetary/circumsubstellar disks (CPD) were carried out. 
One of the first systematic efforts in this direction was presented by \citet{Lazzoni2020a}, who developed a dedicated post-processing technique based on the negative fake companion (NEGFC) method \citep{Marois2010a, Lagrange2010} to suppress the flux of directly imaged companions and probe their immediate surroundings. 


In their analysis of 27 substellar objects, \citet{Lazzoni2020a} distinguished between two categories of companions: bright objects detectable in individual frames, before post-processing techniques are applied, and fainter ones only visible in the final combined image. For the first sample of objects, a frame-by-frame subtraction routine was implemented, estimating the flux and position of the companion in each frame, whereas the second kind of objects was treated by retrieving the parameters in the post-processed image product.
In both cases, the subtraction relied on a single point spread function (PSF) reference, typically an image of the host star acquired either before or after the coronagraphic scientific acquisition. \citet{Lazzoni2020a} showed that, since such models are not co-temporal with the images of the system, spurious residuals due to atmospheric and instrumental variability are frequent, and an improved reference strategy is needed.

In this work, we analyze {12} substellar companions, both planets and brown dwarfs, searching for satellites and CPDs. Each target is bright and sufficiently separated from its host star to allow frame-by-frame PSF subtraction using the method of \citet{Lazzoni2020a}. To improve the modeling of the companion’s PSF, we employed the star-hopping technique, acquiring non-coronagraphic images of a nearby reference star with a similar spectral type. These quasi-simultaneous frames served as input for the NEGFC routine, enabling refined subtraction of the companion’s PSF and mitigating temporal variability. This constitutes a novel application of RDI, applied directly to the companion rather than the host star, reducing residuals and enhancing sensitivity to CPDs and massive satellite candidates. The paper is organized as follows: Section~\ref{sec2} describes the survey and data reduction, Section~\ref{sec3} presents the results with analyses for each companion, and Section~\ref{sec4} summarizes the conclusions.

\section{Observations and Data Reduction}
\label{sec2}

\subsection{Observations}

We selected a sample of {12} systems with known DI planets or brown dwarfs characterized by a high signal-to-noise ratio (S/N $\gtrsim 50$) and separations $\gtrsim 0.5''$ to avoid regions that are strongly contaminated by bright speckles. For each target, we also selected a suitable reference for the star-hopping technique. Star-hopping was first implemented at SPHERE/VLT in 2021 with the aim of using the RDI technique to reduce the speckles noise in the close vicinity of the { coronagraph} \citep{Wahhaj2021}. This is achieved by observing a reference star within $\sim 2$ degrees of the main target a few times during the scientific observation. The main advantage of this technique is the rapid switching between target and reference, with the adaptive optics loop closing immediately at the start of the science observations. The similarities between the main star and its reference, along with their proximity, allow for the acquisition of an image of the reference star in just a few minutes.

Our strategy made a novel use of the star-hopping technique, gathering a library of model PSFs to optimize the subtraction of directly imaged exoplanets and brown dwarfs and to detect putative satellites and disks in the residuals. As shown in \cite{Lazzoni2020a}, variations in the weather conditions and atmosphere parameters strongly affect the shape of the PSFs during acquisition. Standard observations taken with SPHERE usually include an image of the central star, to be used as a model PSF, taken before and after the coronagraphic sequence. However, given the long time required by the coronagraphic observations (1-2 hours), relevant mismatches are often identified between the reference PSF and the planet/brown dwarf, leading to spurious residuals in the subtraction. 

At the end, multiple model PSFs, taken at regular intervals during the coronagraphic sequence with the aid of star hopping, are the perfect strategy to improve the subtraction and retrieve reliable residuals. For each target, we thus selected a reference star within 2 degrees, similar in spectral type to the scientific star and with no evident signs of binarity in both the literature and from the Gaia Renormalized Unit Weight Error (RUWE; \citealt{Lindegren2021}). The only exception was HD 45036, the reference star selected for AB Pic, which has a RUWE of 3.3, probably pointing to a spectroscopic binary.

Observations were planned to be carried out for $\sim 1.5$ hours per target, with multiple concatenations of hops between the scientific and reference systems. HIP 64892 was observed for $\sim40$ minutes due to worsening weather conditions and interruption of acquisitions, while DH Tau was observed for $\sim2$ hours to collect more images of the reference. Unlike standard star-hopping observations, where both stars are kept behind the coronagraph to reproduce similar speckle patterns, we took coronagraphic images of the scientific system only. In this way, we collected a PSF library given by as many reference stars as hops performed in the concatenation, plus two bonus PSFs of the science star taken before and after the coronagraphic sequence. This strategy allows to have a rich library of PSFs for the suppression of the sub-stellar companion. The addition of the science star images to the PSF library is fundamental to overcoming possible issues due to the selection of non-ideal PSF references, as in the case of { HD\,45036}.

SPHERE was used in IRDIFS mode, with IRDIS, the dual-band camera, set in H2H3 bands { ($\lambda_{H2}=1.593\pm0.026\, \mu$m, and  $\lambda_{H3}=1.667\pm0.027\,\mu$m)} to minimize the dimension of the PSF of the known DI exoplanets/BDs investigated for the presence of satellites and disks. IFS, instead, worked in the Y-J bands (0.96-1.34 $\mu$m, R$\sim54$) and was mainly used to constrain the presence of additional companions in the inner regions { (between 20 and 70 au)} of these systems (Bernardi et al., {submitted}).\\
The observations presented in this paper were taken between June 2023 and {July} 2025 (programs 111.24UH.001, 112.25QU.001, and 113.26GK.001: PI Nogueira; program 114.276W.001: PI Dasgupta), and a total of 14 targets were observed. Two systems, PZ Tel and HD984, were excluded from this analysis because their orbital configuration, as estimated in the second paper of this series (Bernardi et al., submitted), leaves a dynamically accessible region for bound features that is too small to be probed with the present data. A summary of the main parameters for each observation is given in Table \ref{obs}, along with the reference adopted for the star-hopping. Note that the DH Tau system was observed twice, the first time on 2024-10-05 and then again on 2025-01-01.
DH Tau b is the only substellar companion with a known candidate satellite \citep{Lazzoni2020a}. However, in the first attempt, we were not able to detect the candidate and, thus, rescheduled the target for a second data acquisition.

\begin{table*}
\centering
\caption{Summary of the SPHERE/IRDIS observations. For each target, we report the date, total field rotation, seeing, reference star used in the star-hopping sequence, and the number of hops performed.}
\begin{tabular}{lccccc}
\hline\hline
  Name  &	Obs date     & Tot FoV Rot      & Seeing & Reference & Num Hops   \\
        &              &   ($^{\circ}$)   &  ('')  &           &         \\  
\hline \\
AB Pic	&   2023-10-26 & 22.2 & 0.46 &    HD 45036       &      5    \\[0.75ex]
TYC 7084-794-1 & 2023-11-30  & 100.3 & 0.58 & CD-36 2678  & 5 \\[0.75ex]
TYC 8047-232-1  & 2024-08-27 & 31.0 & 0.45 & HD 10626 & 5 \\[0.75ex]
CT Cha   & 2025-01-26 & 18.2 & 0.61 & TYC 9410-2449-1 & 5 \\[0.75ex]
GQ Lup & 2025-04-13 & 62.8 &  0.38  & 2MASS J15560250-3643472 & 5\\[0.75ex]
HIP78530 & 2025-04-13 & 11.6 &  0.45  & HD 142805 & 5\\[0.75ex]
DH Tau & 2024-10-06  & 25.0 &  0.79  &  V1320 Tau & 8\\[0.75ex]
       & 2025-01-02  & 38.4 &  0.6  &  V1320 Tau & 11\\[0.75ex]
HIP 64892 & 2025-02-24 & 11.8 &  0.68  & HD 115988 & 2 \\[0.75ex]
RX J1609.5-2105 & 2025-05-09 & 8.8 & 0.41   & 2MASS J16120920-2247504 & 5 \\[0.75ex]
HII 1348  & 2023-12-08 & 19.5 & 0.71 & HD282953 & 5\\[0.75ex]
$\eta$ Tel & 2025-07-22 & 47.9 & 0.66 & $\mu$ Tel & 9\\[0.75ex]
HD 130948 & 2024-03-18 & 23.0 & 0.9 & BD+25 2842 & 5 \\[0.75ex]
\hline
\end{tabular}
\label{obs}
\end{table*}

\subsection{Data reduction}
\label{data_red}

Pre-calibration reduction steps, including dark subtraction, flat-fielding, centering, and wavelength calibration, were performed by the SPHERE High Contrast Data Centre \citep{Delorme2017b, 2018A&A...615A..92G}. For some of the datacubes, frame sorting was required, especially in the case of RX J1609.5-2105, for which the planetary companion position in the field of view coincided with a cluster of bad pixels for half of the acquisition. Post-processing was performed using the Vortex Image Processing package \citep[VIP, ][]{Gonzalez2017,2023JOSS....8.4774C}. To suppress the signal of the companion and investigate potential circumsubstellar features, we adopted the frame-by-frame NEGFC subtraction approach described in \citet{Lazzoni2020a}. However, instead of using a single model PSF, we built a custom PSF library composed of all non-coronagraphic images of the host star and the multiple star-hopping reference frames acquired during the observation. For each science frame, we tested all available PSFs in the library and selected the one yielding the lowest residuals after subtraction, thus optimizing the modeling of the companion’s PSF on a per-frame basis. {The lowest residuals are evaluated by calculating the standard deviation within a region of pixels comprising three FWHM.}
Following the subtraction, we produced three final data cubes for each target: the science cube, the cube containing the models, and the cube of the residuals. These were combined using classical ADI to obtain the final image of the companion, its model, and the residual map. 

Contrast curves were computed following the approach described in \citet{Lazzoni2020a}: we recentered the residual image on the position of the companion and measured the standard deviation of the pixels' flux within concentric annuli of 1 FWHM width, starting at a radial separation of 0.5 FWHM. The 5$\sigma$ detection limits were derived from these standard deviations, corrected for small sample statistics as in \citet{Mawet2014}, and scaled by stellar flux to express the contrast in relative units.

As in \citet{Lazzoni2020a}, we also accounted for algorithm throughput losses by injecting synthetic companions into the raw data cubes and repeating the same PSF subtraction and ADI processing steps. In our case, the throughput correction was optimized frame by frame using the same PSF library selection strategy adopted for the real companion. This allowed us to calibrate the sensitivity more accurately at small separations and minimize biases introduced by temporal PSF mismatches.

{Since the contrast curves presented in this work are computed relative to the substellar companion, rather than with respect to the host star, it is important—particularly for companions at relatively small angular separations ($\lesssim2''$)—to verify whether the stellar contribution could introduce azimuthal asymmetries, potentially affecting the detection limits in regions closer to or farther from the star.}

{To address this point, we performed an azimuthal symmetry check focusing on GQ Lup B, the closest companion in our sample. We found no evidence for systematically shallower contrasts in the regions closer to the star. In particular, we compared contrast curves extracted from pairs of semi-annuli located on opposite sides of the companion, defined by a line perpendicular to the star–companion axis and passing through the companion position. The semi-annuli span radial separations from half the FWHM to half of the Hill radius of the companion and have a radial width of half the FWHM, consistent with the approach adopted for our standard one-dimensional contrast curves.}

{The resulting contrast curves obtained from the two sides are fully consistent with each other and with those derived from full annuli. Given the small spatial extent involved (for GQ Lup B, $R_{Hill}/2$ corresponds to $\sim 4$ pixels), the contrasts at these separations are dominated by residuals from the companion subtraction rather than by asymmetries related to the host star. For completeness, we repeated the same analysis for HIP 64892 B and HII 1348 B, the only other companions at separations smaller than 2'', obtaining similar results. The corresponding contrast curves and a visual illustration of the adopted semi-annuli are presented in Appendix~\ref{appA}.}

In Figure \ref{cc_SHVSsing}, we compare the contrast performance achieved using a single reference PSF, as done in \citet{Lazzoni2020a}, with that obtained using a PSF library composed of non-coronagraphic images of the target star and reference stars observed via the star-hopping technique. We observed varying outcomes when incorporating multiple references.

For several systems, such as AB Pic, GQ Lup, RX J1609.5–2105, $\eta$ Tel and TYC 7084-794-1, we found a consistent improvement in contrast across all separations when using the PSF library. In other cases, such as CT Cha and HIP 78530, the gain was more pronounced at small angular separations, particularly around the location of the substellar companion, while at larger separations the improvement was less significant or absent. In some instances, including the second epoch of DH Tau, HII 1348, and TYC 8047-232-1, while an evident gain was achieved in the inner regions, the contrast at wider separations was actually worse when using the PSF library. Finally, for a few targets such as the first epoch of DH Tau and HIP 64892, the performance of the PSF library compared to the single reference varied with separation, alternating between deeper and shallower contrasts.

Overall, the star-hopping technique, by enabling the construction of PSF libraries with reference images obtained closer in time to the science observations, proved to be effective in improving contrast performance near substellar companions, particularly at small angular separations.

For this sample, we also {tried} a novel method, consisting of a mock ADI, to subtract the contribution of the secondary and test the presence of point-like features in the residuals. In this way, we {could, in principle,} cross-check the residuals obtained with two different techniques, including the one described above. In more details, the mock-ADI method consists of constructing a library of the companion's PSF with a median or a PCA algorithm, using a datacube recentered on the companion itself. A putative satellite would rotate following the parallactic angle, while artifacts mimicking point-like sources, as the low-wind effect \citep{Milli2018}, are static in the datacube. {However, while} this technique is very powerful in subtracting artifacts, it is also aggressive and affects the retrieval of real signals. 

Mock ADI was tested on HIP\,87836's crowded field, presented in Section B.3 from \citet{Lazzoni2020a}, from which we selected an object with two close background stars \citep[P1 in][]{Lazzoni2020a}. The results of the mock ADI and our regular subtraction technique are shown in Fig.~\ref{mock}. The former shows strong self-subtraction wings, even if the final residuals are lower. We also tested the mock ADI on a synthetic dataset mimicking DH Tau b and its satellite. The satellite is not visible in the final image (Fig.~\ref{mockDHTau}), proving that faint sources are easily suppressed by this technique. Since faint and close-in satellites would be canceled out using mock ADI, we preferred to systematically apply the method presented above, and only show the results obtained with the first technique.

\subsection{Astrometry and photometry}

Astrometry and photometry for all companions were derived using the {well-established NEGFC technique implemented in the} \texttt{VIP} package \citep{Gonzalez2017, 2023JOSS....8.4774C}. Preliminary positions and fluxes were obtained with the \texttt{first\_guess()} function, which applies the NEGFC technique combined with the Nelder-Mead minimization algorithm. These values were then refined with the \texttt{mcmc\_negfc\_sampling()} function, running 50 walkers over 1000 iterations. The final astrometric uncertainties include contributions from the centering error behind the coronagraph {($\pm$2.5 mas)}, the fitting error from the MCMC procedure, and calibration terms such as distortion, pixel scale, True-North correction, and pupil zero-point angle \citep{Maire2016b,Maire2021a}. Photometric fluxes were converted into companion masses using the ATMO evolutionary models \citep{Phillips2020}, with system ages listed in Table \ref{stars}. {The errors assumed for the masses of the companions account for the uncertainties in the ages of each system as listed in Table \ref{stars}, the error estimated for the fluxes of the substellar companions as derived by the \texttt{mcmc\_negfc\_sampling()} function (more details for the photometry are presented in Bernardi et al., submitted) and the uncertainties on the fluxes of the stars.} Other contributions{, such as instrumental systematics,} were neglected. The derived separations and position angles are reported in Table~\ref{comps}, along with the corresponding mass estimates.

The astrometric and photometric analysis was not applied to HD 130948, given that its substellar companion is actually a binary-brown dwarf, but the two components, B and C, could not be resolved in our observations. Since the two PSFs are almost merged into one, it was not possible to retrieve precise radial distances and fluxes for the two components. We thus used the parameters for the total mass of the brown dwarf binary and its orbital motion around the star as found in the literature, as discussed in Section~\ref{s:HD130948}.

\section{Analysis of the results} 
\label{sec3}
{After establishing precise astrometry and photometry for each companion, we proceeded to investigate} the residual structures in the PSF-subtracted images, enabling a search for circumsubstellar disks {or circumplanetary disks, referred from now on as CPDs,} and satellite candidates.

The Hill radius { ($R_{Hill}$)} is a fundamental parameter needed to estimate the region around each substellar companion that could be investigated for the presence of satellites or CPDs. In particular, satellites on prograde and circular orbits must reside within half the Hill radius to remain dynamically stable \citep{Domingos2006}, and circumplanetary/circumsubstellar disks are expected to be truncated between { one third and half} of the Hill radius \citep[e.g][]{Ayliffe2009}.  $R_{Hill}$, in turn, depends on the semi-major axis and the eccentricity of the planet/brown dwarf. Unfortunately, these parameters are hardly constrained given the wide separation at which DI companions are detected. In fact, their revolution periods usually range from hundreds to thousands of years, with only a small portion of the orbit covered in the observation baseline. 

To estimate the best orbital parameters for the targets in our sample, we performed a joint orbit fit of all available relative astrometry (this paper and astrometric measurements present in the literature), supplemented by absolute astrometry for the stars with proper motion measurements coming from Hipparcos \citep{Nielsen2020,VanLeeuwen2007} and Gaia EDR3 \citep{GAIA2021} (AB Pic, HIP 64892 and HIP 78530). Our orbit fit is carried out with the \textit{orbitize!} Python package \citep{Blunt2020,Blunt2024}, which uses the parallel-tempered Affine-invariant MCMC algorithm from the Python package \textit{ptemcee} \citep{Foreman-Mackey2013,Vousden2016}. We adopted the default priors detailed in \textit{orbitize!}, which are logarithmic for the semi-major axis, flat for the eccentricity and Gaussian for the total mass of the system (defined as the sum of the stellar mass, listed in Table \ref{stars}, and the companion mass, listed in Table \ref{comps}, with their respective errors). We use a total of 400 walkers, 14 temperatures, and $2.5\cdot10^4$ total steps to sample the parameter space, with a thinning factor of 400. To help the assessment of convergence, we discarded the first 200 steps of each walker as burn-in. The Table presenting all the astrometric measurements included in this analysis, along with a detailed analysis of the fitting procedure adopted and results obtained, is presented in {the second paper of this series} which focuses on the characterization of the companions observed in our survey (Bernardi et al., {submitted}). Here, we present only the derived semi-major axis and eccentricities used to calculate the Hill radii of 11 of the 12 targets in the sample.
For HD 130948, instead, we used the orbital parameters found in the literature (see Section~\ref{s:HD130948}) and we did not include any $R_{Hill}$ given the complicated dynamics of this triple system.

For the suitable substellar companions, we derived the Hill radius
\begin{equation}
R_{Hill}=a_P(1-e_P)\sqrt[3]{\frac{M_P}{3M_{\odot}}},
\end{equation}
using the mass {($M_P$)}, semi-major axis {($a_P$)}, and eccentricity {($e_P$)} presented in columns two, five, and six of Table \ref{comps} and the mass of the star {($M_{\odot}$)} from column four of Table \ref{stars}. {Hill radii are presented in column seven of Table \ref{comps}.}

\begin{figure*}
    \centering

    \begin{subfigure}{0.31\textwidth}
        \includegraphics[width=\linewidth]{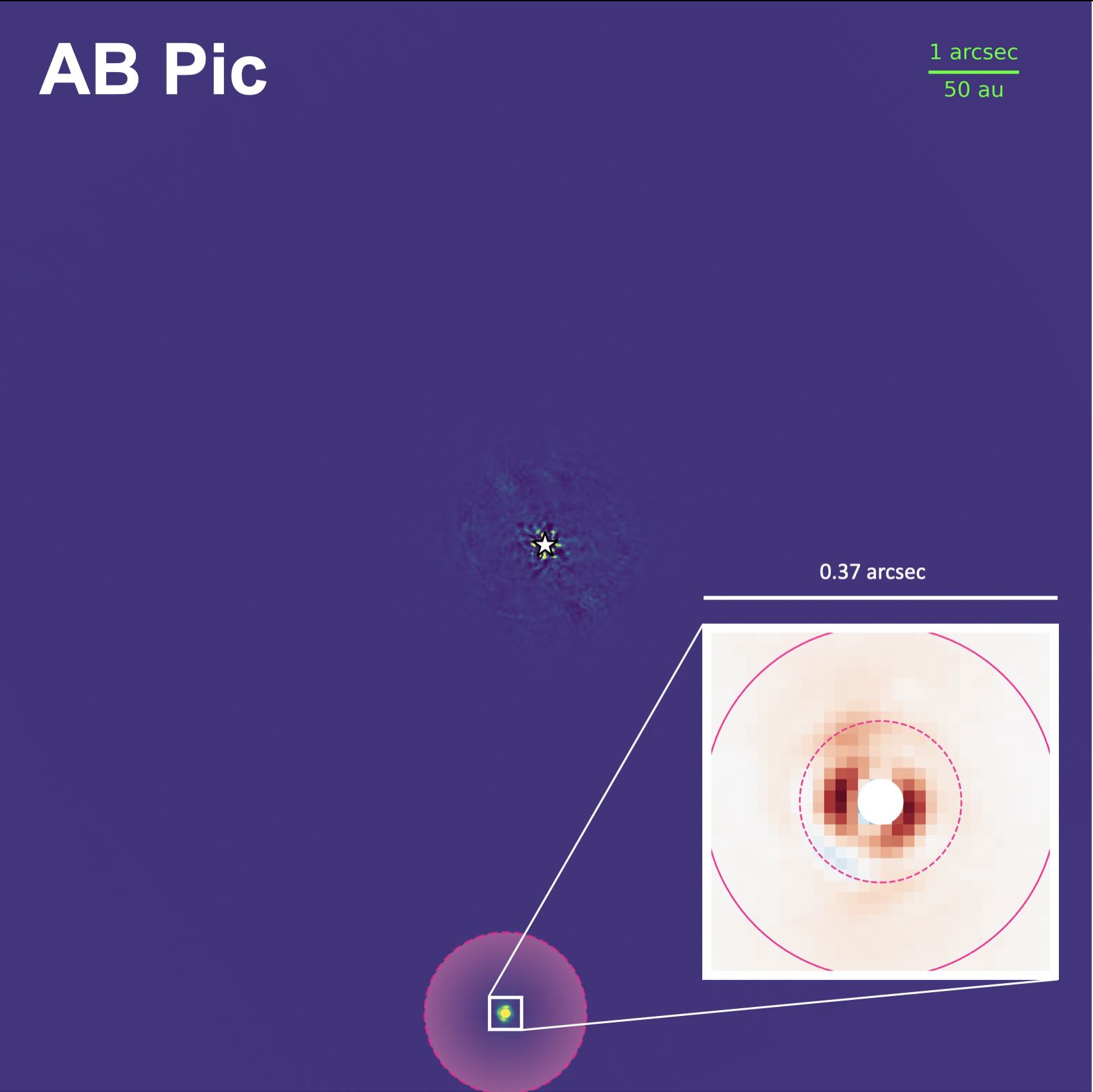}
 
    \end{subfigure}
    \begin{subfigure}{0.31\textwidth}
        \includegraphics[width=\linewidth]{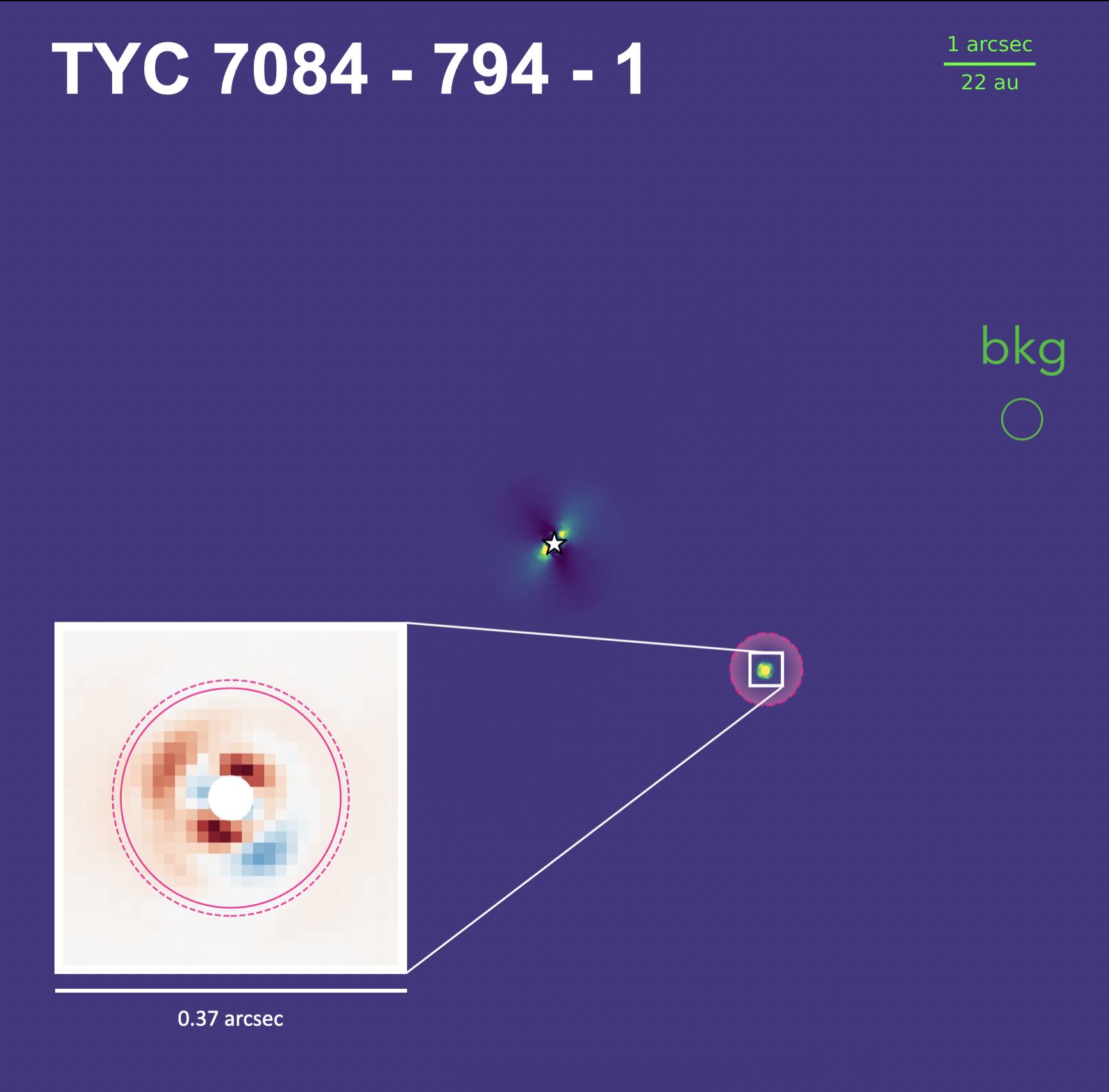}

    \end{subfigure}
    \begin{subfigure}{0.31\textwidth}
        \includegraphics[width=\linewidth]{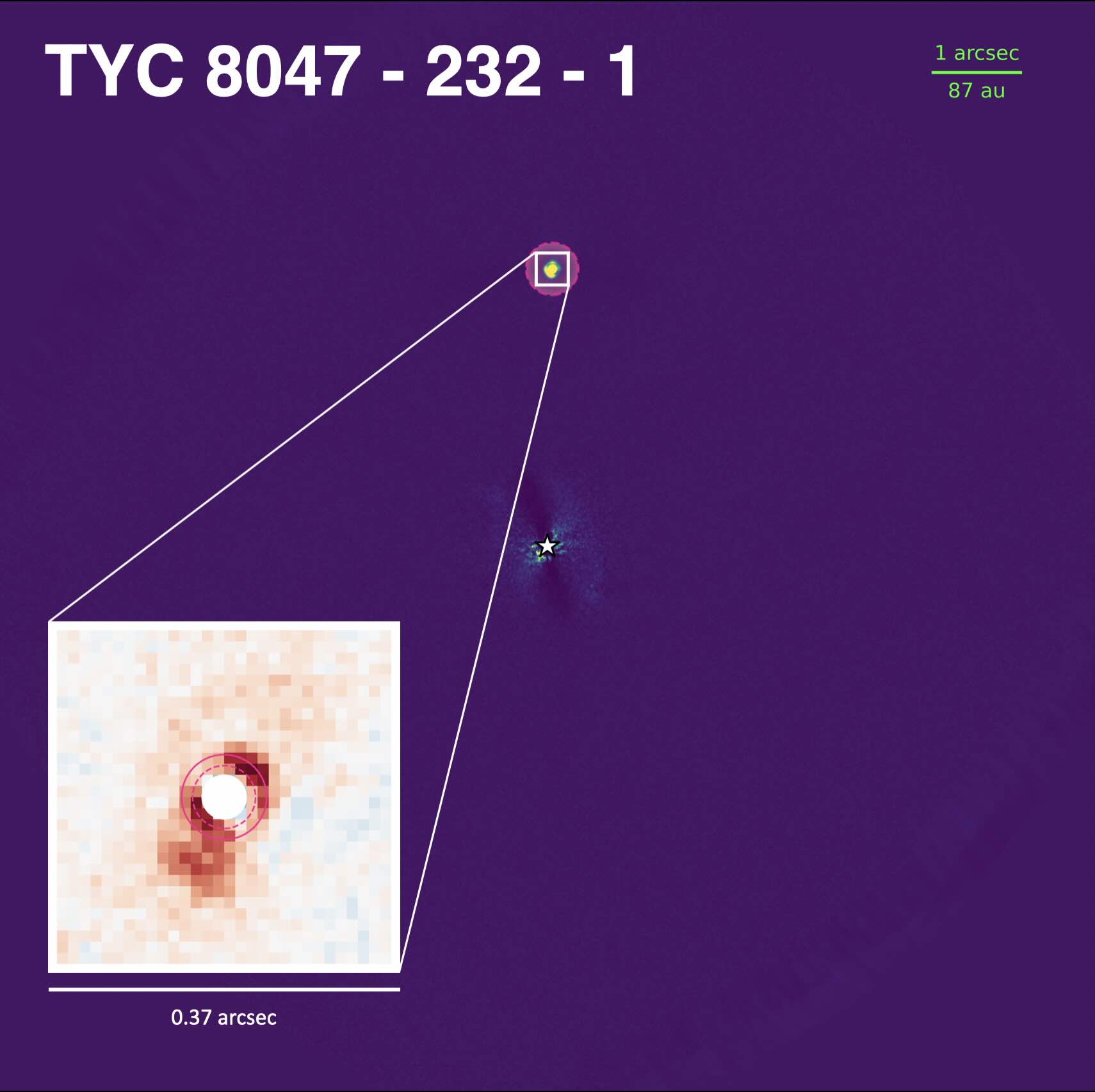}

    \end{subfigure}

    \begin{subfigure}{0.31\textwidth}
        \includegraphics[width=\linewidth]{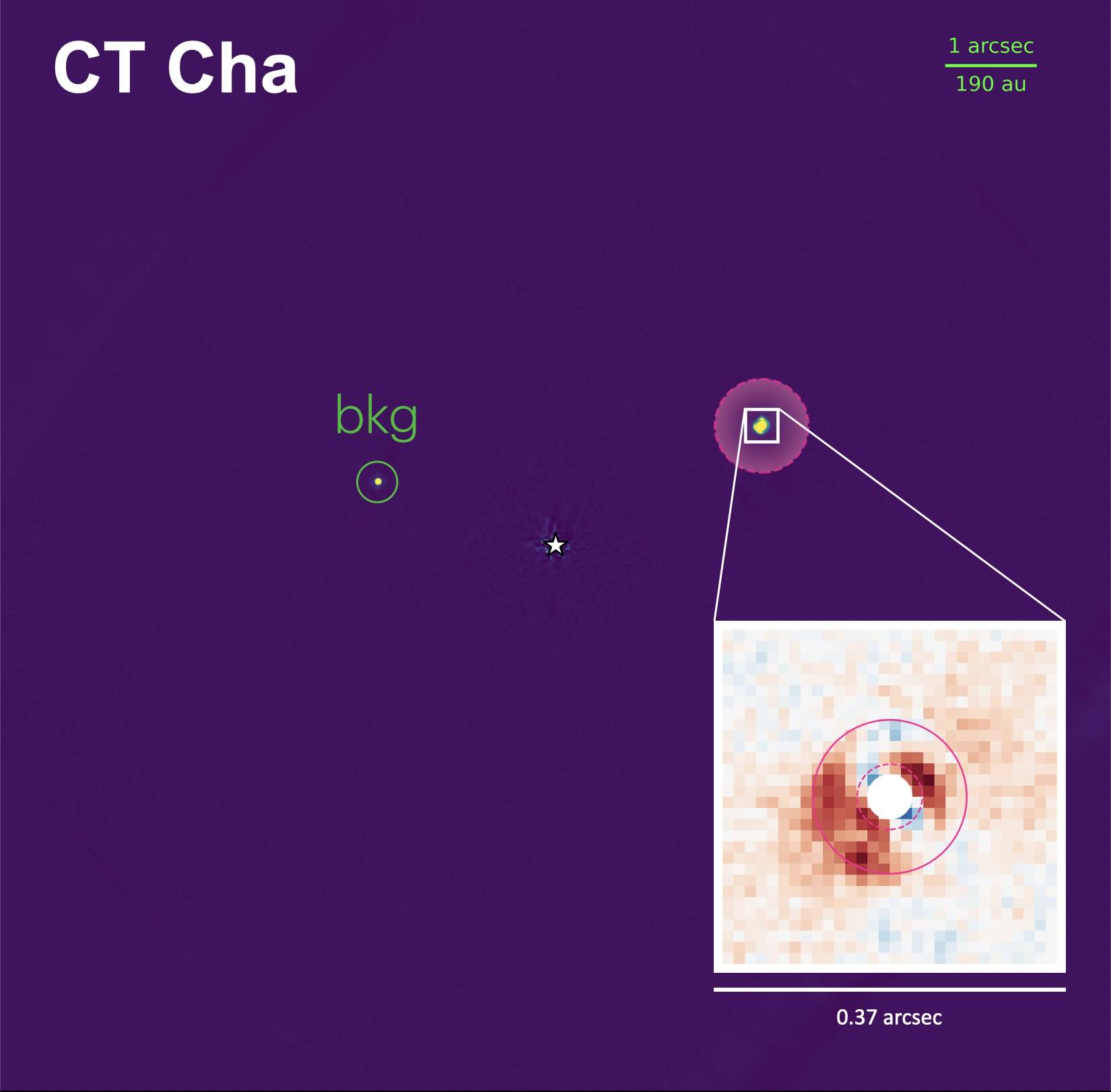}
 
    \end{subfigure}
    \begin{subfigure}{0.31\textwidth}
        \includegraphics[width=\linewidth]{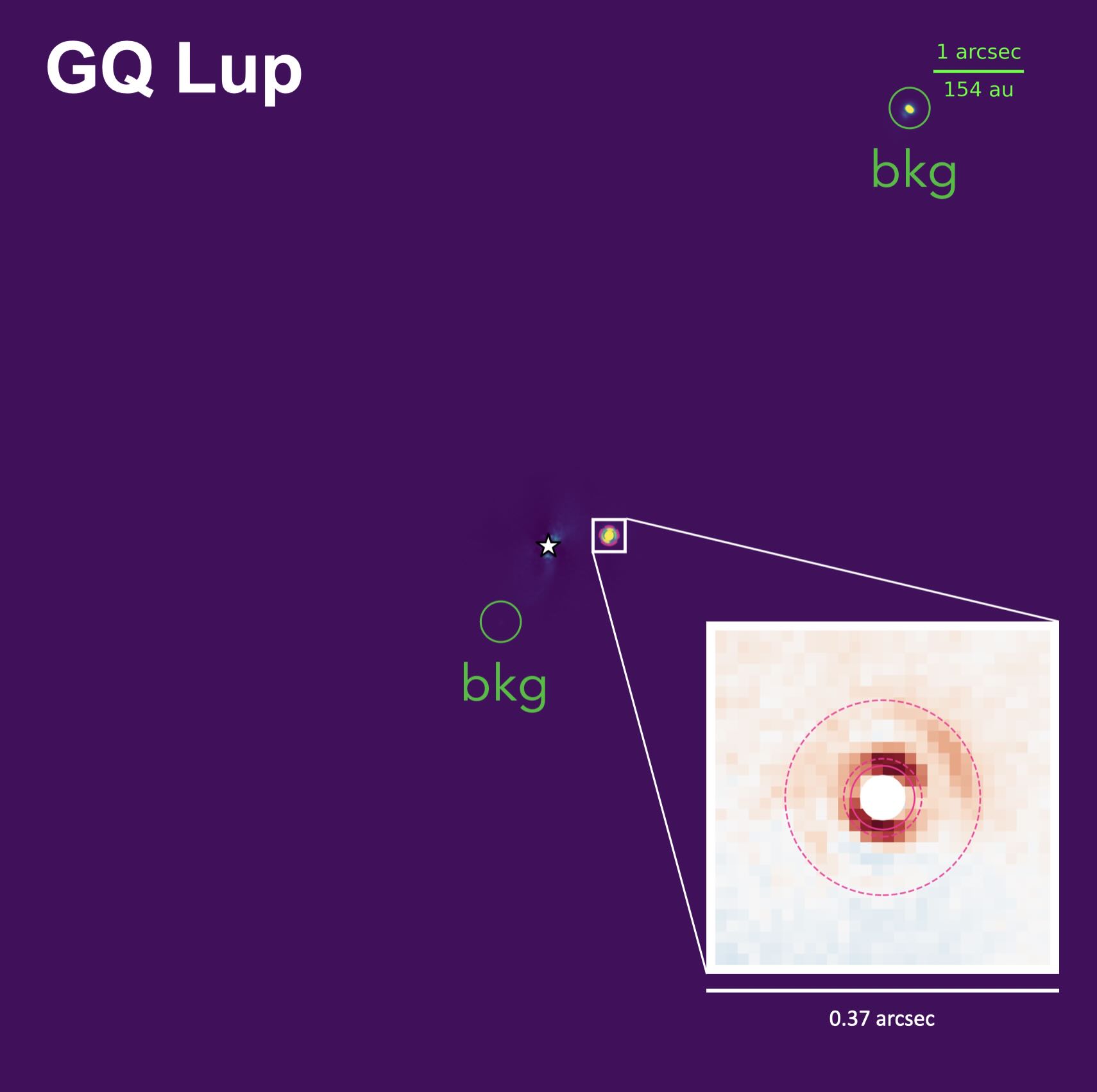}

    \end{subfigure}
    \begin{subfigure}{0.31\textwidth}
        \includegraphics[width=\linewidth]{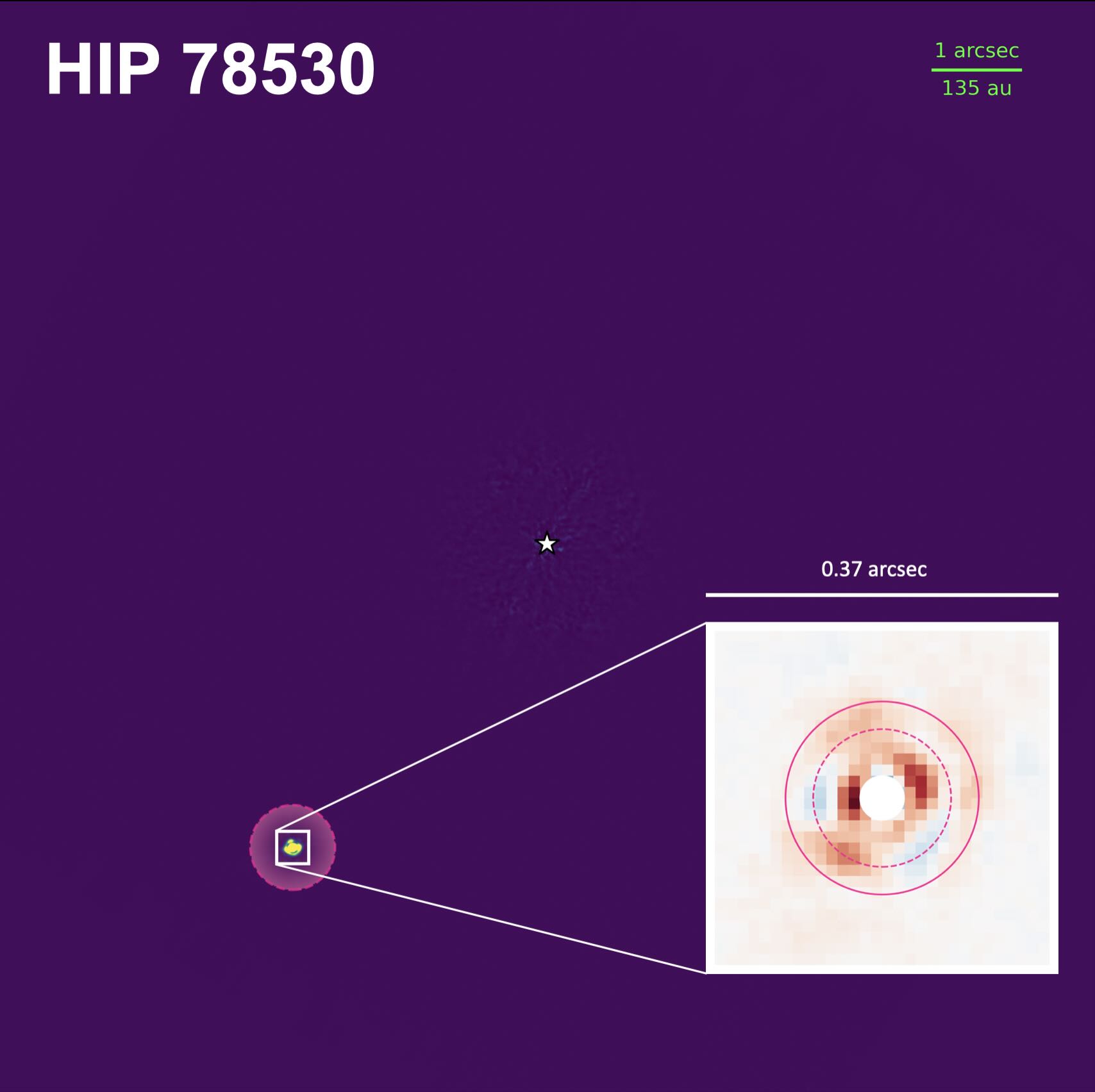}

    \end{subfigure}

    \begin{subfigure}{0.31\textwidth}
        \includegraphics[width=\linewidth]{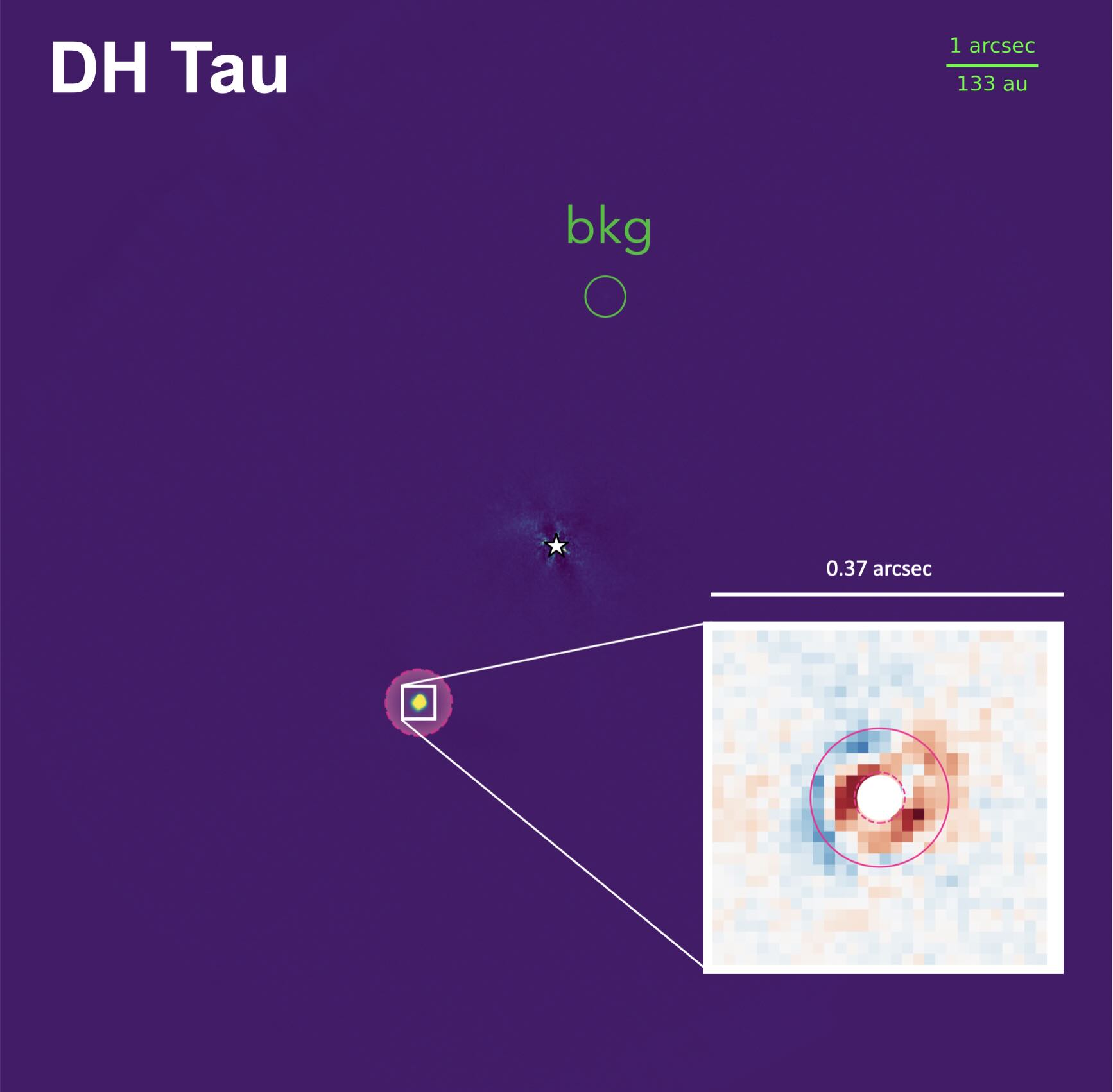}

    \end{subfigure}
    \begin{subfigure}{0.31\textwidth}
        \includegraphics[width=\linewidth]{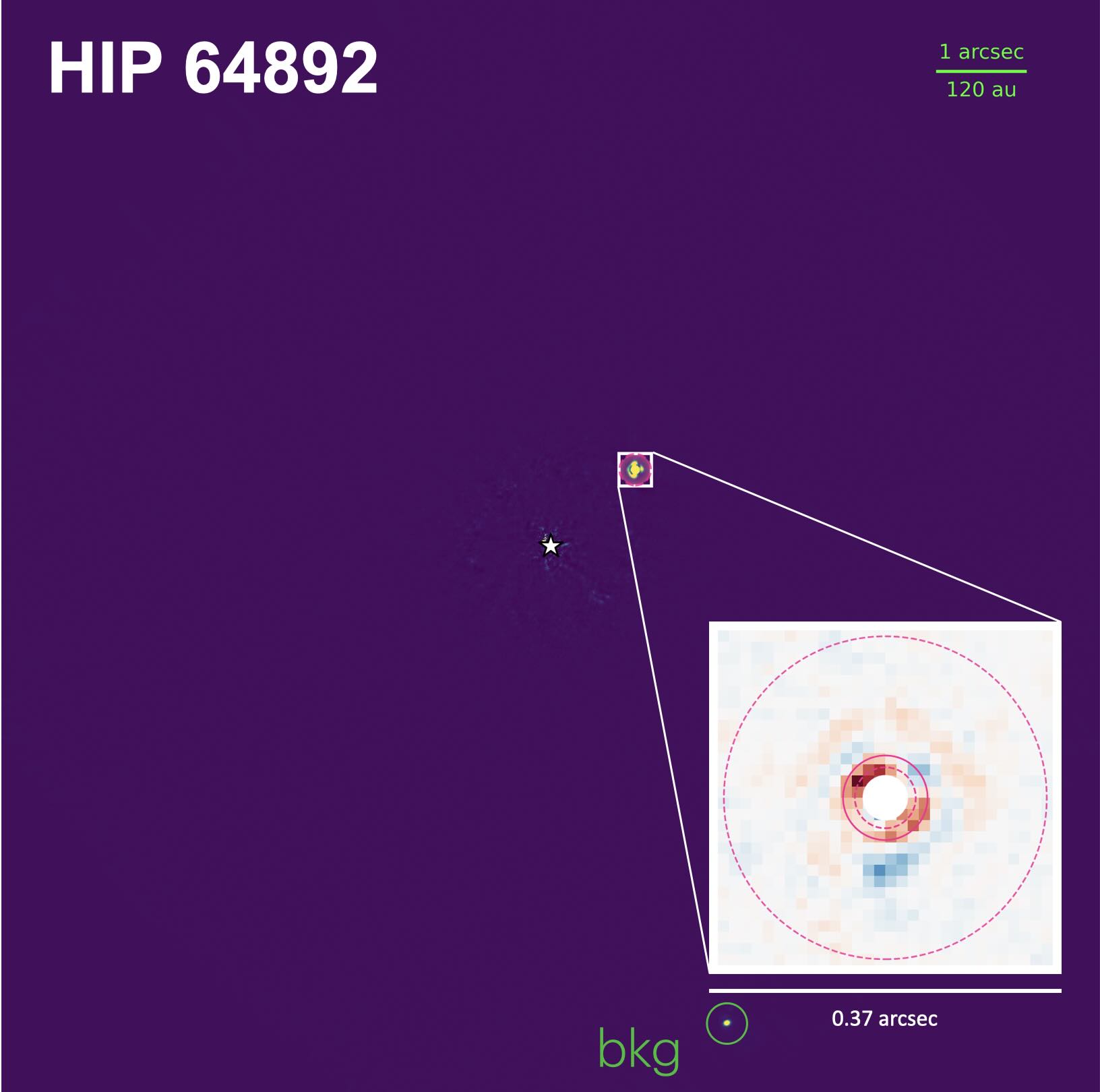}

    \end{subfigure}
    \begin{subfigure}{0.31\textwidth}
        \includegraphics[width=\linewidth]{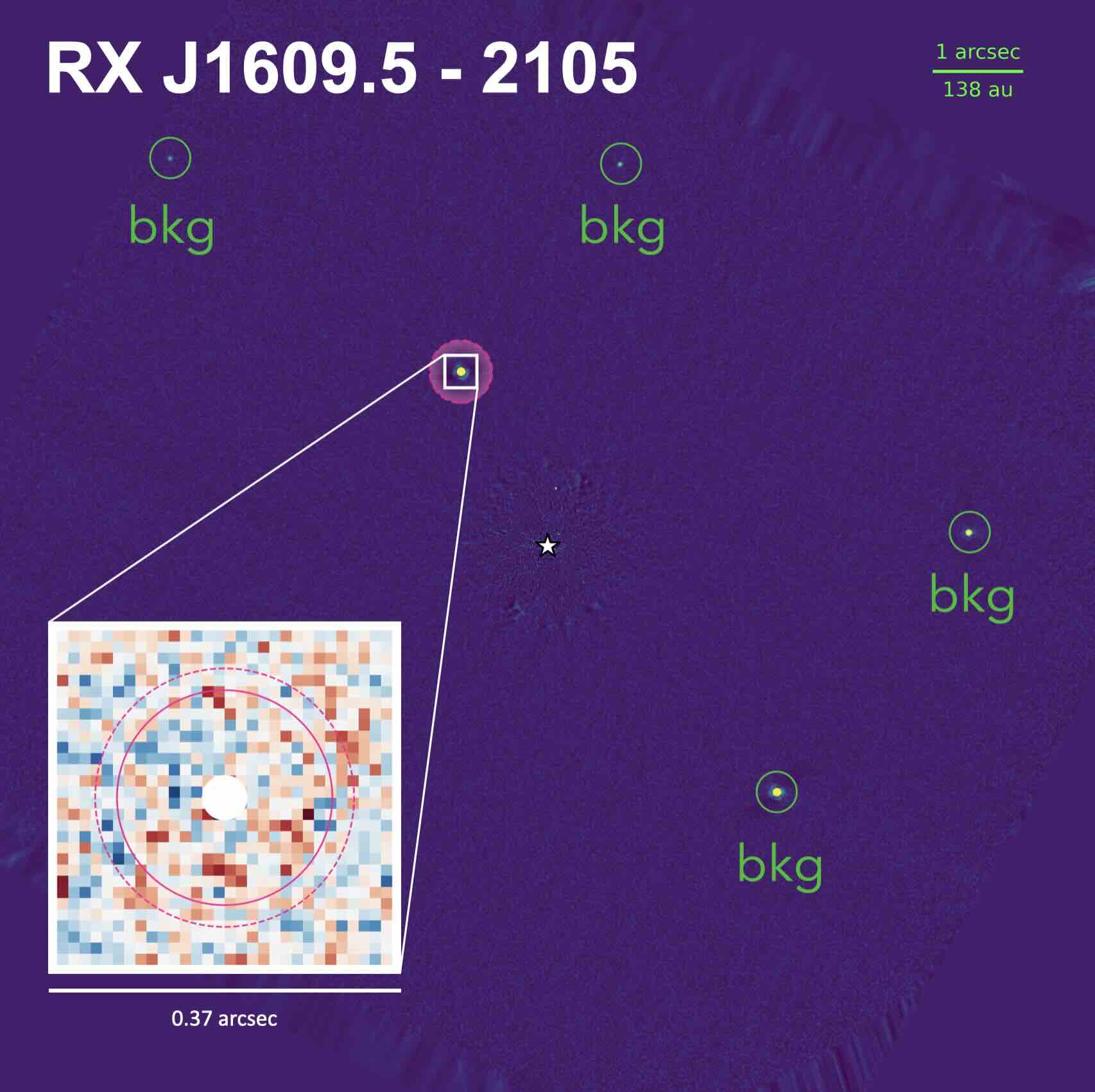}

    \end{subfigure}

    \begin{subfigure}{0.31\textwidth}
        \centering
        \includegraphics[width=\linewidth]{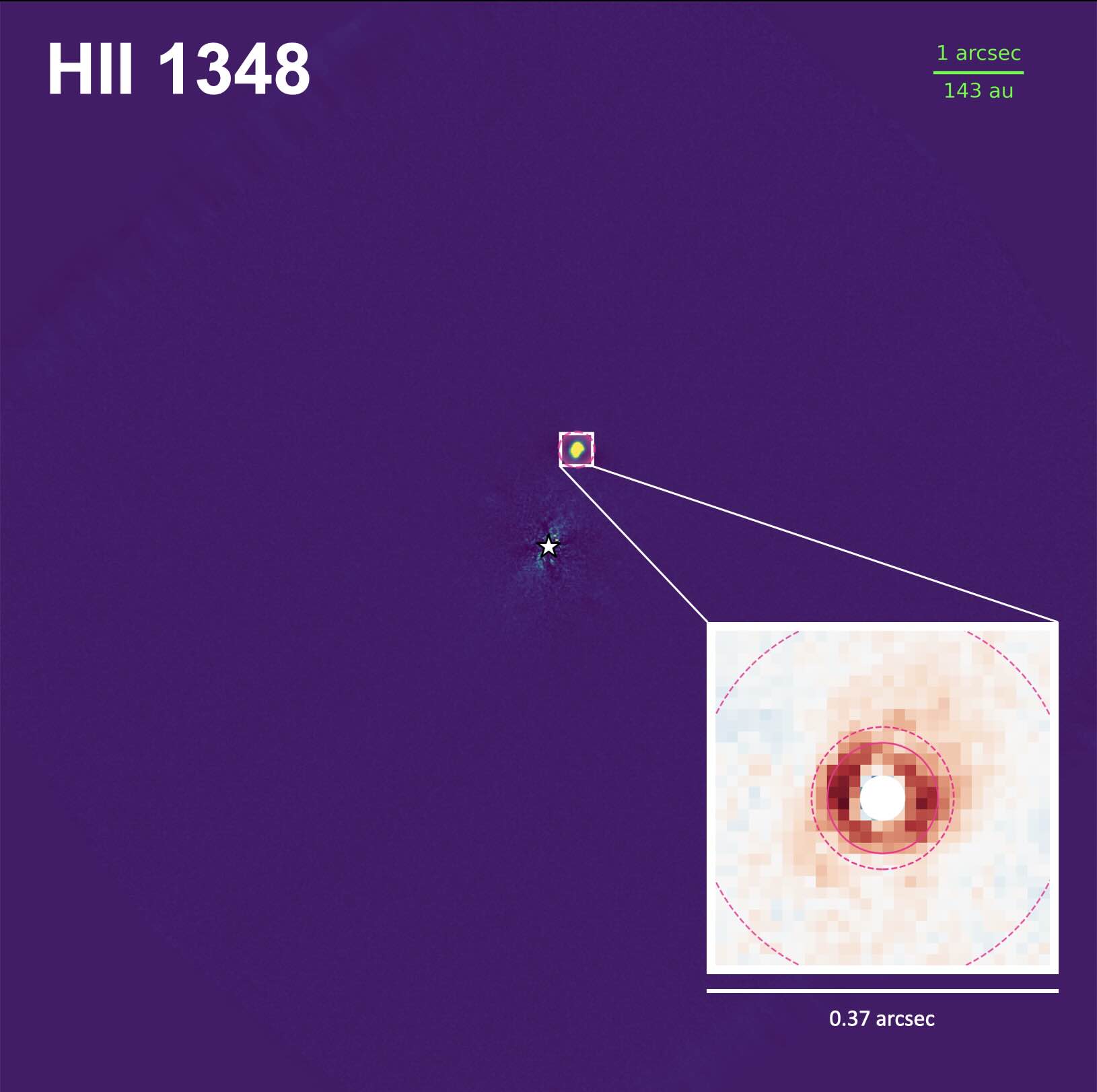}
    \end{subfigure}
    \begin{subfigure}{0.31\textwidth}
        \centering
        \includegraphics[width=\linewidth]{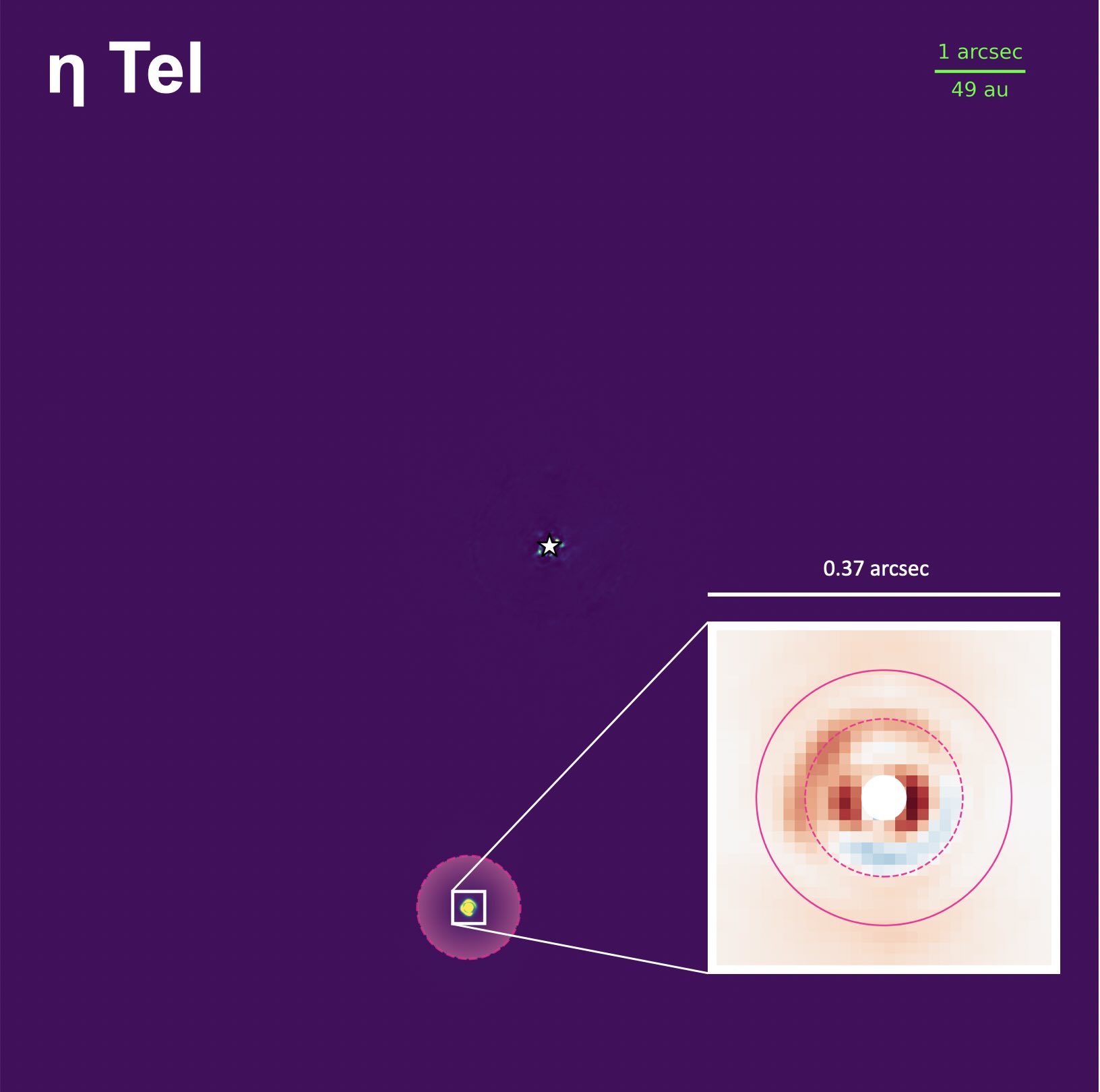}
    \end{subfigure}
    \begin{subfigure}{0.31\textwidth}
        \centering
        \includegraphics[width=\linewidth]{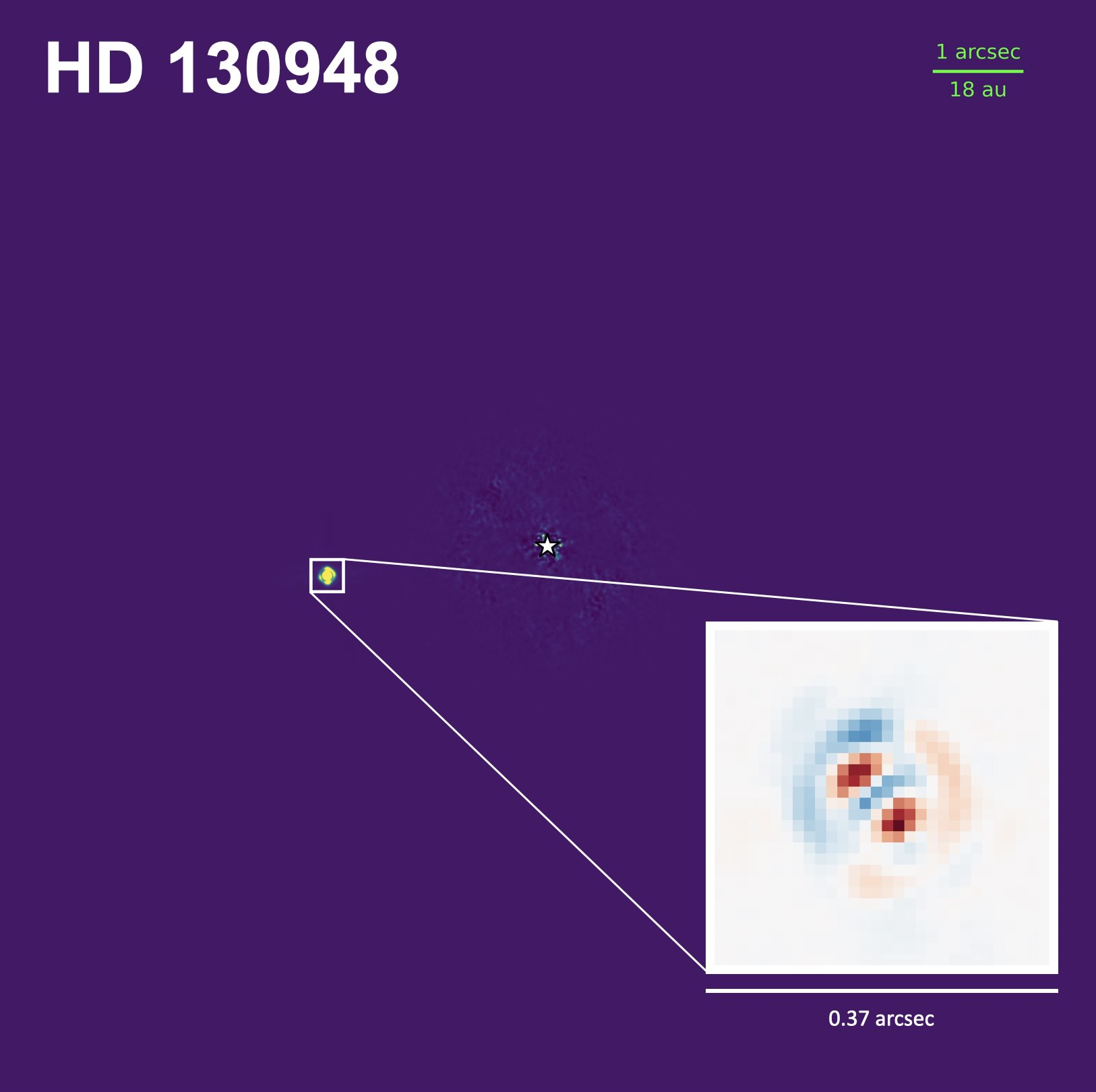}

    \end{subfigure}

    \caption{Final ADI-processed images of each system and the residual outcome of the PSF subtraction. For each target, the large panel shows the full field of view, while the inset presents a $30\times30$ pixel zoom on the subtracted companion and its residuals. Positive residuals are shown in red while negative ones are in blue. Detected background sources are marked in green. Pink shading denotes the range of possible Hill radii, with dashed lines for the extrema and a solid circle for half of the mean Hill radius.}
    \label{mosaic}
\end{figure*}

Figure~\ref{mosaic} presents, for each system, the final ADI-processed image from the datacube, along with a $30\times30$ pixel zoom-in of the residuals centered on the subtracted planet or brown dwarf. In a few cases, other sources were detected in the field of view and encircled in green in the ADI image. In particular, we found one source for {TYC 7084-794-1,} DH Tau, CT Cha, and HIP64892, two sources for GQ Lup and four further sources for RX J1609.5-2105. We verified that all these sources are compatible with background objects when considering the proper motion of the star and their previous positions in archival images. {The astrometry for each background source is given in Table \ref{bgastro}.}
The full span of the possible Hill radii, depending on the uncertainties of the mass and orbital parameters for the companions, is shown as a pink-shaded annulus, with maximum and minimum Hill radii depicted as pink dashed lines. In addition, the mean $R_{Hill}/2$ is highlighted with a solid pink circle, serving as a reference outer boundary for potential bound structures. Finally, in the residual image, with the exception of HD 130948, we masked the central region corresponding to one FWHM, as this inner area does not allow us to resolve, or only marginally resolve, additional sources.

For systems for which previous SPHERE/IRDIS observations were available, we also compared the new residuals with older archival data to see if similar structures emerged after the subtraction of the PSF. For most systems, the residuals obtained at different epochs significantly differ from each other. For { this reason}, we can thus conclude that any excess in the final images comes from differences between the PSF of the substellar companion and the model used. However, some consistencies were derived for three systems, TYC 8047-232-1 B, CT Cha b, and DH Tau b, for which we give a full analysis in the dedicated sections. As a final remark, we note that most substellar companions' PSFs are intrinsically different in the innermost regions from {any} stellar model used, {either the host or reference star}. This fact reflects our difficulties in exploring the closest accessible regions around these objects, leaving an oversubtraction in correspondence with the peak of the PSF and bright artifacts in its surroundings. We stress that the negative residuals are related to the aforementioned PSF–model mismatches and not to ADI self-subtraction wings. 

\begin{figure*}
 \includegraphics[width=0.5\linewidth]{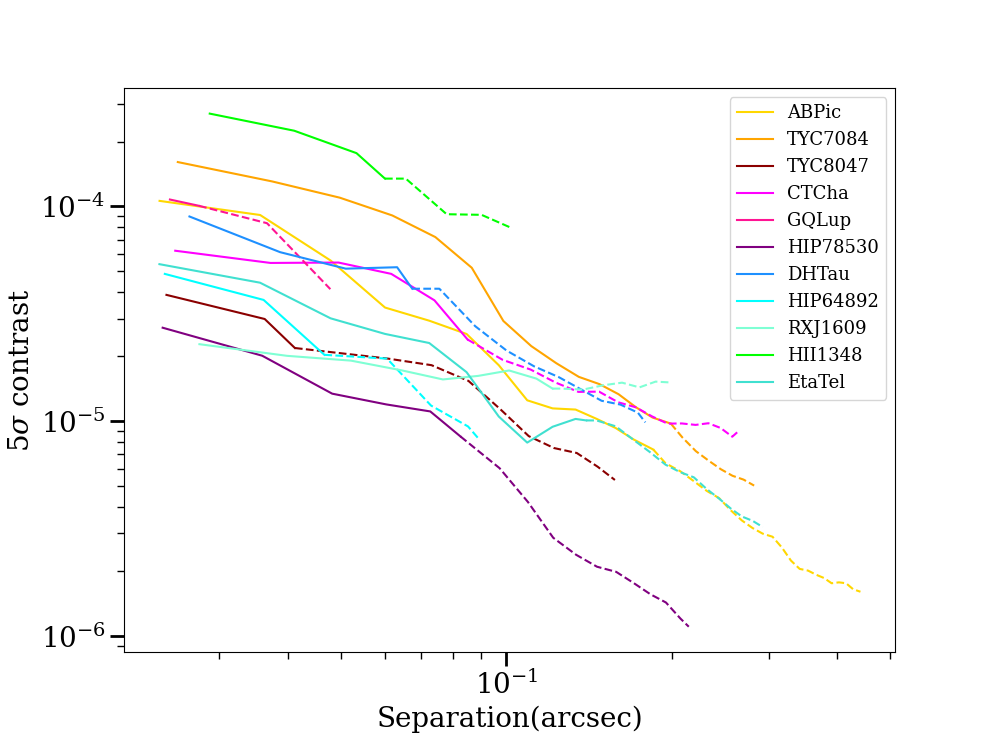}
 \includegraphics[width=0.5\linewidth]{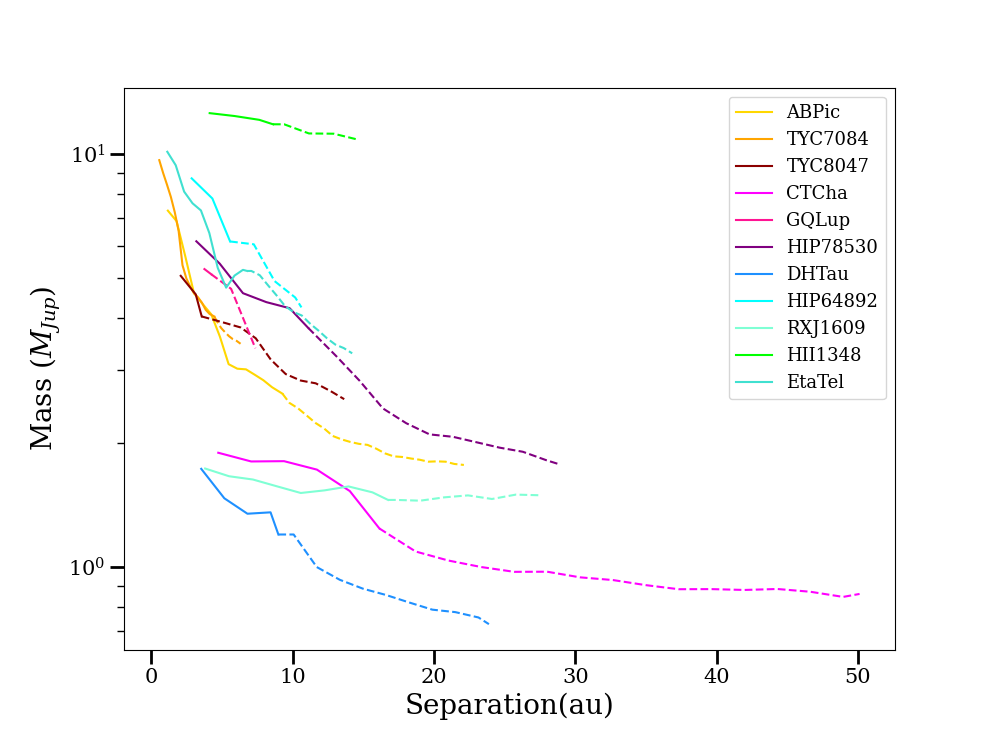}
 \caption{Left: 5$\sigma$ contrast curves {in H2-band} expressed in arcseconds versus contrast relative to the host star, {truncated at half mean (solid curves) and half maximum (dashed curves) Hill radii for dynamical stability}. Right: corresponding detection limits converted into mass using ATMO evolutionary models.}
    \label{cc_mass}

\end{figure*}

The contrast curves derived around each substellar companion {in the H2-band}, expressed in arcseconds versus contrast with respect to the central star, are shown in the left panel of Figure \ref{cc_mass}. The right panel instead shows the detection limits, obtained by converting each contrast curve into mass using the mean ages and distances listed in Table \ref{stars} and the ATMO evolutionary models \citep{Phillips2020}. 

Moreover, each curve was truncated at half of the mean Hill radius of the planet/brown dwarf in order to account only for satellites on dynamically stable orbits. {However, given the significant uncertainties affecting the orbital parameters of these companions, which translate into a wide range of plausible Hill radius values, we also show in the plot the maximum extent of half Hill radius. In Figure \ref{cc_mass}, the mean Hill radius is indicated by solid lines, while its maximum value is represented by dashed lines.}

\begin{table}
\centering
\caption{Stellar parameters of the systems in the sample: distance, age, stellar mass, {and apparent H magnitude} for each host star.}
\begin{tabular}{lcccc}
\hline\hline
  Name  &	Distance     &    Age   & Mass  & {Mag H} \\
        &  (pc)        &   (Myrs)   &  (M\textsubscript{$\odot$})     \\  
\hline \\
AB Pic	        &  50.14  & $28\pm 11$               &      0.8  & {7.088} \\[0.75ex]
TYC 7084-794-1  &  22.36  & $137\pm34$               &      0.4  & {7.28}\\[0.75ex]
TYC 8047-232-1  &  86.63  & $36\pm8$                 &  0.8-0.9  & {8.53}\\[0.75ex]
CT Cha          & 189.95  & $1.41^{+0.38}_{-0.30}$   &      0.8  & {8.944}\\[0.75ex]
GQ Lup          & 154.10  & 2-5                      &     1.03  & {7.702}\\[0.75ex]
HIP 78530       & 134.70  & $11\pm3$                 &      2.5  & {6.946}\\[0.75ex]
DH Tau          & 133.45  & $1.4\pm0.1$              &     0.41  & {8.824}\\[0.75ex]
HIP 64892       & 119.62  & $16\pm2$                 &     2.35  & {6.879}\\[0.75ex]
RX J1609.5-2105 & 138.04  & 5                        & 0.68-0.77 & {9.121}\\[0.75ex]
HII 1348        & 143.29  &  $112\pm5$               &     1.22  & {9.83}\\[0.75ex]
{$\eta$ Tel} & {48.54} & {24$\pm$ 5} & {2.09} & {5.15} \\[0.75ex]
HD 130948       &  18.20  & $550^{+250}_{-150}$      &           & \\[0.75ex]
\hline
\end{tabular}
\label{stars}
\end{table}

\begin{table*}
\centering
\caption{Properties of the substellar companions: mass, separation, position angle, orbital semi-major axis, eccentricity, and derived Hill radius. With the exception of HD 130948 BC, for which we adopted the parameters found in the literature (see Section~\ref{s:HD130948}), we retrieved masses and positions in this work, whereas semi-major axis and eccentricities were taken from Bernardi et al., {submitted} Hill radii were calculated accordingly.}
\begin{tabular}{lcccccc}
\hline\hline
  Name  & Mass & Separation & Pos Angle & Semi-Major Axis & Eccentricity & Hill Radius  \\
        &  (M\textsubscript{Jup})  & (mas)          &  ($^{\circ}$)   & (au)& &(au)    \\  
\hline \\
AB Pic b	&  $12.9^{+0.3}_{{-1.0}}$  & $5379 \pm 6$ & $175.17 \pm 0.06$ &    ${{242}_{-56}^{+109}}$   &   ${0.54}_{-0.30}^{+0.32}$ &    ${19.2}^{+27.2}_{-14.8}$   \\[0.75ex]
TYC 7084-794-1 B & ${33.5}^{+3.8}_{-4.5}$  &  $2777 \pm {4}$ 	&  $239.02 \pm {0.08}$   & ${204}_{-66}^{+48}$  &  ${0.912}_{-0.029}^{+0.015}$ & ${5.4}^{+3.7}_{-2.5}$\\[0.75ex]
TYC 8047-232-1 B & $13.1^{+0.2}_{-0.3}$ &  $3166 \pm {4}$ & $358.91 \pm {0.07}$  & ${264}_{-40}^{+81}$ & ${0.82}_{-0.26}^{+0.10}$ & ${8.1}^{+17.8}_{-5.1}$ \\[0.75ex]
CT Cha b & $9.7^{+1.2}_{{-1.1}}$ &  $2688 \pm {4}$  &  $300.10 \pm {0.08}$  &  ${{500}}_{{-150}}^{{+320}}$ & ${0.59}_{-0.34}^{+0.28}$ & ${32.1}^{{+68.1}}_{{-25.3}}$\\[0.75ex]
GQ Lup B & ${30.1}^{{+1.1}}_{{-1.2}}$ & $706 \pm {3}$  &  $279.6 \pm {0.2}$  & ${93}_{-13}^{+15}$ & ${0.45}_{-0.17}^{+0.15}$ & ${ 10.8}^{{+5.7}}_{{-4.2}}$ \\[0.75ex]
HIP 78530 B & ${19.1}^{{+1.4}}_{{-0.6}}$ & $4525 \pm 5	$  & $140.08 \pm {0.07}$  & ${472}_{-88}^{+206}$ & $0.55\pm0.25$ & ${28.6}^{+36.7}_{-18.4}$\\[0.75ex]
DH Tau b & {11.3}$\pm{2.2}$ & ${2346} \pm 3$  & $138.74 \pm {0.08}$  &  ${{234}}_{{-41}}^{{+100}}$ & ${0.58}_{-0.26}^{+0.32}$ & ${0.3}^{+29.4}_{-16.6}$\\[0.75ex]
&  &  ${2347} \pm 3$ &  ${138.73} \pm {0.08}$ &  &  \\[0.75ex]
HIP 64892 B & ${40.7}^{{+4.0}}_{{-4.7}}$ & $1298 \pm {3}	$ &  ${312.0} \pm {0.1}$  & ${{128}}_{{-31}}^{{+43}}$ & ${{0.51}}_{{-0.19}}^{{+0.25}}$ & ${11.1}^{{+10.1}}_{{-7.1}}$\\[0.75ex]
RX J1609.5-2105 b & ${6.6}\pm{0.3}$ &  $2219 \pm {4}$	&  ${26.4} \pm {0.1}$  & ${305}_{-51}^{+74}$ & ${0.25}_{-0.15}^{+0.20}$ & ${32.6}^{+16.7}_{-13.0}$ \\[0.75ex]
HII 1348 B & ${53.3}^{{+3.7}}_{{-4.0}}$ & $1154 \pm {3}$ &  $343.9 \pm 0.1$  & ${{181}}_{{-32}}^{{+49}}$ & ${{0.60}}_{{-0.12}}^{{+0.08}}$ & ${17.4}^{{+12.0}}_{{-6.2}}$\\[0.75ex]
{$\eta$ Tel B} & ${54.2}^{{+6.3}}_{{-7.3}}$ & ${4223}\pm{5}$ & ${167.43}\pm{0.07}$ & ${143}^{{+43}}_{{-22}}$ & {${0.53}^{{+0.29}}_{{-0.26}}$} & ${13.6}^{{+14.9}}_{{-9.4}}$ \\ [0.75ex]

HD 130948 BC*  & $114$ & &  & 30-3700 & 0-0.96 & \\[0.75ex]
\hline
\end{tabular}
\label{comps}
\end{table*}

\subsection{AB Pic}
AB Pic is a young K1V star located at a distance of $50.14$ pc \citep{GAIA2021} and is a likely member of the Carina association, with an estimated age of $28\pm11$ Myrs \citep{Gratton2024} and a mass of 0.8 M\textsubscript{$\odot$} \citep{Chauvin2005b}. Its substellar companion, AB Pic b, was first discovered by \citet{Chauvin2005b} through high-contrast imaging observations with VLT/NACO at a separation of $\sim 5.5''$ and was initially estimated to be near the deuterium-burning limit ($\sim 13 - 14$ M\textsubscript{Jup}). More recent studies, however, place it within the planetary-mass regime, with an estimated mass of $10\pm1$ M\textsubscript{Jup} \citep{Palma-Bifani2023}. Orbital fitting based on astrometric data from NaCo and SPHERE suggests a semi-major axis of $190^{+200}_{-50}$ au, a near edge-on inclination, and a poorly constrained eccentricity \citep{Palma-Bifani2023}.
As presented in Bernardi et al., {submitted}, including this epoch in the orbital fitting of AB Pic b, the baseline used by \citet{Palma-Bifani2023} is extended by $\sim 8$ years. They obtain possible semi-major axes in the range $[186,351]$ au and eccentricities in the range $[0.24,0.86]$, with peaks at 242 au and 0.54, respectively, in agreement with previous orbital estimates \citep{Palma-Bifani2023}. The Hill radius, thus, spans a variety of values, from 4.4 to 46.4 au, with a mean value around 19.2 au.

Although AB Pic b resides at a wide orbital separation and is an ideal candidate for hosting circumplanetary material, ALMA 1.3 mm observations have not detected any associated dust emission, placing an upper limit on the protolunar disk mass of $5\times10^{-6}$ M\textsubscript{Jup}, or $2.4\times10^{-6}$ M\textsubscript{Jup} under the assumption of a compact disk \citep{Perez2019}. This non-detection may indicate rapid dissipation of the circumplanetary disk or that the system has already reached an advanced stage of satellite formation. Additionally, polarimetric observations in the near-infrared with SPHERE/IRDIS did not detect any significant polarized signal from AB Pic b, yielding upper limits on the polarized-to-total intensity ratio and the disk-integrated polarized flux \citep{VanHolstein2021}. These results suggest that if a circumplanetary or circumsubstellar disk exists, it is either extremely faint, lacks sufficient scattering dust grains, or is viewed at an unfavorable geometry. Combined, these constraints from ALMA and polarimetry point toward a low dust content around AB Pic b at the current epoch.

After the subtraction of the substellar companion, no further features are detected in the residuals (see Figure \ref{mosaic}). Detection limits for satellites were estimated to be 7.3 M\textsubscript{Jup} at the closest separation allowed by the dimension of the PSF (1.2 au) and can reach 2.6  M\textsubscript{Jup} at half of the mean Hill radius.

\subsection{TYC 7084-794-1}
\label{s:TYC}
TYC 7084-794-1 is a nearby M1-type star located at a distance of $\sim 22.36$ pc \citep{GAIA2021} and is a confirmed member of the AB Doradus moving group, implying an age of $137\pm34$ { Myr} \citep{Gratton2024}. Its wide substellar companion, TYC 7084-794-1 B, was discovered through high-contrast imaging with Gemini NICI at a projected separation of $3.14\arcsec$ by \citet{Wahhaj2011}. They estimated the mass of the star to be 0.4 $M_{\odot}$ and a mass for the companion of $31\pm8$ M\textsubscript{Jup}, placing the object within the brown dwarf regime.

Multiple epochs of high-precision astrometry obtained over a decade have confirmed common proper motion and enabled constraints on the orbit of TYC 7084-794-1 B. Although the orbital arc covered remains short, orbit fitting using Monte Carlo techniques yields an estimated semi-major axis of $241^{+85}_{-130}$ au and a high eccentricity of $\sim 0.94$ \citep{Bowler2020}. Since the last epoch included in the \citet{Bowler2020} is from 2018-01-30, we extended the orbital baseline by more than 5 years. The estimates of semi-major axis and eccentricity presented in Bernardi et al., {submitted}, are in good agreement with previous results, with $a_p=204^{+48}_{-66}$ { au} and $e_p=0.912^{+0.015}_{-0.029}$. The Hill radius calculated for TYC 7084-794-1 B is restricted to a few au from the brown dwarf, going from a minimum of 2.9 au to a maximum of 9.1 au.

No significant infrared excess has been detected in the system, and polarimetric observations with SPHERE-IRDIS place upper limits at $0.1-0.3\%$ on the degree of linear polarization of the companion \citep{VanHolstein2021}, disfavoring the presence of a prominent disk around TYC-7084-793-1 B. In our data, once the signal from the substellar companion is subtracted, no additional point or extended sources are detected in the residuals (see Figure \ref{mosaic}). We therefore derived detection limits for potential satellites, which are restricted to 9.7-4.0 M\textsubscript{Jup} at 0.6-4.4 au, respectively.

\subsection{TYC 8047-232-1}
\label{s:TYC8047}

TYC 8047-232-1 is a young K3V-type star \citep{Torres2006} of 0.8-0.9 $M_{\odot}$ \citep{Baraffe1998} located at a distance of $86.63\pm$ pc \citealp{GAIA2021} and a likely member of the Columba Moving Group, implying an estimated age of approximately $36\pm8$ { Myr}  \citep{Gratton2024}.
The system was claimed to host a brown dwarf companion by \cite{Neuhauser2003} at a projected separation of $3.2$ arcsec and with a mass estimated between 20 and 40 M\textsubscript{Jup}. The companion was subsequently confirmed to share common proper motion with TYC 8047-232-1 \citep{Chauvin2005c}.

Although the full orbit was not yet resolved, \cite{Ginski2014b} derived broad posterior distributions for the orbit of the brown dwarf considering a long-term stable system. Their results are consistent with a semi-major axis of $\sim$196-880 au and an eccentricity distribution favoring eccentric orbits in the range 0.28-0.999, with a peak at 0.997. 
Even if more than 11 years have passed since the last epoch considered by \citet{Ginski2014b} (2012-12-03 to 2024-08-27), the companion moved by a few tens of arcseconds, leaving large uncertainties on the orbital parameters. In Bernardi et al., {submitted}, we found an estimate of a semi-major axis of $264^{+81}_{-40}$ { au} and eccentricity of $0.82^{+0.10}_{-0.26}$, with a Hill radius of $8.1^{+17.8}_{-5.1}$ { au}.

\begin{figure*}
    \centering

    \begin{subfigure}{0.49\textwidth}
        \includegraphics[width=\linewidth]{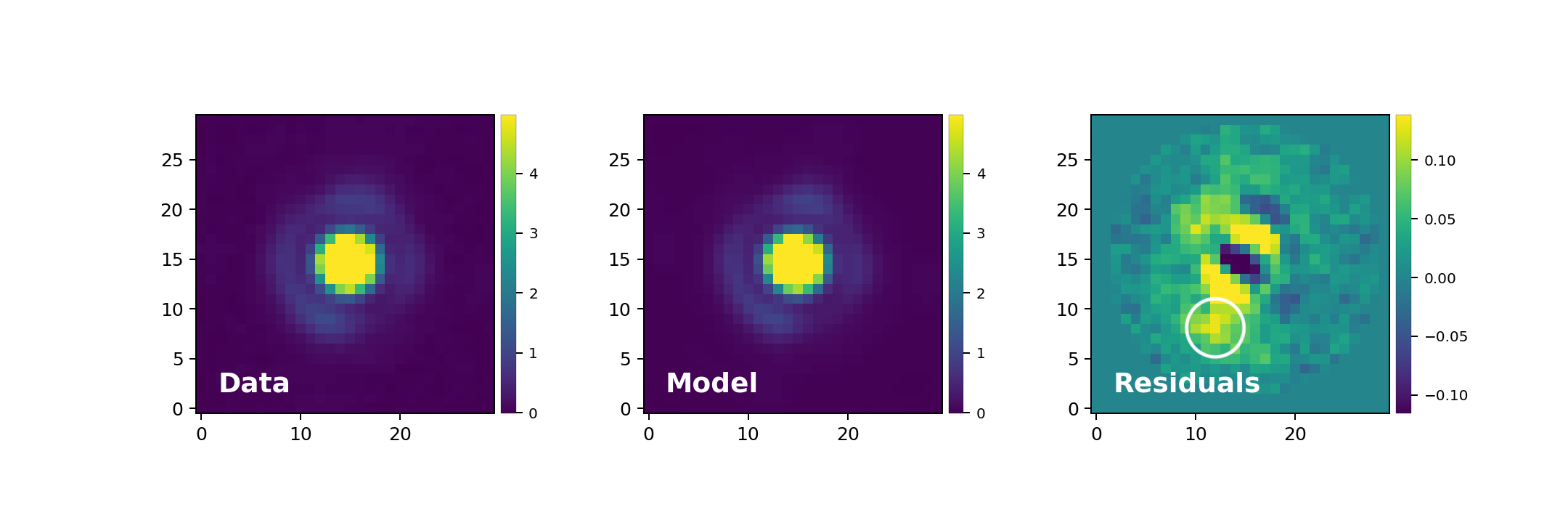}
 
    \end{subfigure}
    \begin{subfigure}{0.49\textwidth}
        \includegraphics[width=\linewidth]{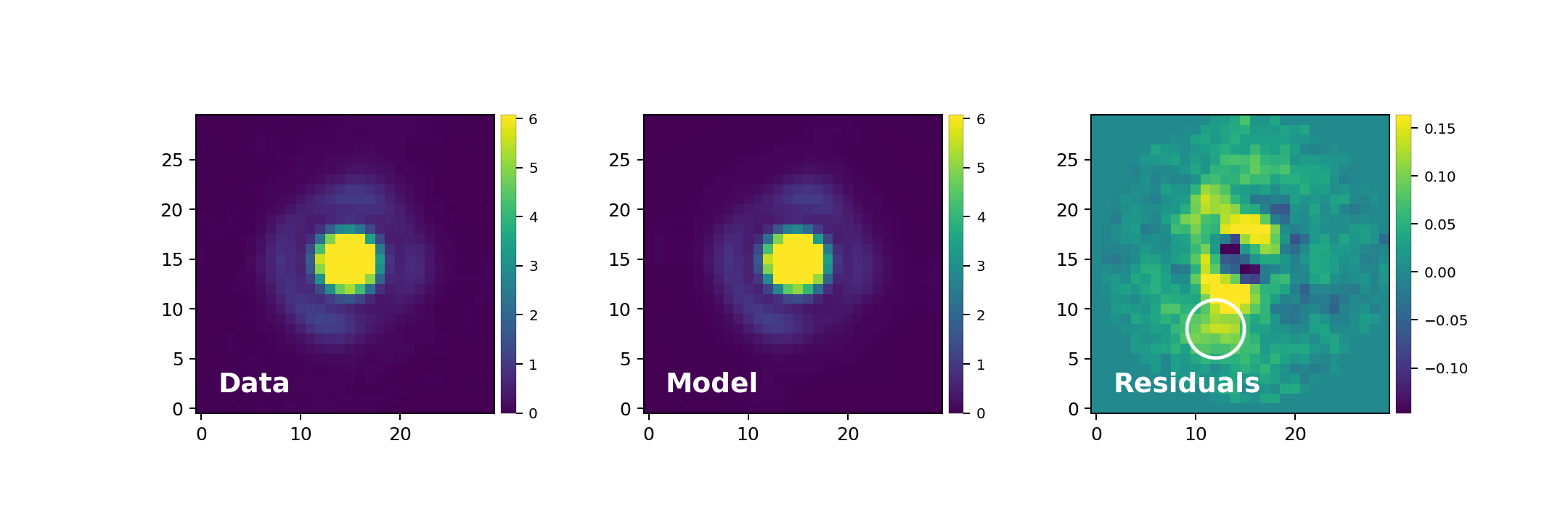}
    \end{subfigure}

    \begin{subfigure}{0.49\textwidth}
        \includegraphics[width=\linewidth]{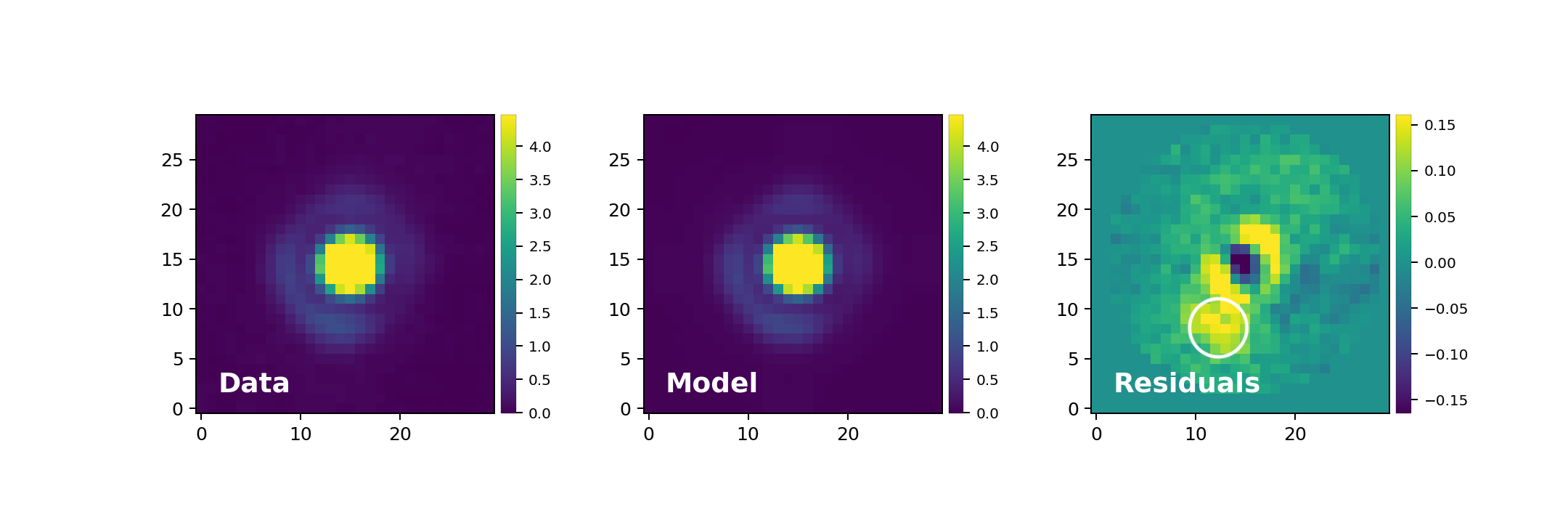}
 
    \end{subfigure}
    \begin{subfigure}{0.49\textwidth}
        \includegraphics[width=\linewidth]{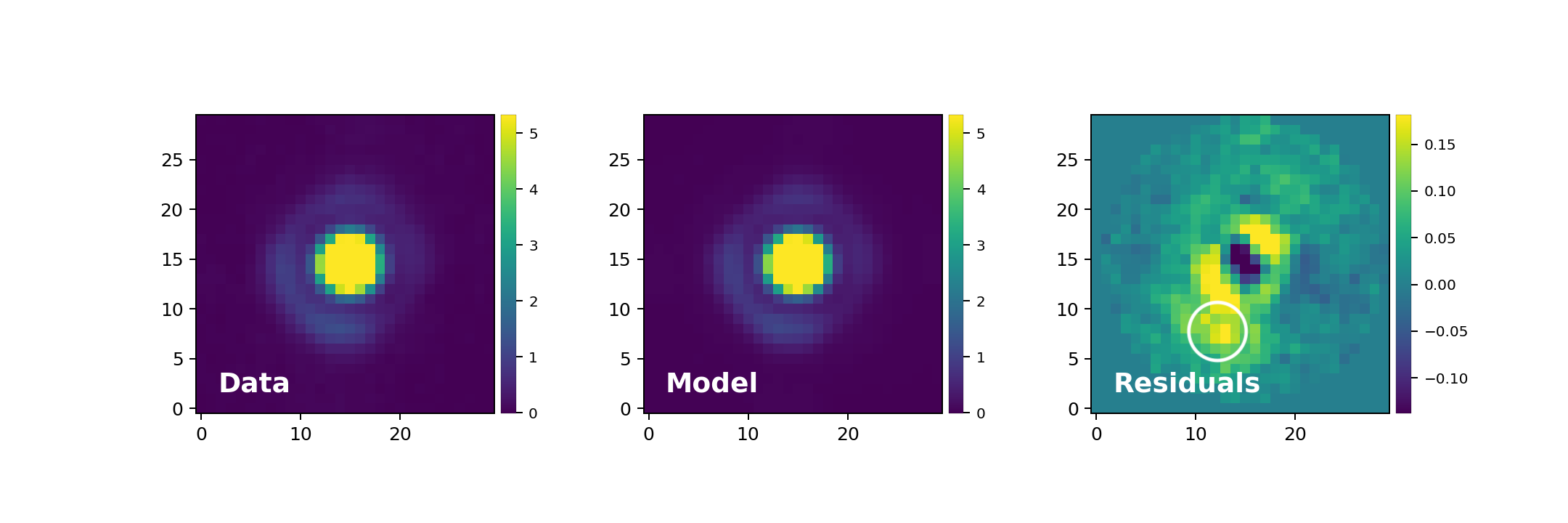}
    \end{subfigure}

    \caption{Residuals around TYC 8047-232-1 B after PSF subtraction for 2023-10-11 (top row) and 2024-08-27 (bottom raw) epochs and for H2 (left side) and H3 (right side) filters. Each set of three panels shows, from left to right, the substellar companion, the model PSF, and the residuals, highlighting a recurrent excess feature with a white circle.}
    \label{TYC8047}
\end{figure*}

Regarding putative features bound to this substellar object, SPHERE-IRDIS polarimetric measurements constrain the linear polarization of the companion to below 0.2–0.3\% \citep{VanHolstein2021}, making the presence of a prominent disk around TYC-8047-232-1 B unlikely.\\
For this target, we also include a further observation carried out at the very beginning of this program (2023-10-01, proposal ID 111.24UH.001) for which the reference star selected was actually a spectroscopic binary too wide to be used for the NEGFC technique. In this case, we only used the two images of the central star gathered before and after the coronagraphic sequence to model and subtract the PSF of the brown dwarf.
Interestingly, we detected similar residuals in both observing epochs, 2023-10-01 and 2024-08-27, as shown in Figure \ref{TYC8047}. Each sub-figure is composed of three panels, showing (from left to right) the PSF of the substellar companion, the model PSF, and the residuals after subtraction. The top and bottom rows correspond to the first (2023-10-01) and second (2024-08-27) epoch, respectively, while the left and right columns show the H2 and H3 filters.

The excess emission is located at a projected separation of $\sim$85 mas and position angle of $159^{\circ}$ from the brown dwarf and it is consistently visible in both epochs and in both filters (see white circles in Figure \ref{TYC8047}. Moreover, we retrieved a S/N map, using the function \texttt{snrmap()} provided by VIP, and centering it on the substellar companion so that the noise is evaluated on the residuals left from the subtraction. As a result, we obtained for the excess a S/N of 3.7 and 4.6 in H2 and 2.7 and 3.3 in H3 for the 2023 and 2024 epochs, respectively). Importantly, the residuals were obtained using either the PSFs of the central star or the reference star for the star-hopping (HD 10626), which strengthens the case for this being a genuine feature bound to TYC-8047-232-1 B. The excess appears slightly extended, particularly in the second epoch and in the H2 filter, suggesting a possible disk component. However, given the polarimetric constraints of \citet{VanHolstein2021}, if a circumsubstellar disk is present, it is unlikely to extend as far as $\sim$7 au, where the residuals are detected.

Alternatively, the residuals may correspond to a further companion orbiting TYC-8047-232-1 B.
{In order to extract the flux of the candidate, we developed a variation of the NEGFC technique that enables the subtraction of two PSFs at the same time, one at the location of the companion and the other at the location of the candidate satellite. The flux and relative errors are estimated by running an MCMC that optimizes the residuals coming from the subtraction.}

\begin{figure}
    \centering
    \includegraphics[width=\linewidth]{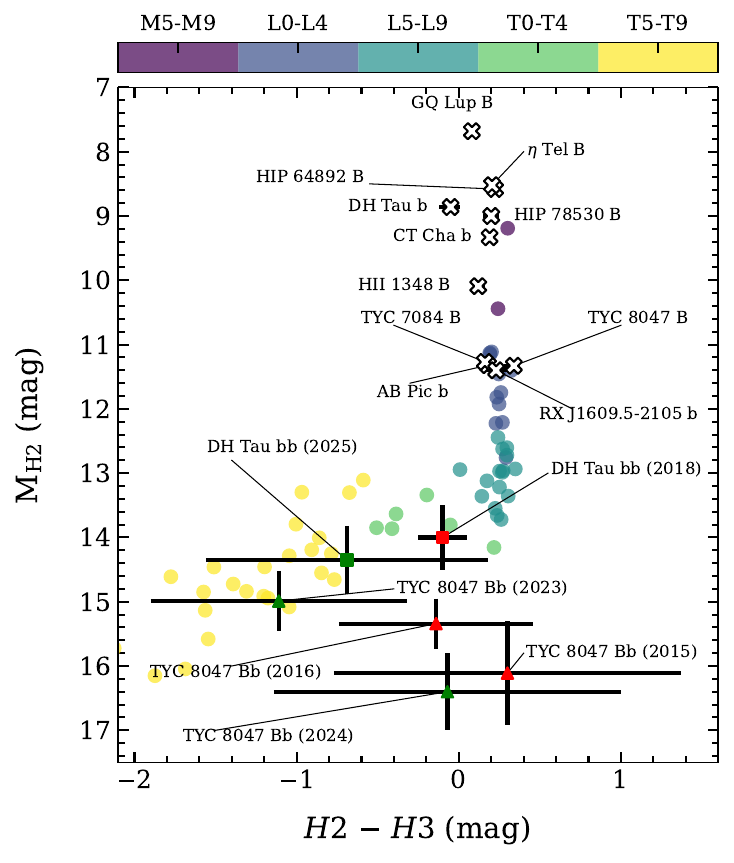}
    \caption{{H2-H3 color magnitude diagram for the objects in our sample, adapted from Bernardi et al submitted. The candidates around DH Tau b and TYC 8047-232-1 B are indicated with squares and triangles, respectively, both for datasets presented in this paper (green) and archival epochs (red).}}
    \label{CMD}
\end{figure}

{As a result, we obtained a} H2-band contrasts of ${1.32}\times10^{-5}\pm 5.7 \times10^{-6}$ (2023-10-01) and ${9.46}\times10^{{-6}}\pm 5.2 \times10^{-6}$ (2024-08-27), which, using the \texttt{atmo} evolutionary models \citep{Phillips2020} and assuming a mean age of 36 Myr, correspond to mean masses of {3.0-3.4} M\textsubscript{Jup}. In H3, we obtain {${1.25}\times10^{-5}\pm 7.2 \times10^{-6}$ for the first epoch and ${8.9}\times10^{{-6}}\pm 7.3 \times10^{-6}$ for the second one, corresponding to mean masses of 6.1-6.5 M\textsubscript{Jup}}. Although the masses derived for the two filters differ by roughly a factor of two, this is consistent with uncertainties introduced by the NEGFC subtraction procedure, the intrinsic faintness of the signal, and the evolutionary models assumed. Moreover, Figure \ref{CMD} shows the color–magnitude diagram of the candidate TYC 8047-232-1 Bb at the two epochs (green triangles) in the available bands (H2–H3 vs $M_{H2}$). The derived color is consistent between the two epochs within the uncertainties, while a discrepancy is observed in the absolute H2 magnitude, with the 2023 measurement appearing systematically brighter. We note, however, that the quoted error bars—despite being relatively large—are likely underestimated, as the flux calculation is performed via MCMC while keeping the positions of both companions fixed and assuming a single model PSF for both companions. A more accurate and self-consistent flux retrieval, allowing for additional degrees of freedom and PSF variability, will be explored in future studies, for instance through the use of neural-network–based approaches.

\begin{figure}
    \centering

        \includegraphics[width=\linewidth]{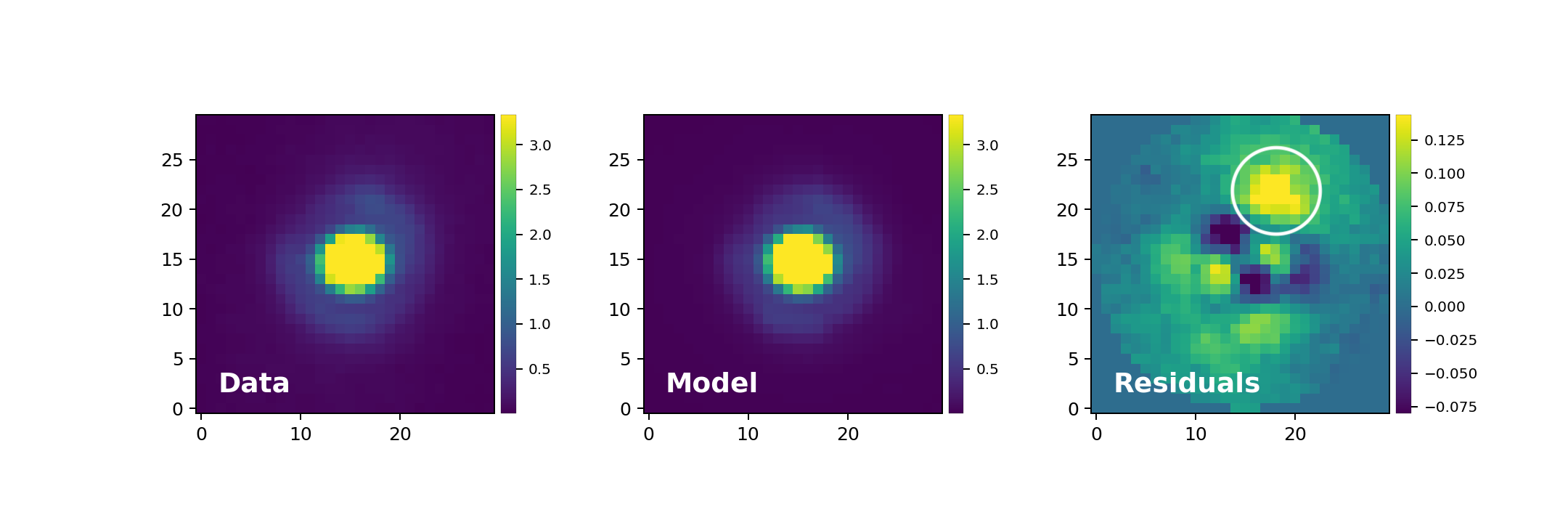}

        \includegraphics[width=\linewidth]{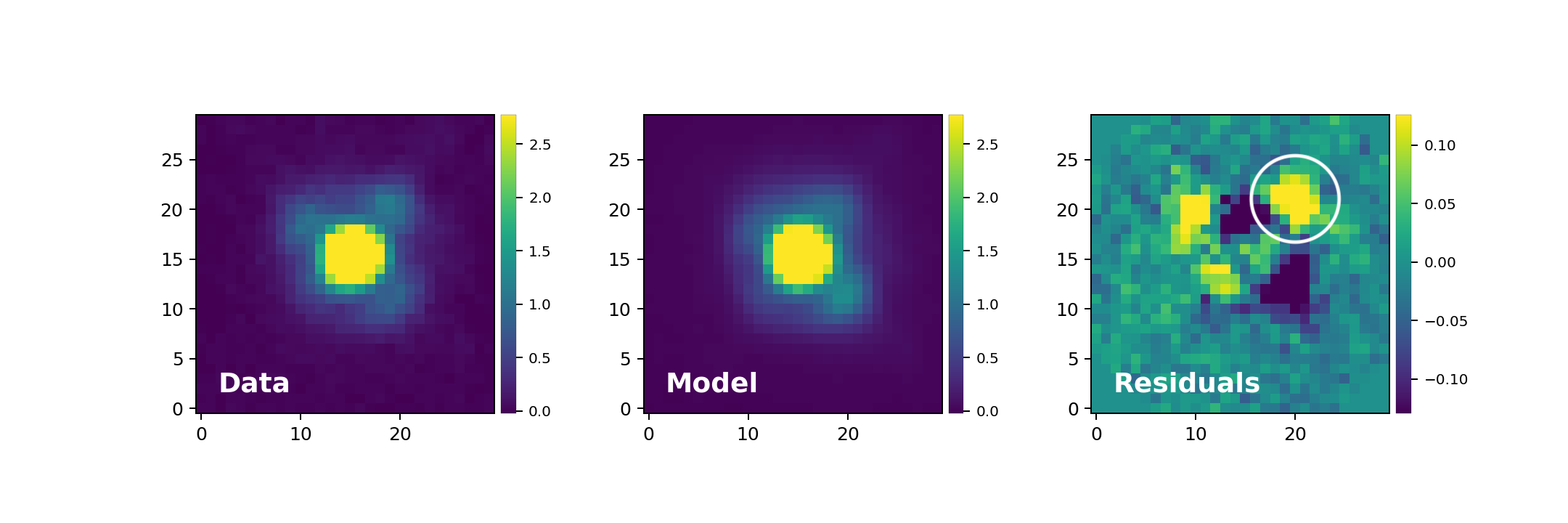}

    \caption{Archival SPHERE/IRDIS observations of TYC 8047-232-1 B from 2015-09-25 (top row) and 2016-01-17 (bottom row) in H2 filter. In each row, from left to right, the substellar companion, the model PSF, and the residuals are displayed. The recurring excess is highlighted with a white circle.}
    \label{TYC8047_old}
\end{figure}

We also note a point source at a S/N in H2 of 6.8 and 4.0 and in H3 of 6.6 and 3.7 in archival SPHERE/IRDIS observations from 2015-09-25 and 2016-01-17, located at {separation and position angle relative to the brown dwarf of $\sim80$ mas and $\sim-23^{\circ}$ for the first epoch and $\sim120$ mas and $\sim-55^{\circ}$ for the second one (see Figure \ref{TYC8047_old})}. The point source was detected in both H2 and H3 filters with contrasts comparable to the ones derived for the candidate in 2023 and 2024 (for 2015 contrast in H2 of $1.24\times10^{-5}\pm6.4\times10^{-6}$ and in H3 of $1.64\times10^{-5}\pm8.8\times10^{-6}$; for 2016 a contrast in H2 of $2.50\times10^{-5}\pm8.9\times10^{-6} $ and in H3 of $2.21\times10^{-5}\pm 9.4\times10^{-6}$). The colors for the tentative point source detected in 2015 and 2016 are plotted in Figure \ref{CMD} (red triangles) and they are broadly consistent with the results obtained in this paper, especially when compared to the 2024 epoch. 

{We note that} both datasets are affected by the low-wind effect \citep{Milli2018}, which is particularly severe in the 2016 data, {and could strongly affect not only the shape of the residuals but also the contrasts retrieved}. Nevertheless, it is useful to assess whether the sources detected in the archival and new datasets could correspond to the same object. 

Based on the system’s proper motion, we can rule out the possibility of a background star. In fact, the source detected in 2015 at $\Delta \mathrm{RA} = 0.0245''$ and $\Delta \mathrm{Dec} = 3.087''$ is not consistent with a background object, since it should have been located in 2023 at $\Delta \mathrm{RA} = 0.425''$ and $\Delta \mathrm{Dec} = 3.170''$, whereas it is instead detected at $\Delta \mathrm{RA} = 0.098''$ and $\Delta \mathrm{Dec} = 3.283''$. If the candidate were bound to TYC 8047-232-1 B, the 2015 and 2016 datasets yields a mass of about $3-4 M_{\mathrm{Jup}}$ (from atmospheric models, assuming an age of 36 Myr). This estimate is consistent with the masses retrieved from the 2023 and 2024 detections. 

Assuming the system is observed face-on and the companion follows a circular orbit, the candidate would have completed a little more than half of its orbit ($\sim 187^{\circ}$ if we consider a clockwise motion) since 2015. From Kepler’s third law, a semi-major axis of $\sim 7$ au and a companion mass of $3$–$6,M_{\mathrm{Jup}}$ imply an orbital period of 130–150 years. Over an 8-year baseline, the expected orbital motion would thus be only $\sim 20^{\circ}$, smaller by a factor of $\sim 9$ than the observed displacement. Since these assumptions provide the shortest possible orbital period, the discrepancy is even more significant. 

Although the photometric properties of the residuals — in particular their location in the color–magnitude diagram — appear broadly consistent across the 2015, 2016, 2023, and 2024 epochs, orbital considerations provide a stronger discriminant. Under reasonable assumptions on the system geometry and companion mass, the displacement observed between the archival and recent datasets is incompatible with the expected orbital motion. This dynamical inconsistency indicates that the features detected in 2015 and 2016 are unlikely to trace a genuine bound companion and are more plausibly explained as residual artifacts, likely exacerbated by the low-wind effect, whereas the excesses seen in 2023 and 2024 may indeed correspond to a real feature associated with TYC 8047-232-1 B, not detected in previous observations.

Even though the nature of the excess remains uncertain, this system could represent a second case, after DH Tau, of a binary-planet configuration {formed through gravitational capture enhanced by the action of tides \citep[see for example][]{Lazzoni2024}}. Deeper observations will be necessary to confirm this candidate, while future instruments such as METIS and MICADO on the ELT will offer stringent constraints on its nature.

\subsection{CT Cha}

CT Cha is a young K7-type star located in the Chamaeleon star-forming region, at a distance of $\sim189.95$ pc \citep{GAIA2021}, with a mass of $\sim0.8 M_{\odot}$ \citep{Sheehan2019} and an estimated age of $1.41^{+0.38}_{-0.30}$ { Myr} \citep{Feiden2016}. The system hosts a substellar companion, CT Cha b, first discovered through high-contrast imaging by \citet{Schmidt2008}, at a projected separation of 2.67 arcsec and with an initially estimated mass of 17 M\textsubscript{Jup}, assuming an age of 2-3 Myrs. Considering the more recent estimate of the age of the system, this substellar companion is much closer to the planetary regime, with a mass around $\sim$ 10 M\textsubscript{Jup}.  

Although its orbital motion is slow due to the wide separation, multi-epoch astrometry confirms that CT Cha b is co-moving with the primary star. Since there is no study in the literature about the orbital parameters for this object, we presented in the second paper of this series (Bernardi et al., {submitted}) the first estimate of the latter using the 17-year baseline available.
We retrieved possible semi-major axes varying from 350 to 820 au and spanning a wide range of eccentricities, from 0.25 to 0.87. Most probable values for these two parameters are 500 au and 0.59, respectively. The Hill radius spans a range of values, from 6.8 to 100.2 au, with a mean of 32.1 au.

A disk surrounding CT Cha B was inferred thanks to the detection of emission lines, such as and Pa$\beta$ and Br$\gamma$, in the spectrum of the brown dwarf, as well as significant dust extinction, indicative of accretion activity \citep{Schmidt2008,Bonnefoy2014,Lachapelle2015}. The presence of the CPD has been confirmed in a recent spectroscopic analysis of JWST data \citep{2025arXiv250915209C}. Moreover, \citet{Wu2015a} report variability in the H$\alpha$ emission and derive a mass accretion rate of $\sim 6\times10^{-10}$ M$_\odot$/yr. Non-detection at sub-millimeter and millimeter wavelengths with ALMA  \citep[0.88 and 1.3 mm,][]{Wu2017a,Wu2020} suggests either a compact disk or a relatively evolved disk with significant grain growth or depletion.

The residuals analyzed in this paper show an extended shape that could be attributed to the presence of the disk known to surround the planet. Therefore, we developed a forward-modeling procedure to characterize the circumsubstellar disk detected around the target. We make use of the GRaTeR \citep{1999A&A...348..557A} implementation in the vip\_hci package, in particular the ScatteredLightDisk class, to generate disk models with parametric density distributions. A double Henyey–Greenstein phase function is adopted to reproduce the anisotropic scattering properties of the dust grains. The disk model, consisting of a single circular component, is defined by a set of parameters describing its geometry (inclination $i$ and position angle $PA$), density distribution (inner and outer power-law slopes, $a_{in}$ and $a_{out}$, and reference radius, $a$), surface brightness ($F$),  and scattering phase function (asymmetry parameters $g_1$, $g_2$ and the weight of the two coefficients $w_{g1}$ and $w_{g2}=1-w_{g1}$).

The synthetic images are then convolved with the PSF extracted from the science data to account for instrumental response. The resulting model is injected at the location derived for CT Cha b in the image where the planet was subtracted. Second-iteration residuals, with the first being the outcome of the subtraction of the PSF of the planet, are then obtained by subtracting the convolved models from the observed science data.

\begin{figure}
    \centering
    \includegraphics[width=\columnwidth]{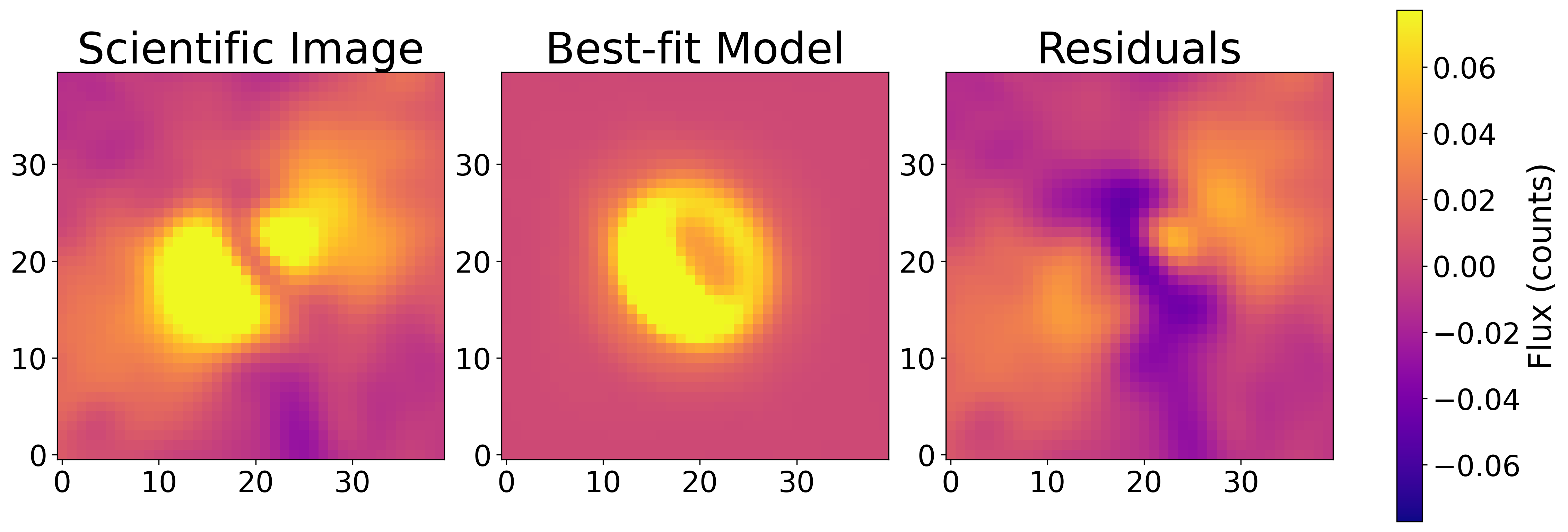}
    \includegraphics[width=\columnwidth]{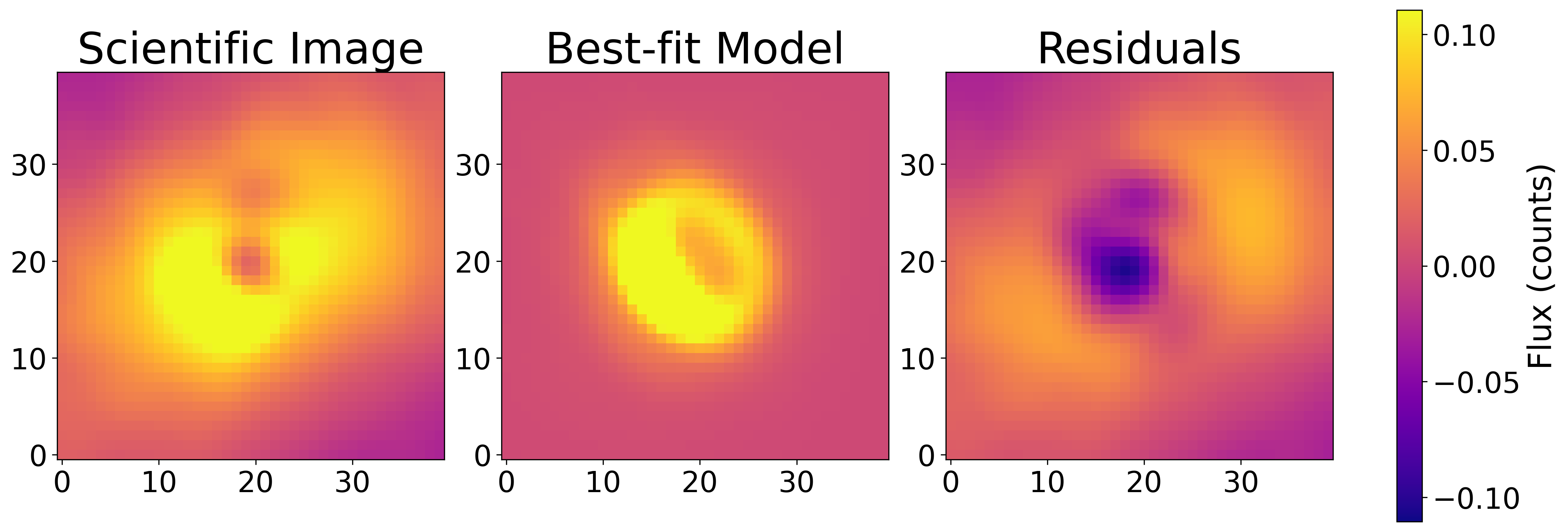}

    \caption{Disk detection around CT Cha b and its modeling for 2025-01-26 (top row) and 2017-03-19 (bottom row) epochs for H2-band. In each row, from left to right, are shown the residuals resulting from the subtraction of the PSF of the planet, best-fit disk model, and second-iteration residuals after disk model subtraction.}
    \label{CTCha}
\end{figure}

To assess the set of disk parameters that best reproduce the observed scattered-light morphology, we iteratively vary the 9 parameters to minimize the residuals, adopting the Powell minimization algorithm \citep{10.1093/comjnl/7.2.155}. Parameter uncertainties were then estimated using an MCMC exploration of the posterior distribution (via \texttt{emcee}), initialized around the best-fit solution. The best-fit is given by a disk with a reference radius  $a=15.4^{+3.1}_{-3.4}$ au, $a_{in}=5.8^{+2.9}_{-3.5}$ and $a_{out}=-13.1^{+7.1}_{-4.8}$, inclination of $41.4^{+1.0}_{-1.0}$ deg with respect to face-on and position angle of $127.9^{+8.5}_{-10.6}$ deg, surface brightness $F=4.3^{+2.9}_{-2.3}$ {counts} and scattering parameters $g_1=0.999^{+0.0006}_{-0.018}$, $g_2=-0.996^{+0.002}_{-0.023}$ and $w_{g1}=0.999^{+0.0004}_{-0.0005}$. In the first row of Figure \ref{CTCha} we show, from left to right, the outcome of the PSF subtraction of the planet (first-iteration residuals), the disk model obtained assuming the best-fit parameters, and the subtraction of the two (second-iteration residuals). It is important to note that the disk modeling is performed on an image where the PSF of the companion has already been subtracted; as a consequence, the residual disk signal may be affected by this preliminary step, and the derived disk parameters should therefore be regarded as highly uncertain.

Similar first-iteration residuals were also derived after reducing archival data obtained with SPHERE on 2017-03-19 (second row, first panel of Figure \ref{CTCha}). Unfortunately, such observations were of poor quality and cannot be used to derive precise disk parameters, given the strongly variable conditions of the atmosphere. However, generating a disk model with the same parameters obtained for the 2025 dataset, we obtain second-iteration residuals with significantly lower noise (second row, last panel of Figure \ref{CTCha}), pointing to the fact that the extended structure detected in the two epochs is oriented in a similar direction and has a similar extension. 
As a last note, the star itself, CT Cha A, hosts a disk which is oriented along a position angle of $59^{\circ}$ and with an inclination of $45.7^{\circ}$ \citep{Ginski2024}. If the residuals detected in this work are tracing the circumplanetary disk around CT Cha b, then the planet and the star have disks with very similar inclination, although they might not be aligned along the same position angle \citep[see for example][]{Zurlo2023}.

Being very young, CT Cha b is also among the substellar companions for which the deepest detection limits have been achieved. If we neglect the absorption coming from the CPD, we can already rule out the presence of Jupiter-like satellites around CT Cha b at separations of about half the mean Hill radius. Overall, given the presence of a CPD where satellite formation is likely ongoing, together with its young age, CT Cha stands out as one of the most promising systems to be investigated with future $\sim40$ meters class telescopes.

\subsection{GQ Lup}

GQ Lup is a classical T Tauri star located in the Lupus star-forming region at a distance of approximately $154.10$ pc \citep{GAIA2021}. The system is young, with an estimated age of 2–5 Myr and a stellar mass of 1.03 M\textsubscript{$\odot$} \citep{MacGregor2017}. Its widely separated substellar companion, GQ Lup B, was first identified via direct imaging at a projected separation of $\sim$ 0.7 arcsec \citep{Neuhauser2005b, Neuhauser2008} and with a mass estimate placing the companion near the brown dwarf/planetary-mass boundary \citep{Neuhauser2008, Patience2012, Stolker2021}.

Astrometric monitoring over nearly 10 years has revealed partial orbital motion consistent with an eccentricity between 0.21 and 0.69 and a semi-major axis of 0.54-0.92 arcsec \citep{Ginski2014a}. Interestingly, the spin period of GQ Lup B appears to be relatively slow compared to field brown dwarfs of similar mass and age, possibly linked to ongoing disk locking or magnetic breaking \citep{Schwarz2016}. 
Bernardi et al., {submitted}, further extended the baseline to nearly two decades, deriving more stringent constraints on the orbit of GQ Lup B, with semi-major axis $93^{+15}_{-13}$ {au} and eccentricity $0.45^{+0.15}_{-0.17}$. The Hill radius instead could vary between 6.6 and 16.5 au.

GQ Lup B exhibits signatures of active accretion, including emission and variability in both the Pa$\beta$ \citep{Seifahrt2007, Demars2023} and H$\alpha$ \citep{Zhou2014} lines. Based on H$\alpha$ photometry and accretion shock models, typical mass accretion rates are in the range of $2-3\times10^{-10}$ M\textsubscript{$\odot$}/yr \citep{Zhou2014,Stolker2021}.
ALMA observations did not reveal any sub-millimeter \citep{MacGregor2017} or millimeter \citep{Wu2017b} emission. The upper limit for the dust mass in the circumsubstellar disk was estimated to be $\sim 0.04$ M\textsubscript{$\oplus$} \citep{MacGregor2017}, consistent with compact and/or low-mass disks. However, recent JWST/MIRI mid-infrared spectroscopy has confirmed the presence of warm dust, providing the first direct detection of the disk around GQ Lup B in this wavelength range \citep{Cugno2024}. These results suggest the presence of small dust grains and thermal emission from a still-active circumsubstellar disk. No polarized scattered light signal has been detected with SPHERE/IRDIS, placing upper limits on the linear polarization of the companion at 0.2–0.3\% \citep{VanHolstein2021}. Moreover, no extended and coherent excess was detected when comparing the observations presented in this paper with archival data. 

It has been suggested that the disk around GQ Lup B might feature an inner cavity, extending from the sublimation radius at 6.6 R\textsubscript{Jup} out to $\sim$40 R\textsubscript{Jup} \citep{Cugno2024}. This cavity could have been carved by forming satellites, but current direct imaging observations cannot yet probe this region. For instance, if we assume that a single massive satellite is forming at the center of such a gap, it would be located at $\sim$23 R\textsubscript{Jup} (0.01 au) from GQ Lup B. Assuming that the dust-cleared region on each side of the satellite ($\sim$17 R\textsubscript{Jup}) corresponds to 3 $R_{\rm Hill}$, the resulting satellite mass would be 1.3 M\textsubscript{Jup}. As shown in Figure \ref{cc_mass}, the innermost separation accessible to our observations is 3.7 au, corresponding to a mass detection limit of 5.3 M\textsubscript{Jup}, which is far above the predicted physical parameters of the satellite. It is also unlikely that a single massive satellite is forming within the circumsubstellar disk, both from theoretical predictions \citep[i.e.][]{Canup2006} and from the RV measurements retrieved for GQ Lup B \citep{Horstman2024}. A more plausible scenario would include multiple smaller satellites, as in the case of the Galilean satellite system of Jupiter, implying even lower masses for each individual satellite. A comparable scenario has been recently investigated for GQ Lup B, where the putative cavity in its circumplanetary disk has been interpreted as potential evidence of exosatellites orbiting therein, and high-resolution radial velocity monitoring was carried out to directly test this hypothesis \citep{Horstman2024}, ruling out exomoons of 0.6\%-2.8\% the mass of GQ Lup B at separations between the Roche limit and 65 R\textsubscript{Jup}.

\subsection{HIP78530}

HIP 78530 is a young, $2.5 M_{\odot}$ B9V-type star \citep{Houk1988} located at a distance of $\sim 134.70$ pc \citep{GAIA2021}, with an estimated age of $11\pm3$ Myrs (Squicciarini et al, in prep). The star hosts a wide substellar companion, HIP 78530 B, discovered with ADONIS/SHARPII+ at a projected separation of $\sim4.5\arcsec$ \citep{Kouwenhoven2005}, and then confirmed through common proper motion, with a mass estimate of $\sim 20$ M\textsubscript{Jup} \citep{Lafreniere2011}. 

The orbital parameters of HIP 78530 B are hard to constrain due to the long orbital period, and only a small portion of the orbit has been covered in the past 20 years. From the orbital modeling by Bernardi et al., {submitted}, we retrieved a semi-major axis of $472^{+206}_{-88}$ { au} and eccentricity of $0.55^{+0.25}_{-0.25}$, with an estimated Hill radius of $28.6^{+36.7}_{-18.4}$ { au}.

We do not detect any significant excess from the residuals (see Figure \ref{mosaic}) and we restricted the detection limits for further companions around the brown dwarf to 6.1-3.8 M\textsubscript{Jup} at 3.2-11.1 au, respectively.

\subsection{DH Tau}

DH Tau is a young system located in the Taurus star-forming region at a distance of $\sim$133.45 pc \citep{GAIA2021}. The central star has an estimated mass of $\sim0.41$ M\textsubscript{$\odot$} and an age of 1.4 Myr \citep{Feiden2016}. Its widely separated substellar companion, DH Tau b, was discovered at a projected separation of $\sim$2.3 arcsec through direct imaging \citep{Itoh2005}. Previous works have placed DH Tau b near the planetary–brown dwarf mass boundary, with mass estimates ranging from 8–22 M\textsubscript{Jup} depending on evolutionary models \citep{Bonnefoy2014, Zhou2014, Itoh2005}.

Astrometric monitoring over more than a decade has confirmed common proper motion between DH Tau A and b \citep{Ginski2014a}, although the orbital parameters remain poorly constrained due to the wide separation and long period. Here we report the first attempt to constrain the orbit of DH Tau b, using a baseline of nearly 20 years. In Bernardi et al., {submitted}, we estimated a semi-major axis of $234^{+100}_{-41}$ { au} and eccentricity of $0.58^{+0.32}_{-0.26}$, with a Hill radius of $20.3^{+29.4}_{-16.6}$ { au}. 

Evidence for ongoing accretion on the substellar companion has been observed in multiple emission lines, including strong H$\alpha$ \citep{Zhou2014}, Pa$\beta$ \citep{Bonnefoy2014}, and Br$\gamma$ \citep{Bonnefoy2014}, supporting the presence of a circumsubstellar disk. Mid-infrared photometry from Spitzer/IRAC suggests a flux excess consistent with disk emission \citep{Martinez2022}, while ALMA observations at 0.88 mm places an upper limit on the dust mass of $\sim0.09$ M\textsubscript{$\oplus$}, indicating that any disk around DH Tau b must be compact and low-mass \citep{Wu2020, Wolff2017}. Moreover, polarimetric observations with SPHERE/IRDIS revealed a polarized scattered light signal from the companion, interpreted as direct evidence of a dusty disk \citep{VanHolstein2021}.

\citet{Lazzoni2020a} reported possible evidence for an additional companion in the vicinity of DH Tau b, identified as a point-like source {with a S/N in H2 of $\sim 2.7-2.3$} at a projected separation of $\sim$10 au. If confirmed, such an object could represent a massive satellite of $\sim$1 M\textsubscript{Jup}, providing a rare case of a hierarchical system composed of a young star and binary planets. However, the nature of this candidate remains debated due to observational challenges. The presence of a Jupiter-like companion to DH Tau b could also explain the compact circumsubstellar disk, whose radius is truncated to $\leq 3$ au \citep{VanHolstein2021}, well within 1/3 $R_{Hill}$ ($\sim 7$ au).

\begin{figure}
    \centering
    \includegraphics[width=\columnwidth]{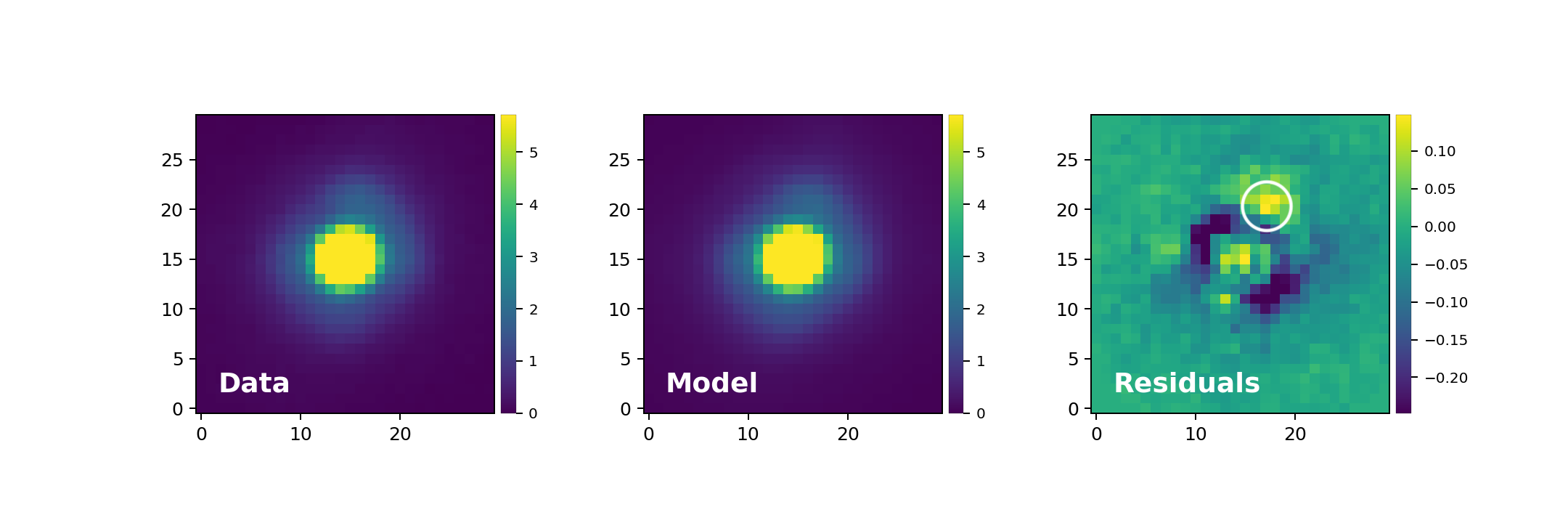}
    \includegraphics[width=\columnwidth]{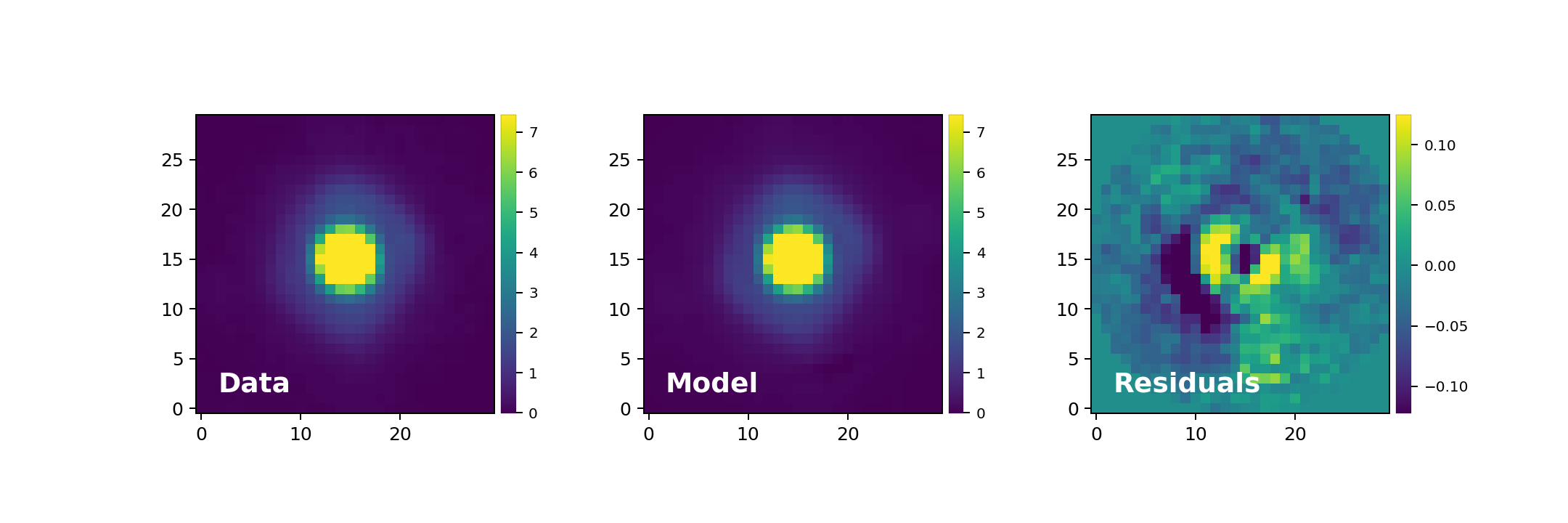}
    \includegraphics[width=\columnwidth]{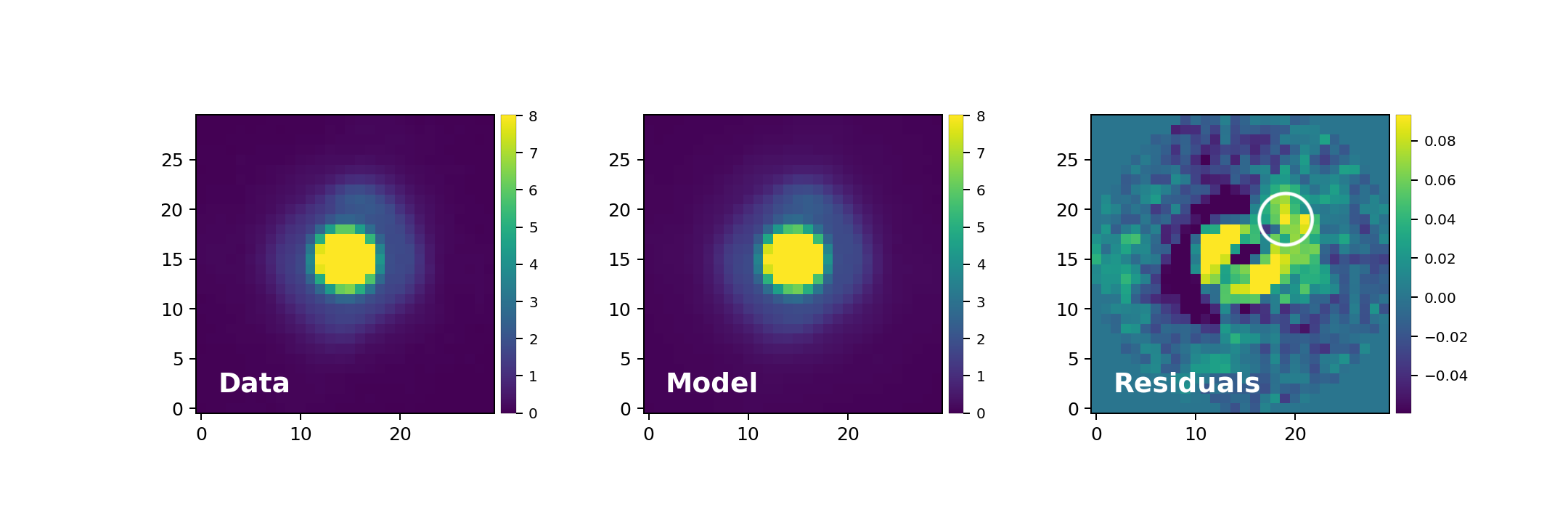}
    \caption{Reductions for DH Tau b for 2018-12-16 (top row) 2024-10-06 (mid row) and 2025-01-02 (bottom row) epochs. Panels show, from left to right, the observed PSF, the model PSF, and residuals {for the H2 band}. In the 2018 and 2025 data, a white circle highlights the marginal detection of a point-like source attributed to the satellite candidate DH Tau bb.}
    \label{DHTau}
\end{figure}

We present in Figure \ref{DHTau} the reductions of the observations obtained in 2024 mid row) and 2025 bottom row). As for TYC 8047-232-1 B, the panels from left to right show the PSF of DH Tau b, the model PSF, and the residuals obtained after subtraction. In the first epoch, no clear excess is detected in the residuals, while in the second epoch a point-like source is marginally recovered (S/N of 2.3 in H2 and 2.8 in H3) at 72 mas and $-42.5^\circ$ relative to the planet (see the white circle in the third row of Figure \ref{DHTau}). {The contrasts for the candidate were retrieved with the same procedure used for TYC 8047-232-1 Bb candidate (see Section~\ref{s:TYC8047}) and resulted in a contrast of $3.46\times10^{-5}\pm 1.69\times10^{-5}$ and of $1.82\times10^{-5}\pm1.16\times10^{-5}$ in the H2 and H3 bands, respectively ($\sim 1.1$ and $\sim1.6$ M\textsubscript{Jup} for H2 and H3 bands, respectively, considering mean values for the contrasts and age).}

In order to have a meaningful comparison for the properties retrieved here for DH Tau bb candidate, we re-reduced the latest H-band datasets (2018-12-16) presented in \citet{Lazzoni2020a} with the most recent tools developed in this paper. In particular, the 2018 epoch is particularly suitable for such comparison not only because instrumental biases were strongly reduced after 2017 (see forthcoming discussion), but also because three off-axis PSFs of the central star were acquired during the observation, two standard ones before and after the coronagraphic sequence and a further one in the middle of the scientific exposure. We thus created a PSF library with three models and applied the procedure described in Section 2.2. The new residuals for H2 band are presented in the top panel of Figure \ref{DHTau}, where, in the third panel, we encircled the candidate satellite DH Tau bb in white. Using the new precise astrometry for DH Tau b from the 2018 epoch (see Bernardi et al, sub.) and this new reduction we retrieved a separation of 77 mas and position angle of $-24.3^{\circ}$ for the candidate with respect to DH Tau b, and a contrast with respect to the star of $3.22\times10^{-5}\pm 2.07\times10^{-5}$ and $8.65\times10^{-6}\pm 7.70\times10^{-6}$ in H2 and H3 bands, respectively. 

Comparing the 2018 and 2025 epochs, the candidate satellite has consistent contrasts between the two observations in both bands, translating to a mass of $\sim 1$ M\textsubscript{Jup}. In terms of astrometry instead, the candidate has shifted of 5 mas inward and moved of $18.2^{\circ}$ clock-wise.
Ignoring the small inward radial shift, we can assume a circular orbit with radius $\sim 10$ au. The orbital period would then be $\sim 320$ yr. Over a 6-year baseline, the expected motion would be $\sim 10^\circ$ ($\sim 1$ pixel), a factor of $\lesssim 2$ smaller than the measured displacement. This is, however, not implausible given the large uncertainties on the masses of the two substellar companions and, especially, their relative position. These uncertainties are significantly affected by observational quality, including variable atmospheric conditions and mismatches between the model PSF and the PSF of the substellar companion, as well as by assumptions in evolutionary models and in the system’s age.

We note that the apparent azimuthal evolution of the tentative candidate DH Tau bb is not strictly monotonic when considering all available epochs (2015–2025). However, we believe that this behavior can be explained by the heterogeneous quality of the data and by systematic effects affecting some of the observations.

{In particular, the first epoch acquired on 2015-10-26 was strongly influenced by the low-wind effect, which produced pronounced azimuthally extended residuals around the substellar companion. We therefore consider it likely that, in some frames, the superposition of these low-wind residuals with the candidate signal biased its apparent position toward the east. This interpretation is consistent with the analysis of \citep{Lazzoni2020a}, where this epoch was excluded from the mass estimates of the candidate due to the anomalously high retrieved flux compared to all other observations.}

{For the 2020-01-05 epoch, the detection of the candidate in the K band is extremely shallow, and the PSF of DH Tau b is significantly broader than in the H band observations. This makes the precise determination of the candidate’s centroid particularly challenging and increases the associated astrometric uncertainties.}

{Taking these limitations into account, we argue that a meaningful assessment of the candidate’s azimuthal motion should primarily rely on the H-band epochs acquired on 2015-12-19, 2018-12-16, and 2025-01-02, which are characterized by more homogeneous data quality and a better-controlled PSF subtraction. Considering only these epochs, the candidate appears to evolve azimuthally around DH Tau b in a clockwise direction, and with an angular displacement that is broadly consistent with expectations for a bound object on a long-period orbit, given the large uncertainties.}

\begin{figure}
    \centering
    \includegraphics[width=\columnwidth]{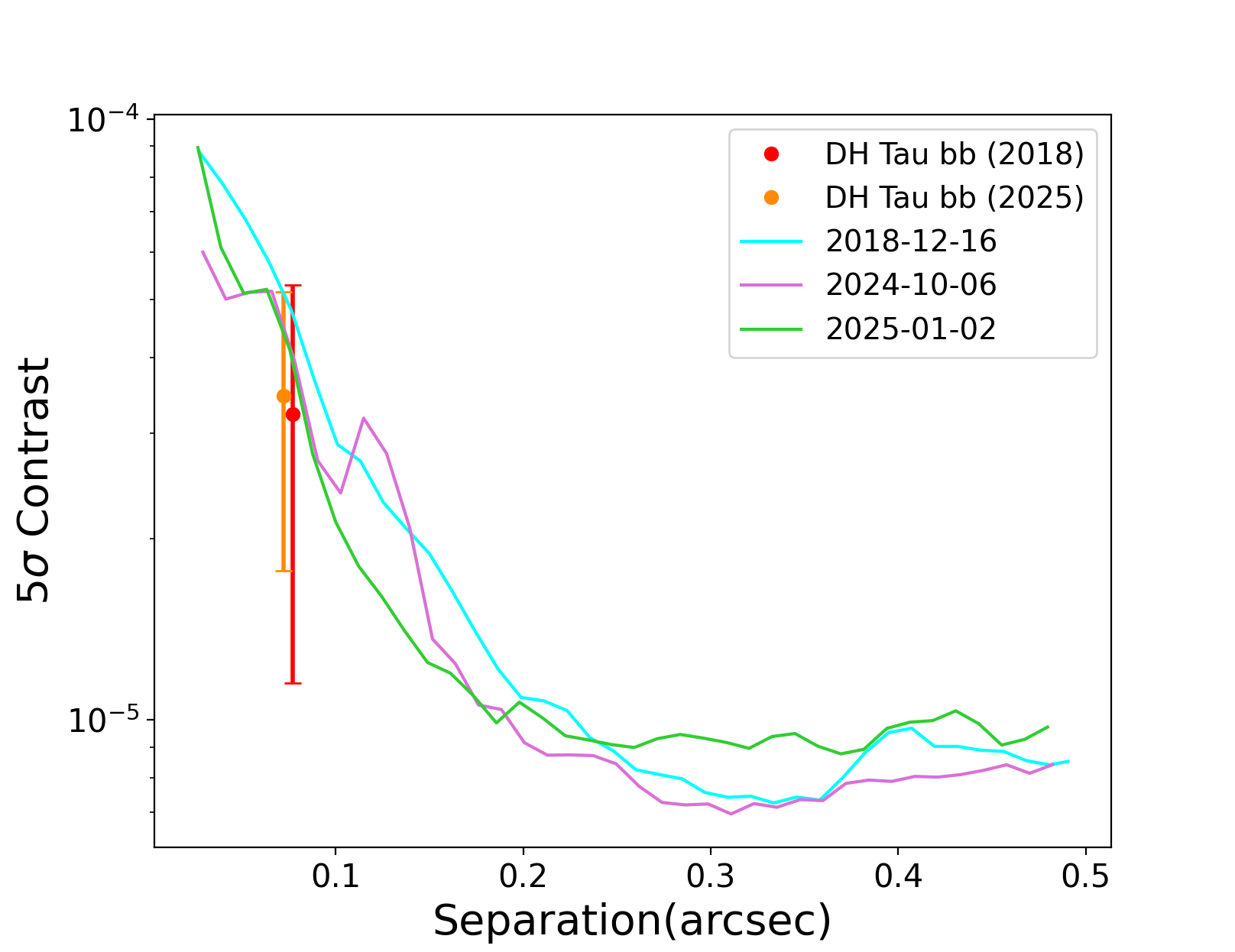}
    \caption{{H2-band} $5\sigma$ contrast curves for DH Tau b in the 2018-12-16 (light-blue), 2024-10-06 (green) and 2025-01-02 (pink) epochs. The red circle marks the 2018 candidate parameters and the orange circle indicates the 2025 detection.}
    \label{DHTaucc}
\end{figure}

In Figure \ref{DHTaucc} we show the $5\sigma$ contrast curves derived for the two epochs analyzed in this work and for the archival epoch of 2018-12-16. The red circle marks the contrast and position re-evaluated here for the 2018-12-06 epoch, while the orange circle shows the parameters obtained for the 2025-01-02 epoch.  The absence of a detection in 2024 and the marginal detection in 2025 are consistent with the candidate lying very close to the contrast limit. Regarding the H2-H3 color, we obtain $-1.43^{+1.3}_{-2.44}$ for the 2018 epoch, using new estimations from this paper, whereas from 2025 observation we obtain a color of $-0.69\pm0.87$. We show in Figure \ref{CMD} the color-magnitude diagram for the candidate DH Tau bb in 2018 observations (red square) and as found in this survey (green square). These results seem to suggest that, the candidate retrieved in the archival data and in this work exhibits broadly similar physical properties.


{Overall, while the persistence of the residuals over a time span of nearly ten years and their repeated recovery at comparable separations and position angles naturally lend themselves to an interpretation in terms of a candidate companion, we emphasize that this scenario cannot be considered conclusive. Although the low-wind effect has been strongly mitigated on SPHERE since 2018 and is expected to play a reduced role in the most recent epochs, we cannot fully exclude that the detected signal is linked to instrumental artifacts. In particular, the presence of strong negative signals in the processed images suggests that at least part of the signal may arise from data reduction effects rather than from an astrophysical source. The interpretation of the satellite candidate DH Tau bb, therefore, remains tentative and should be regarded with caution until confirmed by future observations with independent instruments.}

\subsection{HIP\,64892}

HIP 64892 is a young B9-type star \citep{Houk1988} of 2.35 $M_{\odot}$ \citep{Bressan2012} located at a distance of $\sim119.62$ pc \citep{GAIA2021} and with an age estimated at $16\pm2$ Myrs (Squicciarini et al, in prep). The star hosts a directly imaged substellar companion, HIP 64892 B, discovered with VLT/SPHERE at a projected separation of $1.27\arcsec$ and with an inferred mass above the deuterium burning limit \citep{Cheetham2018}.

Given the long orbital timescale and the relatively recent discovery of this brown dwarf, only a small portion of the orbit has been observed, leading to a semi-major axis of $128^{+43}_{-31}$ { au} and eccentricity of $0.51^{+0.25}_{-0.19}$ {(Bernardi et al., submitted)}. The Hill radius of HIP 64892 B thus spans a range of values, from 4.0 to 21.2 au, peaking at 11.1 au.

No evidence of a circumsubstellar disk or infrared excess has been reported to date, though deeper follow-up with ALMA or polarimetry may provide additional constraints on the presence of dust.
In our observations, we did not detect any relevant excesses attributed to a disk or further companions around HIP64892 B (see Figure \ref{mosaic}). From the retrieved detection limits, we can exclude the presence of satellites with masses down to 8.7-6.1 M\textsubscript{Jup} at 2.8-5.6 au, respectively.

\subsection{RX J1609.5-2105}

RX J1609-2105 is a young K7-type star located at a distance of 138.04 pc \citep{GAIA2021}, with an estimated age of 5 Myrs \citep{Lafreniere2010} and a mass of 0.68-0.77 M\textsubscript{$\odot$} \citep{Ireland2011}. Its substellar companion, RXJ1609.5-2105 b, was first identified by \citet{Lafreniere2010} through direct imaging at a projected separation of $\sim2.2$ arcsec. Mass estimates for the companion range from 8 to 14 M\textsubscript{Jup}, placing it at the boundary between the planetary and brown dwarf regimes \citep{Lafreniere2010, Wu2015b}.

Astrometric monitoring between 2010 and 2014 has confirmed common proper motion between the star and companion, but no orbital motion was detected in the four-year baseline \citep{Ginski2014a}. Adding the astrometric point acquired in 2025, in Bernardi et al., {submitted}, we derived an orbit compatible with a semi-major axis of $305^{+74}_{-51}$ au and eccentricity of $0.25^{+0.20}_{-0.15}$ and, thus, Hill radius of $32.6^{+16.7}_{-13.0}$ { au}.

The presence of a circumsubstellar disk around RXJ1609-2105 b remains debated. Early observations suggested possible disk-related features. For instance, \citet{Wu2015b} found evidence for reddening, in the optical bands $z'$ and $Y_s$, consistent with the presence of circumsubstellar dust surrounding the companion. However, this interpretation is challenged by several more recent studies. Mid-infrared photometry at 3.6-8.0 $\mu$m revealed no excess above the expected photospheric levels, ruling out significant warm dust emission around the companion \citep{Martinez2022}. Likewise, \citet{Bailey2013} found no 3-4 $\mu$m excess in either the primary or companion, although they did report a modest 24 $\mu$m excess for the system, potentially attributable to warm dust around the primary and/or the companion. Millimeter observations with ALMA at 0.88 mm further failed to detect any continuum emission at the companion's location, placing stringent 3$\sigma$ upper limits of 120-210 $\mu$Jy on the dust emission \citep{Wu2020}, implying a lack of massive cold disks. In addition, polarimetric observations with SPHERE/IRDIS revealed no detection of polarized scattered light, setting upper limits on the linear polarization at levels below $\sim$0.2\% \citep{VanHolstein2021}. Finally, residuals derived after the subtraction of the PSF of the planet did not reveal either extended or point-like sources. We can nevertheless exclude the presence of further companions around RXJ1609 b with masses in the range 1.7-1.4 M\textsubscript{Jup} between separations of 3.8-16.7 au.

\subsection{HII\,1348}

HII 1348 is located at a distance of $143.29$ pc \citep{GAIA2021}, with an estimated age of $112\pm5$ Myrs \citep{Dahm2015}. The primary is a K-type double-lined spectroscopic binary star with a mass of approximately 1.22 M$_\odot$ \citep{Geissler2012}. A faint companion, HII 1348 B, was first identified by \citet{Bouvier2007} as a background star but then confirmed as a common proper motion object by \citet{Geissler2012} through high-resolution adaptive optics imaging. First estimates for the mass placed HII 1348 B within the brown dwarf regime, having $\sim 60$ M\textsubscript{Jup}.

Astrometric monitoring for more than two decades has revealed some orbital motion despite the long orbital period ($\gtrsim 1000$ yrs). Recent orbital modeling by \citet{Weible2025} suggests that HII 1348 B follows an eccentric orbit with $e = 0.78^{+0.12}_{-0.29}$ and a semi-major axis of $\sim140^{+130}_{-30}$ au. The last epoch considered by \citet{Weible2025} for the orbital fitting traces back to 2019. Bernardi et al., in prep., thus extended the baseline by four more years and retrieved values for the semi-major axis and eccentricity of $181^{+49}_{-32}$ au and $0.60^{+0.08}_{-0.12}$, further increasing the precision on the orbital parameters.

No evidence of the presence of a circumsubstellar disk is reported in the literature, nor did we retrieve significant excesses from the dataset presented in this paper. Detection limits for the presence of satellites can only exclude massive companions around HII1348 B, with masses above $\sim$ 12 M\textsubscript{Jup} in the discovery area (4-9 au).

\subsection{$\eta$\,Tel}
{$\eta$ Tel is a young star located at a distance of $48.54$ pc \citep{GAIA2021}. The system is a confirmed member of the $\beta$ Pictoris moving group, with an estimated age of $24 \pm 5$ Myr and the stellar mass is estimated to be $2.09\,M_{\odot}$ \citep{Desidera2021}.}

{$\eta$ Tel hosts a wide brown dwarf companion, $\eta$ Tel B, originally discovered through near-infrared imaging and subsequently confirmed via proper motion and spectroscopic follow-up \citep{Guenther2001, Neuhauser2011}. The companion is located at a projected separation of $\sim4.2''$ and has been the subject of extensive astrometric and photometric monitoring over the past two decades \citep{Neuhauser2011, Nogueira2024}.}

{Due to the wide separation and long orbital period, the orbital parameters of $\eta$ Tel B remain poorly constrained. From the orbital modeling presented in Bernardi et al. (submitted), we retrieved a semi-major axis of $143^{+43}_{-22}$ au and an eccentricity of $0.53^{+0.29}_{-0.26}$. Based on these orbital parameters and the estimated masses of the system, we derive a Hill radius for $\eta$ Tel B of $13.6^{+14.9}_{-9.4}$ au.}

{No significant excess emission attributable to additional companions/disks is detected in the residual images (see Figure~\ref{mosaic}). We therefore place constraints on the presence of further companions orbiting $\eta$ Tel B, excluding masses larger than $\sim10.1$–$5.2$ M\textsubscript{Jup} at separations of $1.1$-$6.8$ au.}

\subsection{HD\,130948}
\label{s:HD130948}

HD 130948 is solar-type G2V star located at a distance of $\sim18.20$ pc \citep{GAIA2021}, with an estimated age of $550^{+250}_{-150}$ Myrs \citep{Vigan2017}. It hosts a remarkable brown dwarf-binary companion, HD 130948 BC, first discovered by \citet{Potter2002} via adaptive optics imaging with Gemini/Hokupa‘a. The pair comprises two nearly equal-mass brown dwarfs, with a total dynamical mass of $114$  M\textsubscript{Jup}, orbiting each other at a separation of $120$ mas, with an eccentricity of 0.167 and orbital period of $\sim 10$ yrs \citep{Dupuy2009}.

The orbit of HD 130948BC around the primary star has also been partially constrained by \citet{Ginski2013} through a combination of astrometric measurements coming from multiple instruments. The brown dwarf binary orbits HD 130948A at a projected separation of $2.6$ arcsec, with orbital fitting suggesting a semi-major axis of between 30 and 1112 au, peaking at { 47 au}, and an eccentricity between 0 and 0.96, peaking at 0.83. Moreover, the fitting resulted in a highly inclined system with $i$ ranging between 91 and 105 degrees.

In our observations, the two brown dwarfs are unresolved, and we were therefore unable to retrieve precise astrometric positions. For this reason, we report in Table \ref{comps} only the tentative orbital parameters available in the literature. In a triple system composed of the primary star and a brown dwarf binary, the orbital stability of potential satellites and the corresponding Hill radius are significantly more difficult to constrain. In particular, the half–$R_{\rm Hill}$ criterion adopted for the other systems does not apply, as the complex dynamics of the binary prevent a straightforward estimate of satellite stability. Consequently, we do not display the Hill radius in the figure and instead present only the residuals obtained after subtracting the model in the last panel of Figure \ref{mosaic}. In these residuals, the subtraction of a single PSF produces a bilobed structure, which can be naturally interpreted as the imprint of the two unresolved PSFs of the brown dwarfs. Finally, since this subtraction is based on an intrinsically inaccurate model, using a single PSF to represent the unresolved brown dwarf pair, we also do not report contrast curves for HD 130948 BC, since they would not be accurate.

\section{Conclusions}
\label{sec4}
In the first paper of this series, we presented the SaNDi-SHoP survey, designed to explore the circumplanetary environment of directly imaged substellar companions. The main objective was to search for satellites and circumsubstellar disks around a sample of 12 planets and brown dwarfs with VLT/SPHERE, using a novel application of the star-hopping technique. By repeatedly observing nearby reference stars of similar brightness and spectral type, we constructed rich PSF libraries that allowed us to apply frame-by-frame NEGFC subtraction of each companion. This approach effectively mitigated temporal PSF variability and strongly reduced spurious residuals, yielding cleaner images of the regions where faint satellites and CPDs are expected.

The data reduction pipeline combined standard calibration steps with customized PSF selection for each frame, and sensitivity was assessed through contrast curves converted into mass detection limits using evolutionary models. The stability regions were defined through Hill radii estimated from orbital fits presented in the second paper of the SaNDi-SHoP survey, providing dynamical boundaries for possible satellites and disks.

Across the full sample, our analysis constrains the presence of massive satellites to above the $\sim$ 1–10 M\textsubscript{Jup} level at separations of a few au, depending on system age, distance, and Hill radius. Most systems show no residuals compatible with bound companions or disks, but the achieved limits already exclude giant-mass satellites in wide orbits. The measured Hill radii range from a few au to several tens of au, and we verified that current instrumentation probes only a fraction of these stable regions.

Interestingly, three systems stand out from the survey for revealing compatible excesses throughout multiple observations. The first one is TYC 8047-232-1 B, which shows repeated excesses {at S/N of between 3 and 4}  across multiple epochs and filters, at projected separations of $\sim$85 mas, corresponding to masses of 3–6 M\textsubscript{Jup}. While the nature of the feature remains uncertain, the repeatability of the detection makes this system a strong candidate for hosting a low-mass companion. 

CT Cha b shows extended residuals consistent with the known presence of an accreting circumplanetary disk. Forward modeling of the residuals suggests a compact disk with a radius of $\sim$15 au, inclined by $\sim 40^{\circ}$, and compatible with independent spectroscopic evidence of accretion. This confirms that our method can recover genuine circumplanetary features in scattered light.

DH Tau b displays marginal point-like residuals in our second epoch  {at S/N of between 2 and 3}, relatively consistent with the candidate $\sim$ 1 M\textsubscript{Jup} satellite reported by \cite{Lazzoni2020a}. {However, significant uncertainties remain, related to PSF mismatches, heterogeneous data quality across epochs, non-monotonic apparent azimuthal motion when considering all epochs, and the possible impact of low-wind effect in several observations. While the repeated presence of residuals at similar separations over multiple epochs is intriguing, the available data do not allow us to unambiguously confirm the existence of a bound satellite.}

The constraints derived in this work are limited by the contrast and angular resolution of current facilities. The next generation of $\sim  40$ meters telescopes will drastically expand the accessible parameter space. In particular, second generation instruments such as the Planetary Camera and Spectrograph \citep[PCS,][]{Kasper2021} on the ELT are expected to achieve contrasts of $10^{-9}-10^{-10}$, compared to current limits of $10^{-6}-10^{-7}$, corresponding to an improvement of 2–3 orders of magnitude in detectable companion masses. At the same time, the increase in telescope aperture from 8 meters (VLT) to 39 m (ELT) will reduce the diffraction limit by a factor of $\sim$ 5. In the H2 band ($\lambda$ $\sim$ 1.6 $\mu$m), the angular resolution will improve from $\lambda$/D $\sim$ 0.04'' for SPHERE/VLT to $\lambda$/D  $\sim$ 0.008'' for PCS/ELT. This gain will allow us to probe separations down to <1 au for nearby systems, well inside the Hill spheres of most companions studied here. With such capabilities, it will become possible to detect satellites with masses of a few Earth masses or below, to access the innermost regions where moons are expected to form, and to directly resolve CPD whose presence is currently only inferred.

In addition to instrumental advances, further progress is expected on the methodological side. Statistically driven, detection-oriented approaches, such as the PACO framework \citep{Flasseur2018}, provide a powerful complement to classical high-contrast post-processing. By operating directly on the statistical properties of residual speckle noise, these methods can improve sensitivity at small angular separations and offer an alternative way to assess the robustness of faint signals \citep[see for example, ][]{Chomez2023}. Extending such frameworks to the circumplanetary regime explored in this work could provide an independent pathway to validate candidate satellites and circumplanetary material around directly imaged planets and brown dwarfs. In combination with the enhanced contrast and stability expected from next-generation facilities, these approaches may significantly improve the reliability of future detections and help open the path toward identifying hierarchical substructures at progressively smaller scales.

In conclusion, paper I of the SaNDi-SHoP survey has delivered the most stringent direct constraints to date on circumplanetary environments of directly imaged companions, recovering signals for three particularly promising systems. Future high-contrast imagers on the ELT will transform this exploratory effort into a systematic characterization of the processes of moon and disk formation around young planets and brown dwarfs.

\begin{acknowledgements}
{ The authors thank the referee for their comments and suggestions; the manuscript has been significantly improved after their revision.} We gratefully acknowledge support from the “Programma
di Ricerca Fondamentale INAF 2023” of the Italian National Institute of Astrophysics (INAF Large Grant 2023 “NextSTEPS”). Cecilia Lazzoni acknowledges the financial contribution from PRIN MUR 2022 (code 2022YP5ACE) funded by the European Union – NextGenerationEU. 
The authors acknowledge support from ANID -- Millennium Science Initiative Program -- Center Code NCN2024\_001. A.Z. acknowledges support from the Fondecyt Regular grant number 1250249.
This work has made use of the High Contrast Data Centre, jointly operated by OSUG/IPAG (Grenoble), PYTHEAS/LAM/CeSAM (Marseille), OCA/Lagrange (Nice), Observatoire de Paris/LESIA (Paris), and Observatoire de Lyon/CRAL, and supported by a grant from Labex OSUG@2020 (Investissements d’avenir – ANR10 LABX56).

\end{acknowledgements}
\bibliographystyle{aa}
\bibliography{bibliography}

@ARTICLE{Ayliffe2009,
       author = {{Ayliffe}, Ben A. and {Bate}, Matthew R.},
        title = "{Circumplanetary disc properties obtained from radiation hydrodynamical simulations of gas accretion by protoplanets}",
      journal = {\mnras},
         year = 2009,
        month = aug,
       volume = {397},
       number = {2},
        pages = {657-665},
          doi = {10.1111/j.1365-2966.2009.15002.x},
archivePrefix = {arXiv},
       eprint = {0904.4884},
 primaryClass = {astro-ph.EP},
       adsurl = {https://ui.adsabs.harvard.edu/abs/2009MNRAS.397..657A}
}

@INPROCEEDINGS{Bailey2013,
       author = {{Bailey}, Vanessa P. and {Hinz}, P. and {Currie}, T.~M. and {Su}, K.~Y. and {Esposito}, S. and {Fabrycky}, D.~C. and {Hill}, J.~M. and {Hoffmann}, W.~F. and {Jones}, T.~J. and {Kim}, J. and {Leisenring}, J. and {Meyer}, M.~R. and {Murray-Clay}, R. and {Nelson}, M. and {Pinna}, E. and {Puglisi}, A. and {Rieke}, G. and {Rodigas}, T. and {Skemer}, A. and {Skrutskie}, M.~F. and {Vaitheeswaran}, V. and {Wilson}, J.~C.},
        title = "{3-5 {\ensuremath{\mu}}m Imaging of Three Wide Brown Dwarf Companions to Upper Scorpius Stars: Searching for Hot Disks}",
    booktitle = {American Astronomical Society Meeting Abstracts \#221},
         year = 2013,
       series = {American Astronomical Society Meeting Abstracts},
       volume = {221},
        month = jan,
          eid = {158.17},
        pages = {158.17},
       adsurl = {https://ui.adsabs.harvard.edu/abs/2013AAS...22115817B}
}

@ARTICLE{Baraffe1998,
       author = {{Baraffe}, I. and {Chabrier}, G. and {Allard}, F. and {Hauschildt}, P.~H.},
        title = "{Evolutionary models for solar metallicity low-mass stars: mass-magnitude relationships and color-magnitude diagrams}",
      journal = {\aap},
         year = 1998,
        month = sep,
       volume = {337},
        pages = {403-412},
          doi = {10.48550/arXiv.astro-ph/9805009},
archivePrefix = {arXiv},
       eprint = {astro-ph/9805009},
 primaryClass = {astro-ph},
       adsurl = {https://ui.adsabs.harvard.edu/abs/1998A&A...337..403B}
}

@ARTICLE{Bardalez2025,
       author = {{Bardalez Gagliuffi}, Daniella C. and {Balmer}, William O. and {Pueyo}, Laurent and {Brandt}, Timothy D. and {Giovinazzi}, Mark R. and {Millholland}, Sarah and {Black}, Brennen and {Lu}, Tiger and {Rice}, Malena and {Mang}, James and {Morley}, Caroline and {Lacy}, Brianna and {Girard}, Julien H. and {Matthews}, Elisabeth C. and {Carter}, Aarynn L. and {Bowler}, Brendan P. and {Faherty}, Jacqueline K. and {Fontanive}, Clemence and {Rickman}, Emily},
        title = "{JWST Coronagraphic Images of 14 Her c: A Cold Giant Planet in a Dynamically Hot Multiplanet System}",
      journal = {\apjl},
         year = 2025,
        month = jul,
       volume = {988},
       number = {1},
          eid = {L18},
        pages = {L18},
          doi = {10.3847/2041-8213/ade30f},
archivePrefix = {arXiv},
       eprint = {2506.09201},
 primaryClass = {astro-ph.EP},
       adsurl = {https://ui.adsabs.harvard.edu/abs/2025ApJ...988L..18B}
}

@ARTICLE{Benisty2021,
       author = {{Benisty}, Myriam and {Bae}, Jaehan and {Facchini}, Stefano and {Keppler}, Miriam and {Teague}, Richard and {Isella}, Andrea and {Kurtovic}, Nicolas T. and {P{\'e}rez}, Laura M. and {Sierra}, Anibal and {Andrews}, Sean M. and {Carpenter}, John and {Czekala}, Ian and {Dominik}, Carsten and {Henning}, Thomas and {Menard}, Francois and {Pinilla}, Paola and {Zurlo}, Alice},
        title = "{A Circumplanetary Disk around PDS70c}",
      journal = {\apjl},
         year = 2021,
        month = jul,
       volume = {916},
       number = {1},
          eid = {L2},
        pages = {L2},
          doi = {10.3847/2041-8213/ac0f83},
archivePrefix = {arXiv},
       eprint = {2108.07123},
 primaryClass = {astro-ph.EP},
       adsurl = {https://ui.adsabs.harvard.edu/abs/2021ApJ...916L...2B}
}

@ARTICLE{Beuzit2019,
       author = {{Beuzit}, J. -L. and {Vigan}, A. and {Mouillet}, D. and {Dohlen}, K. and
         {Gratton}, R. and {Boccaletti}, A. and {Sauvage}, J. -F. and
         {Schmid}, H.~M. and {Langlois}, M. and {Petit}, C. and {Baruffolo}, A. and
         {Feldt}, M. and {Milli}, J. and {Wahhaj}, Z. and {Abe}, L. and
         {Anselmi}, U. and {Antichi}, J. and {Barette}, R. and {Baudrand}, J. and
         {Baudoz}, P. and {Bazzon}, A. and {Bernardi}, P. and {Blanchard}, P. and
         {Brast}, R. and {Bruno}, P. and {Buey}, T. and {Carbillet}, M. and
         {Carle}, M. and {Cascone}, E. and {Chapron}, F. and {Charton}, J. and
         {Chauvin}, G. and {Claudi}, R. and {Costille}, A. and {De Caprio}, V. and
         {de Boer}, J. and {Delboulb{\'e}}, A. and {Desidera}, S. and
         {Dominik}, C. and {Downing}, M. and {Dupuis}, O. and {Fabron}, C. and
         {Fantinel}, D. and {Farisato}, G. and {Feautrier}, P. and
         {Fedrigo}, E. and {Fusco}, T. and {Gigan}, P. and {Ginski}, C. and
         {Girard}, J. and {Giro}, E. and {Gisler}, D. and {Gluck}, L. and
         {Gry}, C. and {Henning}, T. and {Hubin}, N. and {Hugot}, E. and
         {Incorvaia}, S. and {Jaquet}, M. and {Kasper}, M. and {Lagadec}, E. and
         {Lagrange}, A. -M. and {Le Coroller}, H. and {Le Mignant}, D. and
         {Le Ruyet}, B. and {Lessio}, G. and {Lizon}, J. -L. and {Llored}, M. and
         {Lundin}, L. and {Madec}, F. and {Magnard}, Y. and {Marteaud}, M. and
         {Martinez}, P. and {Maurel}, D. and {M{\'e}nard}, F. and {Mesa}, D. and
         {M{\"o}ller-Nilsson}, O. and {Moulin}, T. and {Moutou}, C. and
         {Orign{\'e}}, A. and {Parisot}, J. and {Pavlov}, A. and {Perret}, D. and
         {Pragt}, J. and {Puget}, P. and {Rabou}, P. and {Ramos}, J. and
         {Reess}, J. -M. and {Rigal}, F. and {Rochat}, S. and {Roelfsema}, R. and
         {Rousset}, G. and {Roux}, A. and {Saisse}, M. and {Salasnich}, B. and
         {Santambrogio}, E. and {Scuderi}, S. and {Segransan}, D. and
         {Sevin}, A. and {Siebenmorgen}, R. and {Soenke}, C. and {Stadler}, E. and
         {Suarez}, M. and {Tiph{\`e}ne}, D. and {Turatto}, M. and {Udry}, S. and
         {Vakili}, F. and {Waters}, L.~B.~F.~M. and {Weber}, L. and {Wildi}, F. and
         {Zins}, G. and {Zurlo}, A.},
        title = "{SPHERE: the exoplanet imager for the Very Large Telescope}",
      journal = {A\&A},
         year = "2019",
        month = "Nov",
       volume = {631},
          eid = {A155},
        pages = {A155},
          doi = {10.1051/0004-6361/201935251},
       adsurl = {https://ui.adsabs.harvard.edu/abs/2019A&A...631A.155B}
}

@ARTICLE{Bonnefoy2014,
       author = {{Bonnefoy}, M. and {Chauvin}, G. and {Lagrange}, A. -M. and {Rojo}, P. and
         {Allard}, F. and {Pinte}, C. and {Dumas}, C. and {Homeier}, D.},
        title = "{A library of near-infrared integral field spectra of young M-L dwarfs}",
      journal = {A\&A},
         year = "2014",
        month = "Feb",
       volume = {562},
          eid = {A127},
        pages = {A127},
          doi = {10.1051/0004-6361/201118270},
archivePrefix = {arXiv},
       eprint = {1306.3709},
 primaryClass = {astro-ph.SR},
       adsurl = {https://ui.adsabs.harvard.edu/abs/2014A&A...562A.127B}
}

@INPROCEEDINGS{Bouvier2007,
       author = {{Bouvier}, J. and {Alencar}, S.~H.~P. and {Harries}, T.~J. and
         {Johns-Krull}, C.~M. and {Romanova}, M.~M.},
        title = "{Magnetospheric Accretion in Classical T Tauri Stars}",
    booktitle = {Protostars and Planets V},
         year = "2007",
       editor = {{Reipurth}, Bo and {Jewitt}, David and {Keil}, Klaus},
        month = "Jan",
        pages = {479},
archivePrefix = {arXiv},
       eprint = {astro-ph/0603498},
 primaryClass = {astro-ph},
       adsurl = {https://ui.adsabs.harvard.edu/abs/2007prpl.conf..479B}
}

@ARTICLE{Bowler2020,
       author = {{Bowler}, Brendan P. and {Blunt}, Sarah C. and {Nielsen}, Eric L.},
        title = "{Population-level Eccentricity Distributions of Imaged Exoplanets and Brown Dwarf Companions: Dynamical Evidence for Distinct Formation Channels}",
      journal = {\aj},
         year = 2020,
        month = feb,
       volume = {159},
       number = {2},
          eid = {63},
        pages = {63},
          doi = {10.3847/1538-3881/ab5b11},
archivePrefix = {arXiv},
       eprint = {1911.10569},
 primaryClass = {astro-ph.EP},
       adsurl = {https://ui.adsabs.harvard.edu/abs/2020AJ....159...63B}
}

@ARTICLE{Bressan2012,
       author = {{Bressan}, Alessandro and {Marigo}, Paola and {Girardi}, L{\'e}o. and {Salasnich}, Bernardo and {Dal Cero}, Claudia and {Rubele}, Stefano and {Nanni}, Ambra},
        title = "{PARSEC: stellar tracks and isochrones with the PAdova and TRieste Stellar Evolution Code}",
      journal = {\mnras},
         year = 2012,
        month = nov,
       volume = {427},
       number = {1},
        pages = {127-145},
          doi = {10.1111/j.1365-2966.2012.21948.x},
archivePrefix = {arXiv},
       eprint = {1208.4498},
 primaryClass = {astro-ph.SR},
       adsurl = {https://ui.adsabs.harvard.edu/abs/2012MNRAS.427..127B}
}

@article{10.1093/comjnl/7.2.155,
    author = {Powell, M. J. D.},
    title = {An efficient method for finding the minimum of a function of several variables without calculating derivatives},
    journal = {The Computer Journal},
    volume = {7},
    number = {2},
    pages = {155-162},
    year = {1964},
    month = {01},
    issn = {0010-4620},
    doi = {10.1093/comjnl/7.2.155},
    url = {https://doi.org/10.1093/comjnl/7.2.155},
    eprint = {https://academic.oup.com/comjnl/article-pdf/7/2/155/959784/070155.pdf},
}

@ARTICLE{1999A&A...348..557A,
       author = {{Augereau}, J.~C. and {Lagrange}, A.~M. and {Mouillet}, D. and {Papaloizou}, J.~C.~B. and {Grorod}, P.~A.},
        title = "{On the HR 4796 A circumstellar disk}",
      journal = {\aap},
         year = 1999,
        month = aug,
       volume = {348},
        pages = {557-569},
          doi = {10.48550/arXiv.astro-ph/9906429},
archivePrefix = {arXiv},
       eprint = {astro-ph/9906429},
 primaryClass = {astro-ph},
       adsurl = {https://ui.adsabs.harvard.edu/abs/1999A&A...348..557A}
}

@ARTICLE{2018A&A...615A..92G,
       author = {{Galicher}, R. and {Boccaletti}, A. and {Mesa}, D. and {Delorme}, P. and {Gratton}, R. and {Langlois}, M. and {Lagrange}, A.-M. and {Maire}, A.-L. and {Le Coroller}, H. and {Chauvin}, G. and {Biller}, B. and {Cantalloube}, F. and {Janson}, M. and {Lagadec}, E. and {Meunier}, N. and {Vigan}, A. and {Hagelberg}, J. and {Bonnefoy}, M. and {Zurlo}, A. and {Rocha}, S. and {Maurel}, D. and {Jaquet}, M. and {Buey}, T. and {Weber}, L.},
        title = "{Astrometric and photometric accuracies in high contrast imaging: The SPHERE speckle calibration tool (SpeCal)}",
      journal = {\aap},
         year = 2018,
        month = jul,
       volume = {615},
          eid = {A92},
        pages = {A92},
          doi = {10.1051/0004-6361/201832973},
archivePrefix = {arXiv},
       eprint = {1805.04854},
 primaryClass = {astro-ph.IM},
       adsurl = {https://ui.adsabs.harvard.edu/abs/2018A&A...615A..92G}
}

@ARTICLE{2023JOSS....8.4774C,
       author = {{Christiaens}, Valentin and {Gonzalez}, Carlos and {Farkas}, Ralf and {Dahlqvist}, Carl-Henrik and {Nasedkin}, Evert and {Milli}, Julien and {Absil}, Olivier and {Ngo}, Henry and {Cantero}, Carles and {Rainot}, Alan and {Hammond}, Iain and {Bonse}, Markus and {Cantalloube}, Faustine and {Vigan}, Arthur and {Kompella}, Vijay and {Hancock}, Paul},
        title = "{VIP: A Python package for high-contrast imaging}",
      journal = {The Journal of Open Source Software},
         year = 2023,
        month = jan,
       volume = {8},
       number = {81},
          eid = {4774},
        pages = {4774},
          doi = {10.21105/joss.04774},
       adsurl = {https://ui.adsabs.harvard.edu/abs/2023JOSS....8.4774C}
}

@ARTICLE{2025arXiv250915209C,
       author = {{Cugno}, Gabriele and {Grant}, Sierra L.},
        title = "{A carbon-rich disk surrounding a planetary-mass companion}",
      journal = {arXiv e-prints},
         year = 2025,
        month = sep,
          eid = {arXiv:2509.15209},
        pages = {arXiv:2509.15209},
          doi = {10.48550/arXiv.2509.15209},
archivePrefix = {arXiv},
       eprint = {2509.15209},
 primaryClass = {astro-ph.EP},
       adsurl = {https://ui.adsabs.harvard.edu/abs/2025arXiv250915209C}
}

@ARTICLE{Canup2006,
       author = {{Canup}, Robin M. and {Ward}, William R.},
        title = "{A common mass scaling for satellite systems of gaseous planets}",
      journal = {\nat},
         year = 2006,
        month = jun,
       volume = {441},
       number = {7095},
        pages = {834-839},
          doi = {10.1038/nature04860},
       adsurl = {https://ui.adsabs.harvard.edu/abs/2006Natur.441..834C}
}

@ARTICLE{Carter2023,
       author = {{Carter}, Aarynn L. and {Hinkley}, Sasha and {Kammerer}, Jens and {Skemer}, Andrew and {Biller}, Beth A. and {Leisenring}, Jarron M. and {Millar-Blanchaer}, Maxwell A. and {Petrus}, Simon and {Stone}, Jordan M. and {Ward-Duong}, Kimberly and {Wang}, Jason J. and {Girard}, Julien H. and {Hines}, Dean C. and {Perrin}, Marshall D. and {Pueyo}, Laurent and {Balmer}, William O. and {Bonavita}, Mariangela and {Bonnefoy}, Mickael and {Chauvin}, Gael and {Choquet}, Elodie and {Christiaens}, Valentin and {Danielski}, Camilla and {Kennedy}, Grant M. and {Matthews}, Elisabeth C. and {Miles}, Brittany E. and {Patapis}, Polychronis and {Ray}, Shrishmoy and {Rickman}, Emily and {Sallum}, Steph and {Stapelfeldt}, Karl R. and {Whiteford}, Niall and {Zhou}, Yifan and {Absil}, Olivier and {Boccaletti}, Anthony and {Booth}, Mark and {Bowler}, Brendan P. and {Chen}, Christine H. and {Currie}, Thayne and {Fortney}, Jonathan J. and {Grady}, Carol A. and {Greebaum}, Alexandra Z. and {Henning}, Thomas and {Hoch}, Kielan K.~W. and {Janson}, Markus and {Kalas}, Paul and {Kenworthy}, Matthew A. and {Kervella}, Pierre and {Kraus}, Adam L. and {Lagage}, Pierre-Olivier and {Liu}, Michael C. and {Macintosh}, Bruce and {Marino}, Sebastian and {Marley}, Mark S. and {Marois}, Christian and {Matthews}, Brenda C. and {Mawet}, Dimitri and {McElwain}, Michael W. and {Metchev}, Stanimir and {Meyer}, Michael R. and {Molliere}, Paul and {Moran}, Sarah E. and {Morley}, Caroline V. and {Mukherjee}, Sagnick and {Pantin}, Eric and {Quirrenbach}, Andreas and {Rebollido}, Isabel and {Ren}, Bin B. and {Schneider}, Glenn and {Vasist}, Malavika and {Worthen}, Kadin and {Wyatt}, Mark C. and {Briesemeister}, Zackery W. and {Bryan}, Marta L. and {Calissendorff}, Per and {Cantalloube}, Faustine and {Cugno}, Gabriele and {De Furio}, Matthew and {Dupuy}, Trent J. and {Factor}, Samuel M. and {Faherty}, Jacqueline K. and {Fitzgerald}, Michael P. and {Franson}, Kyle and {Gonzales}, Eileen C. and {Hood}, Callie E. and {Howe}, Alex R. and {Kuzuhara}, Masayuki and {Lagrange}, Anne-Marie and {Lawson}, Kellen and {Lazzoni}, Cecilia and {Lew}, Ben W.~P. and {Liu}, Pengyu and {Llop-Sayson}, Jorge and {Lloyd}, James P. and {Martinez}, Raquel A. and {Mazoyer}, Johan and {Palma-Bifani}, Paulina and {Quanz}, Sascha P. and {Redai}, Jea Adams and {Samland}, Matthias and {Schlieder}, Joshua E. and {Tamura}, Motohide and {Tan}, Xianyu and {Uyama}, Taichi and {Vigan}, Arthur and {Vos}, Johanna M. and {Wagner}, Kevin and {Wolff}, Schuyler G. and {Ygouf}, Marie and {Zhang}, Xi and {Zhang}, Keming and {Zhang}, Zhoujian},
        title = "{The JWST Early Release Science Program for Direct Observations of Exoplanetary Systems I: High-contrast Imaging of the Exoplanet HIP 65426 b from 2 to 16 {\ensuremath{\mu}}m}",
      journal = {\apjl},
         year = 2023,
        month = jul,
       volume = {951},
       number = {1},
          eid = {L20},
        pages = {L20},
          doi = {10.3847/2041-8213/acd93e},
archivePrefix = {arXiv},
       eprint = {2208.14990},
 primaryClass = {astro-ph.EP},
       adsurl = {https://ui.adsabs.harvard.edu/abs/2023ApJ...951L..20C}
}

@ARTICLE{Chauvin2005b,
       author = {{Chauvin}, G. and {Lagrange}, A. -M. and {Zuckerman}, B. and {Dumas}, C. and {Mouillet}, D. and {Song}, I. and {Beuzit}, J. -L. and {Lowrance}, P. and {Bessell}, M.~S.},
        title = "{A companion to AB Pic at the planet/brown dwarf boundary}",
      journal = {\aap},
         year = 2005,
        month = aug,
       volume = {438},
       number = {3},
        pages = {L29-L32},
          doi = {10.1051/0004-6361:200500111},
archivePrefix = {arXiv},
       eprint = {astro-ph/0504658},
 primaryClass = {astro-ph},
       adsurl = {https://ui.adsabs.harvard.edu/abs/2005A&A...438L..29C}
}

@ARTICLE{Chauvin2005c,
       author = {{Chauvin}, G. and {Lagrange}, A. -M. and {Lacombe}, F. and {Dumas}, C. and {Mouillet}, D. and {Zuckerman}, B. and {Gendron}, E. and {Song}, I. and {Beuzit}, J. -L. and {Lowrance}, P. and {Fusco}, T.},
        title = "{Astrometric and spectroscopic confirmation of a brown dwarf companion to GSC 08047-00232.  VLT/NACO deep imaging and spectroscopic observations}",
      journal = {\aap},
         year = 2005,
        month = feb,
       volume = {430},
        pages = {1027-1033},
          doi = {10.1051/0004-6361:20041353},
       adsurl = {https://ui.adsabs.harvard.edu/abs/2005A&A...430.1027C}
}

@ARTICLE{Cheetham2018,
       author = {{Cheetham}, A. and {Bonnefoy}, M. and {Desidera}, S. and {Langlois}, M. and
         {Vigan}, A. and {Schmidt}, T. and {Olofsson}, J. and {Chauvin}, G. and
         {Klahr}, H. and {Gratton}, R. and {D'Orazi}, V. and {Henning}, T. and
         {Janson}, M. and {Biller}, B. and {Peretti}, S. and {Hagelberg}, J. and
         {S{\'e}gransan}, D. and {Udry}, S. and {Mesa}, D. and {Sissa}, E. and
         {Kral}, Q. and {Schlieder}, J. and {Maire}, A. -L. and {Mordasini}, C. and
         {Menard}, F. and {Zurlo}, A. and {Beuzit}, J. -L. and {Feldt}, M. and
         {Mouillet}, D. and {Meyer}, M. and {Lagrange}, A. -M. and
         {Boccaletti}, A. and {Keppler}, M. and {Kopytova}, T. and {Ligi}, R. and
         {Rouan}, D. and {Le Coroller}, H. and {Dominik}, C. and {Lagadec}, E. and
         {Turatto}, M. and {Abe}, L. and {Antichi}, J. and {Baruffolo}, A. and
         {Baudoz}, P. and {Blanchard}, P. and {Buey}, T. and {Carbillet}, M. and
         {Carle}, M. and {Cascone}, E. and {Claudi}, R. and {Costille}, A. and
         {Delboulb{\'e}}, A. and {De Caprio}, V. and {Dohlen}, K. and
         {Fantinel}, D. and {Feautrier}, P. and {Fusco}, T. and {Giro}, E. and
         {Gluck}, L. and {Hubin}, N. and {Hugot}, E. and {Jaquet}, M. and
         {Kasper}, M. and {Llored}, M. and {Madec}, F. and {Magnard}, Y. and
         {Martinez}, P. and {Maurel}, D. and {Le Mignant}, D. and
         {M{\"o}ller-Nilsson}, O. and {Moulin}, T. and {Orign{\'e}}, A. and
         {Pavlov}, A. and {Perret}, D. and {Petit}, C. and {Pragt}, J. and
         {Puget}, P. and {Rabou}, P. and {Ramos}, J. and {Rigal}, F. and
         {Rochat}, S. and {Roelfsema}, R. and {Rousset}, G. and {Roux}, A. and
         {Salasnich}, B. and {Sauvage}, J. -F. and {Sevin}, A. and {Soenke}, C. and
         {Stadler}, E. and {Suarez}, M. and {Weber}, L. and {Wildi}, F.},
        title = "{Discovery of a brown dwarf companion to the star HIP 64892}",
      journal = {A\&A},
         year = "2018",
        month = "Aug",
       volume = {615},
          eid = {A160},
        pages = {A160},
          doi = {10.1051/0004-6361/201832650},
archivePrefix = {arXiv},
       eprint = {1803.02725},
 primaryClass = {astro-ph.EP},
       adsurl = {https://ui.adsabs.harvard.edu/abs/2018A&A...615A.160C}
}

@ARTICLE{Chomez2023,
       author = {{Chomez}, A. and {Squicciarini}, V. and {Lagrange}, A.-M. and {Delorme}, P. and {Viswanath}, G. and {Janson}, M. and {Flasseur}, O. and {Chauvin}, G. and {Langlois}, M. and {Rubini}, P. and {Bergeon}, S. and {Albert}, D. and {Bonnefoy}, M. and {Desidera}, S. and {Engler}, N. and {Gratton}, R. and {Henning}, T. and {Mamajek}, E.~E. and {Marleau}, G.-D. and {Meyer}, M.~R. and {Reffert}, S. and {Ringqvist}, S.~C. and {Samland}, M.},
        title = "{An imaged 15 M$_{Jup}$ companion within a hierarchical quadruple system}",
      journal = {\aap},
         year = 2023,
        month = aug,
       volume = {676},
          eid = {L10},
        pages = {L10},
          doi = {10.1051/0004-6361/202347044},
archivePrefix = {arXiv},
       eprint = {2307.01195},
 primaryClass = {astro-ph.EP},
       adsurl = {https://ui.adsabs.harvard.edu/abs/2023A&A...676L..10C}
}

@ARTICLE{Cugno2024,
       author = {{Cugno}, Gabriele and {Patapis}, Polychronis and {Banzatti}, Andrea and {Meyer}, Michael and {Dannert}, Felix A. and {Stolker}, Tomas and {MacDonald}, Ryan J. and {Pontoppidan}, Klaus M.},
        title = "{Mid-infrared Spectrum of the Disk around the Forming Companion GQ Lup B Revealed by JWST/MIRI}",
      journal = {\apjl},
         year = 2024,
        month = may,
       volume = {966},
       number = {1},
          eid = {L21},
        pages = {L21},
          doi = {10.3847/2041-8213/ad3cbc},
archivePrefix = {arXiv},
       eprint = {2404.07086},
 primaryClass = {astro-ph.EP},
       adsurl = {https://ui.adsabs.harvard.edu/abs/2024ApJ...966L..21C}
}

@ARTICLE{Dahm2015,
       author = {{Dahm}, S.~E.},
        title = "{Reexamining the Lithium Depletion Boundary in the Pleiades and the Inferred Age of the Cluster}",
      journal = {\apj},
         year = 2015,
        month = nov,
       volume = {813},
       number = {2},
          eid = {108},
        pages = {108},
          doi = {10.1088/0004-637X/813/2/108},
       adsurl = {https://ui.adsabs.harvard.edu/abs/2015ApJ...813..108D}
}

@INPROCEEDINGS{Delorme2017b,
      author = {{Delorme}, P. and {Meunier}, N. and {Albert}, D. and
{Lagadec}, E. and
        {Le Coroller}, H. and {Galicher}, R. and {Mouillet}, D. and
        {Boccaletti}, A. and {Mesa}, D. and {Meunier}, J. -C. and
        {Beuzit}, J. -L. and {Lagrange}, A. -M. and {Chauvin}, G. and
        {Sapone}, A. and {Langlois}, M. and {Maire}, A. -L. and
        {Montarg{\`e}s}, M. and {Gratton}, R. and {Vigan}, A. and
{Surace}, C.},
       title = "{The SPHERE Data Center: a reference for high
contrast imaging processing}",
   booktitle = {SF2A-2017: Proceedings of the Annual meeting of the
French Society of Astronomy and Astrophysics},
        year = "2017",
       month = "Dec",
       pages = {Di},
archivePrefix = {arXiv},
      eprint = {1712.06948},
primaryClass = {astro-ph.IM},
      adsurl = {https://ui.adsabs.harvard.edu/abs/2017sf2a.conf..347D}
}

@ARTICLE{Demars2023,
       author = {{Demars}, D. and {Bonnefoy}, M. and {Dougados}, C. and {Aoyama}, Y. and {Thanathibodee}, T. and {Marleau}, G. -D. and {Tremblin}, P. and {Delorme}, P. and {Palma-Bifani}, P. and {Petrus}, S. and {Bowler}, B.~P. and {Chauvin}, G. and {Lagrange}, A. -M.},
        title = "{Emission line variability of young 10-30 M$_{Jup}$ companions. I. The case of GQ Lup b and GSC 06214-00210 b}",
      journal = {\aap},
         year = 2023,
        month = aug,
       volume = {676},
          eid = {A123},
        pages = {A123},
          doi = {10.1051/0004-6361/202346221},
archivePrefix = {arXiv},
       eprint = {2305.09460},
 primaryClass = {astro-ph.EP},
       adsurl = {https://ui.adsabs.harvard.edu/abs/2023A&A...676A.123D}
}

@ARTICLE{DeRosa2023,
       author = {{De Rosa}, Robert J. and {Nielsen}, Eric L. and {Wahhaj}, Zahed and {Ruffio}, Jean-Baptiste and {Kalas}, Paul G. and {Peck}, Anne E. and {Hirsch}, Lea A. and {Roberson}, William},
        title = "{Direct imaging discovery of a super-Jovian around the young Sun-like star AF Leporis}",
      journal = {\aap},
         year = 2023,
        month = apr,
       volume = {672},
          eid = {A94},
        pages = {A94},
          doi = {10.1051/0004-6361/202345877},
archivePrefix = {arXiv},
       eprint = {2302.06332},
 primaryClass = {astro-ph.EP},
       adsurl = {https://ui.adsabs.harvard.edu/abs/2023A&A...672A..94D}
}

@ARTICLE{Desidera2021,
       author = {{Desidera}, S. and {Chauvin}, G. and {Bonavita}, M. and {Messina}, S. and {LeCoroller}, H. and {Schmidt}, T. and {Gratton}, R. and {Lazzoni}, C. and {Meyer}, M. and {Schlieder}, J. and {Cheetham}, A. and {Hagelberg}, J. and {Bonnefoy}, M. and {Feldt}, M. and {Lagrange}, A-M. and {Langlois}, M. and {Vigan}, A. and {Tan}, T.~G. and {Hambsch}, F. -J. and {Millward}, M. and {Alcala}, J. and {Benatti}, S. and {Brandner}, W. and {Carson}, J. and {Covino}, E. and {Delorme}, P. and {D'Orazi}, V. and {Janson}, M. and {Rigliaco}, E. and {Beuzit}, J. -L. and {Biller}, B. and {Boccaletti}, A. and {Dominik}, C. and {Cantalloube}, F. and {Fontaniv}, C. and {Galicher}, R. and {Henning}, Th. and {Lagadec}, E. and {Ligi}, R. and {Maire}, A-L. and {Menard}, F. and {Mesa}, D. and {Muller}, A. and {Samland}, M. and {Schmid}, H.~M. and {Sissa}, E. and {Turatto}, M. and {Udry}, S. and {Asensio-Torres}, A. Zurlo R. and {Kopytova}, T. and {Rickman}, E. and {Abe}, L. and {Antichi}, J. and {Baruffolo}, A. and {Baudoz}, P. and {Baudrand}, J. and {Blanchard}, P. and {Bazzon}, A. and {Buey}, T. and {Carbillet}, M. and {Carle}, M. and {Charton}, J. and {Cascone}, E. and {Claudi}, R. and {Costille}, A. and {Deboulbe}, A. and {De Caprio}, V. and {Dohlen}, K. and {Fantinel}, D. and {Feautrier}, P. and {Fusco}, T. and {Gigan}, P. and {Giro}, E. and {Gisler}, D. and {Gluck}, L. and {Hubin}, N. and {Hugot}, E. and {Jaquet}, M. and {Kasper}, M. and {Madec}, F. and {Magnard}, Y. and {Martinez}, P. and {Maurel}, D. and {Le Mignant}, D. and {Moller-Nilsson}, O. and {Llored}, M. and {Moulin}, T. and {Origne}, A. and {Pavlov}, A. and {Perret}, D. and {Petit}, C. and {Pragt}, J. and {Puget}, P. and {Rabou}, P. and {Ramon}, J. and {Rigal}, F. and {Rochat}, S. and {Roelfsema}, R. and {Rousset}, G. and {Roux}, A. and {Salasnich}, B. and {Sauvage}, J. -F. and {Sevin}, A. and {Soenke}, C. and {Stadler}, E. and {Suarez}, M. and {Weber}, L. and {Wildi}, F.},
        title = "{The SPHERE infrared survey for exoplanets (SHINE)- I Sample definition and target characterization}",
      journal = {arXiv e-prints},
         year = 2021,
        month = mar,
          eid = {arXiv:2103.04366},
        pages = {arXiv:2103.04366},
archivePrefix = {arXiv},
       eprint = {2103.04366},
 primaryClass = {astro-ph.EP},
       adsurl = {https://ui.adsabs.harvard.edu/abs/2021arXiv210304366D}
}

@ARTICLE{Domingos2006,
       author = {{Domingos}, R.~C. and {Winter}, O.~C. and {Yokoyama}, T.},
        title = "{Stable satellites around extrasolar giant planets}",
      journal = {\mnras},
         year = 2006,
        month = dec,
       volume = {373},
       number = {3},
        pages = {1227-1234},
          doi = {10.1111/j.1365-2966.2006.11104.x},
       adsurl = {https://ui.adsabs.harvard.edu/abs/2006MNRAS.373.1227D}
}

@ARTICLE{Dupuy2009,
       author = {{Dupuy}, Trent J. and {Liu}, Michael C. and {Ireland}, Michael J.},
        title = "{Dynamical Mass of the Substellar Benchmark Binary HD 130948BC}",
      journal = {\apj},
         year = 2009,
        month = feb,
       volume = {692},
       number = {1},
        pages = {729-752},
          doi = {10.1088/0004-637X/692/1/729},
archivePrefix = {arXiv},
       eprint = {0807.2450},
 primaryClass = {astro-ph},
       adsurl = {https://ui.adsabs.harvard.edu/abs/2009ApJ...692..729D}
}

@ARTICLE{Feiden2016,
       author = {{Feiden}, Gregory A.},
        title = "{Magnetic inhibition of convection and the fundamental properties of low-mass stars. III. A consistent 10 Myr age for the Upper Scorpius OB association}",
      journal = {A\&A},
         year = "2016",
        month = "Sep",
       volume = {593},
          eid = {A99},
        pages = {A99},
          doi = {10.1051/0004-6361/201527613},
archivePrefix = {arXiv},
       eprint = {1604.08036},
 primaryClass = {astro-ph.SR},
       adsurl = {https://ui.adsabs.harvard.edu/abs/2016A&A...593A..99F}
}

@ARTICLE{Flasseur2018,
       author = {{Flasseur}, Olivier and {Denis}, Lo{\"\i}c and {Thi{\'e}baut}, {\'E}ric and {Langlois}, Maud},
        title = "{Exoplanet detection in angular differential imaging by statistical learning of the nonstationary patch covariances. The PACO algorithm}",
      journal = {\aap},
         year = 2018,
        month = oct,
       volume = {618},
          eid = {A138},
        pages = {A138},
          doi = {10.1051/0004-6361/201832745},
       adsurl = {https://ui.adsabs.harvard.edu/abs/2018A&A...618A.138F}
}

@ARTICLE{Foreman-Mackey2013,
       author = {{Foreman-Mackey}, Daniel and {Hogg}, David W. and {Lang}, Dustin and
         {Goodman}, Jonathan},
        title = "{emcee: The MCMC Hammer}",
      journal = {PASP},
         year = 2013,
        month = mar,
       volume = {125},
       number = {925},
        pages = {306},
          doi = {10.1086/670067},
archivePrefix = {arXiv},
       eprint = {1202.3665},
 primaryClass = {astro-ph.IM},
       adsurl = {https://ui.adsabs.harvard.edu/abs/2013PASP..125..306F}
}

@ARTICLE{Franson2023,
       author = {{Franson}, Kyle and {Bowler}, Brendan P. and {Zhou}, Yifan and {Pearce}, Tim D. and {Bardalez Gagliuffi}, Daniella C. and {Biddle}, Lauren I. and {Brandt}, Timothy D. and {Crepp}, Justin R. and {Dupuy}, Trent J. and {Faherty}, Jacqueline and {Jensen-Clem}, Rebecca and {Morgan}, Marvin and {Sanghi}, Aniket and {Theissen}, Christopher A. and {Tran}, Quang H. and {Wolf}, Trevor N.},
        title = "{Astrometric Accelerations as Dynamical Beacons: A Giant Planet Imaged inside the Debris Disk of the Young Star AF Lep}",
      journal = {\apjl},
         year = 2023,
        month = jun,
       volume = {950},
       number = {2},
          eid = {L19},
        pages = {L19},
          doi = {10.3847/2041-8213/acd6f6},
archivePrefix = {arXiv},
       eprint = {2302.05420},
 primaryClass = {astro-ph.EP},
       adsurl = {https://ui.adsabs.harvard.edu/abs/2023ApJ...950L..19F}
}

@ARTICLE{GAIA2021,
       author = {{Gaia Collaboration} and {Brown}, A.~G.~A. and {Vallenari}, A. and {Prusti}, T. and {de Bruijne}, J.~H.~J. and {Babusiaux}, C. and {Biermann}, M. and {Creevey}, O.~L. and {Evans}, D.~W. and {Eyer}, L. and {Hutton}, A. and {Jansen}, F. and {Jordi}, C. and {Klioner}, S.~A. and {Lammers}, U. and {Lindegren}, L. and {Luri}, X. and {Mignard}, F. and {Panem}, C. and {Pourbaix}, D. and {Randich}, S. and {Sartoretti}, P. and {Soubiran}, C. and {Walton}, N.~A. and {Arenou}, F. and {Bailer-Jones}, C.~A.~L. and {Bastian}, U. and {Cropper}, M. and {Drimmel}, R. and {Katz}, D. and {Lattanzi}, M.~G. and {van Leeuwen}, F. and {Bakker}, J. and {Cacciari}, C. and {Casta{\~n}eda}, J. and {De Angeli}, F. and {Ducourant}, C. and {Fabricius}, C. and {Fouesneau}, M. and {Fr{\'e}mat}, Y. and {Guerra}, R. and {Guerrier}, A. and {Guiraud}, J. and {Jean-Antoine Piccolo}, A. and {Masana}, E. and {Messineo}, R. and {Mowlavi}, N. and {Nicolas}, C. and {Nienartowicz}, K. and {Pailler}, F. and {Panuzzo}, P. and {Riclet}, F. and {Roux}, W. and {Seabroke}, G.~M. and {Sordo}, R. and {Tanga}, P. and {Th{\'e}venin}, F. and {Gracia-Abril}, G. and {Portell}, J. and {Teyssier}, D. and {Altmann}, M. and {Andrae}, R. and {Bellas-Velidis}, I. and {Benson}, K. and {Berthier}, J. and {Blomme}, R. and {Brugaletta}, E. and {Burgess}, P.~W. and {Busso}, G. and {Carry}, B. and {Cellino}, A. and {Cheek}, N. and {Clementini}, G. and {Damerdji}, Y. and {Davidson}, M. and {Delchambre}, L. and {Dell'Oro}, A. and {Fern{\'a}ndez-Hern{\'a}ndez}, J. and {Galluccio}, L. and {Garc{\'\i}a-Lario}, P. and {Garcia-Reinaldos}, M. and {Gonz{\'a}lez-N{\'u}{\~n}ez}, J. and {Gosset}, E. and {Haigron}, R. and {Halbwachs}, J. -L. and {Hambly}, N.~C. and {Harrison}, D.~L. and {Hatzidimitriou}, D. and {Heiter}, U. and {Hern{\'a}ndez}, J. and {Hestroffer}, D. and {Hodgkin}, S.~T. and {Holl}, B. and {Jan{\ss}en}, K. and {Jevardat de Fombelle}, G. and {Jordan}, S. and {Krone-Martins}, A. and {Lanzafame}, A.~C. and {L{\"o}ffler}, W. and {Lorca}, A. and {Manteiga}, M. and {Marchal}, O. and {Marrese}, P.~M. and {Moitinho}, A. and {Mora}, A. and {Muinonen}, K. and {Osborne}, P. and {Pancino}, E. and {Pauwels}, T. and {Petit}, J. -M. and {Recio-Blanco}, A. and {Richards}, P.~J. and {Riello}, M. and {Rimoldini}, L. and {Robin}, A.~C. and {Roegiers}, T. and {Rybizki}, J. and {Sarro}, L.~M. and {Siopis}, C. and {Smith}, M. and {Sozzetti}, A. and {Ulla}, A. and {Utrilla}, E. and {van Leeuwen}, M. and {van Reeven}, W. and {Abbas}, U. and {Abreu Aramburu}, A. and {Accart}, S. and {Aerts}, C. and {Aguado}, J.~J. and {Ajaj}, M. and {Altavilla}, G. and {{\'A}lvarez}, M.~A. and {{\'A}lvarez Cid-Fuentes}, J. and {Alves}, J. and {Anderson}, R.~I. and {Anglada Varela}, E. and {Antoja}, T. and {Audard}, M. and {Baines}, D. and {Baker}, S.~G. and {Balaguer-N{\'u}{\~n}ez}, L. and {Balbinot}, E. and {Balog}, Z. and {Barache}, C. and {Barbato}, D. and {Barros}, M. and {Barstow}, M.~A. and {Bartolom{\'e}}, S. and {Bassilana}, J. -L. and {Bauchet}, N. and {Baudesson-Stella}, A. and {Becciani}, U. and {Bellazzini}, M. and {Bernet}, M. and {Bertone}, S. and {Bianchi}, L. and {Blanco-Cuaresma}, S. and {Boch}, T. and {Bombrun}, A. and {Bossini}, D. and {Bouquillon}, S. and {Bragaglia}, A. and {Bramante}, L. and {Breedt}, E. and {Bressan}, A. and {Brouillet}, N. and {Bucciarelli}, B. and {Burlacu}, A. and {Busonero}, D. and {Butkevich}, A.~G. and {Buzzi}, R. and {Caffau}, E. and {Cancelliere}, R. and {C{\'a}novas}, H. and {Cantat-Gaudin}, T. and {Carballo}, R. and {Carlucci}, T. and {Carnerero}, M.~I. and {Carrasco}, J.~M. and {Casamiquela}, L. and {Castellani}, M. and {Castro-Ginard}, A. and {Castro Sampol}, P. and {Chaoul}, L. and {Charlot}, P. and {Chemin}, L. and {Chiavassa}, A. and {Cioni}, M. -R.~L. and {Comoretto}, G. and {Cooper}, W.~J. and {Cornez}, T. and {Cowell}, S. and {Crifo}, F. and {Crosta}, M. and {Crowley}, C. and {Dafonte}, C. and {Dapergolas}, A. and {David}, M. and {David}, P.},
        title = "{Gaia Early Data Release 3. Summary of the contents and survey properties}",
      journal = {\aap},
         year = 2021,
        month = may,
       volume = {649},
          eid = {A1},
        pages = {A1},
          doi = {10.1051/0004-6361/202039657},
archivePrefix = {arXiv},
       eprint = {2012.01533},
 primaryClass = {astro-ph.GA},
       adsurl = {https://ui.adsabs.harvard.edu/abs/2021A&A...649A...1G}
}

@ARTICLE{Geissler2012,
       author = {{Gei{\ss}ler}, Kerstin and {Metchev}, Stanimir A. and {Pham}, Alfonse and {Larkin}, James E. and {McElwain}, Michael and {Hillenbrand}, Lynne A.},
        title = "{A Substellar Common Proper-motion Companion to the Pleiad H II 1348}",
      journal = {\apj},
         year = 2012,
        month = feb,
       volume = {746},
       number = {1},
          eid = {44},
        pages = {44},
          doi = {10.1088/0004-637X/746/1/44},
archivePrefix = {arXiv},
       eprint = {1112.3191},
 primaryClass = {astro-ph.SR},
       adsurl = {https://ui.adsabs.harvard.edu/abs/2012ApJ...746...44G}
}

@ARTICLE{Ginski2013,
       author = {{Ginski}, C. and {Neuh{\"a}user}, R. and {Mugrauer}, M. and {Schmidt}, T.~O.~B. and {Adam}, C.},
        title = "{Orbital motion of the binary brown dwarf companions HD 130948 BC around their host star}",
      journal = {\mnras},
         year = 2013,
        month = sep,
       volume = {434},
       number = {1},
        pages = {671-683},
          doi = {10.1093/mnras/stt1059},
       adsurl = {https://ui.adsabs.harvard.edu/abs/2013MNRAS.434..671G}
}

@article{Ginski2014a,
	Adsurl = {http://adsabs.harvard.edu/abs/2014MNRAS.444.2280G},
	Archiveprefix = {arXiv},
	Author = {{Ginski}, C. and {Schmidt}, T.~O.~B. and {Mugrauer}, M. and {Neuh{\"a}user}, R. and {Vogt}, N. and {Errmann}, R. and {Berndt}, A.},
	Doi = {10.1093/mnras/stu1586},
	Eprint = {1409.1850},
	Journal = {MNRAS},
	Month = nov,
	Pages = {2280-2302},
	Primaryclass = {astro-ph.EP},
	Title = {{Astrometric follow-up observations of directly imaged sub-stellar companions to young stars and brown dwarfs}},
	Volume = 444,
	Year = 2014}

@ARTICLE{Ginski2014b,
       author = {{Ginski}, C. and {Mugrauer}, M. and {Neuh{\"a}user}, R. and {Schmidt}, T.~O.~B.},
        title = "{Astrometric monitoring and orbit constraint of the GSC 08047-00232 system with VLT/NaCo}",
      journal = {\mnras},
         year = 2014,
        month = feb,
       volume = {438},
       number = {2},
        pages = {1102-1113},
          doi = {10.1093/mnras/stt2263},
       adsurl = {https://ui.adsabs.harvard.edu/abs/2014MNRAS.438.1102G}
}

@ARTICLE{Ginski2024,
       author = {{Ginski}, C. and {Garufi}, A. and {Benisty}, M. and {Tazaki}, R. and {Dominik}, C. and {Ribas}, {\'A}. and {Engler}, N. and {Birnstiel}, T. and {Chauvin}, G. and {Columba}, G. and {Facchini}, S. and {Goncharov}, A. and {Hagelberg}, J. and {Henning}, T. and {Hogerheijde}, M. and {van Holstein}, R.~G. and {Huang}, J. and {Muto}, T. and {Pinilla}, P. and {Kanagawa}, K. and {Kim}, S. and {Kurtovic}, N. and {Langlois}, M. and {Manara}, C. and {Milli}, J. and {Momose}, M. and {Orihara}, R. and {Pawellek}, N. and {Pinte}, C. and {Rab}, C. and {Schmidt}, T.~O.~B. and {Snik}, F. and {Wahhaj}, Z. and {Williams}, J. and {Zurlo}, A.},
        title = "{The SPHERE view of the Chamaeleon I star-forming region. The full census of planet-forming disks with GTO and DESTINYS programs}",
      journal = {\aap},
         year = 2024,
        month = may,
       volume = {685},
          eid = {A52},
        pages = {A52},
          doi = {10.1051/0004-6361/202244005},
archivePrefix = {arXiv},
       eprint = {2403.02149},
 primaryClass = {astro-ph.GA},
       adsurl = {https://ui.adsabs.harvard.edu/abs/2024A&A...685A..52G}
}

@ARTICLE{Gonzalez2017,
       author = {{Gomez Gonzalez}, Carlos Alberto and {Wertz}, Olivier and
         {Absil}, Olivier and {Christiaens}, Valentin and {Defr{\`e}re}, Denis and
         {Mawet}, Dimitri and {Milli}, Julien and {Absil}, Pierre-Antoine and
         {Van Droogenbroeck}, Marc and {Cantalloube}, Faustine and
         {Hinz}, Philip M. and {Skemer}, Andrew J. and {Karlsson}, Mikael and
         {Surdej}, Jean},
        title = "{VIP: Vortex Image Processing Package for High-contrast Direct Imaging}",
      journal = {AJ},
         year = "2017",
        month = "Jul",
       volume = {154},
       number = {1},
          eid = {7},
        pages = {7},
          doi = {10.3847/1538-3881/aa73d7},
archivePrefix = {arXiv},
       eprint = {1705.06184},
 primaryClass = {astro-ph.IM},
       adsurl = {https://ui.adsabs.harvard.edu/abs/2017AJ....154....7G}
}

@ARTICLE{Gratton2024,
       author = {{Gratton}, R. and {Bonavita}, M. and {Mesa}, D. and {Desidera}, S. and {Zurlo}, A. and {Marino}, S. and {D'Orazi}, V. and {Rigliaco}, E. and {Nascimbeni}, V. and {Barbato}, D. and {Columba}, G. and {Squicciarini}, V.},
        title = "{Stellar companions and Jupiter-like planets in young associations}",
      journal = {\aap},
         year = 2024,
        month = may,
       volume = {685},
          eid = {A119},
        pages = {A119},
          doi = {10.1051/0004-6361/202348393},
archivePrefix = {arXiv},
       eprint = {2402.02148},
 primaryClass = {astro-ph.EP},
       adsurl = {https://ui.adsabs.harvard.edu/abs/2024A&A...685A.119G}
}

@ARTICLE{Guenther2001,
       author = {{Guenther}, E.~W. and {Neuh{\"a}user}, R. and {Hu{\'e}lamo}, N. and {Brandner}, W. and {Alves}, J.},
        title = "{Infrared spectrum and proper motion of the brown dwarf companion of HR 7329 in Tucanae}",
      journal = {\aap},
         year = 2001,
        month = jan,
       volume = {365},
        pages = {514-518},
          doi = {10.1051/0004-6361:20000051},
       adsurl = {https://ui.adsabs.harvard.edu/abs/2001A&A...365..514G}
}

@ARTICLE{Haffert2019,
       author = {{Haffert}, S.~Y. and {Bohn}, A.~J. and {de Boer}, J. and
         {Snellen}, I.~A.~G. and {Brinchmann}, J. and {Girard}, J.~H. and
         {Keller}, C.~U. and {Bacon}, R.},
        title = "{Two accreting protoplanets around the young star PDS 70}",
      journal = {Nature Astronomy},
         year = "2019",
        month = "Jun",
       volume = {3},
        pages = {749-754},
          doi = {10.1038/s41550-019-0780-5},
archivePrefix = {arXiv},
       eprint = {1906.01486},
 primaryClass = {astro-ph.EP},
       adsurl = {https://ui.adsabs.harvard.edu/abs/2019NatAs...3..749H}
}

@ARTICLE{Horstman2024,
       author = {{Horstman}, Katelyn and {Ruffio}, Jean-Baptiste and {Batygin}, Konstantin and {Mawet}, Dimitri and {Baker}, Ashley and {Hsu}, Chih-Chun and {Wang}, Jason J. and {Wang}, Ji and {Blunt}, Sarah and {Xuan}, Jerry W. and {Xin}, Yinzi and {Liberman}, Joshua and {Agrawal}, Shubh and {Konopacky}, Quinn M. and {Blake}, Geoffrey A. and {Do {\'O}}, Clarissa R. and {Bartos}, Randall and {Bond}, Charlotte Z. and {Calvin}, Benjamin and {Cetre}, Sylvain and {Delorme}, Jacques-Robert and {Doppmann}, Greg and {Echeverri}, Daniel and {Finnerty}, Luke and {Fitzgerald}, Michael P. and {Jovanovic}, Nemanja and {L{\'o}pez}, Ronald and {Martin}, Emily C. and {Morris}, Evan and {Pezzato}, Jacklyn and {Ruane}, Garreth and {Sappey}, Ben and {Schofield}, Tobias and {Skemer}, Andrew and {Venenciano}, Taylor and {Wallace}, J. Kent and {Wallack}, Nicole L. and {Wizinowich}, Peter},
        title = "{RV Measurements of Directly Imaged Brown Dwarf GQ Lup B to Search for Exosatellites}",
      journal = {\aj},
         year = 2024,
        month = oct,
       volume = {168},
       number = {4},
          eid = {175},
        pages = {175},
          doi = {10.3847/1538-3881/ad73d8},
archivePrefix = {arXiv},
       eprint = {2408.10299},
 primaryClass = {astro-ph.EP},
       adsurl = {https://ui.adsabs.harvard.edu/abs/2024AJ....168..175H}
}

@BOOK{Houk1988,
       author = {{Houk}, N. and {Smith-Moore}, M.},
        title = "{Michigan Catalogue of Two-dimensional Spectral Types for the HD Stars. Volume 4, Declinations -26{\textdegree}.0 to -12{\textdegree}.0.}",
         year = 1988,
       volume = {4},
       adsurl = {https://ui.adsabs.harvard.edu/abs/1988mcts.book.....H}
}

@ARTICLE{Ireland2011,
       author = {{Ireland}, M.~J. and {Kraus}, A. and {Martinache}, F. and {Law}, N. and {Hillenbrand}, L.~A.},
        title = "{Two Wide Planetary-mass Companions to Solar-type Stars in Upper Scorpius}",
      journal = {\apj},
         year = 2011,
        month = jan,
       volume = {726},
       number = {2},
          eid = {113},
        pages = {113},
          doi = {10.1088/0004-637X/726/2/113},
archivePrefix = {arXiv},
       eprint = {1011.2201},
 primaryClass = {astro-ph.SR},
       adsurl = {https://ui.adsabs.harvard.edu/abs/2011ApJ...726..113I}
}

@ARTICLE{Isella2019,
       author = {{Isella}, Andrea and {Benisty}, Myriam and {Teague}, Richard and
         {Bae}, Jaehan and {Keppler}, Miriam and {Facchini}, Stefano and
         {P{\'e}rez}, Laura},
        title = "{Detection of Continuum Submillimeter Emission Associated with Candidate Protoplanets}",
      journal = {ApJ},
         year = "2019",
        month = "Jul",
       volume = {879},
       number = {2},
          eid = {L25},
        pages = {L25},
          doi = {10.3847/2041-8213/ab2a12},
archivePrefix = {arXiv},
       eprint = {1906.06308},
 primaryClass = {astro-ph.EP},
       adsurl = {https://ui.adsabs.harvard.edu/abs/2019ApJ...879L..25I}
}

@ARTICLE{Itoh2005,
       author = {{Itoh}, Yoichi and {Hayashi}, Masahiko and {Tamura}, Motohide and
         {Tsuji}, Takashi and {Oasa}, Yumiko and {Fukagawa}, Misato and
         {Hayashi}, Saeko S. and {Naoi}, Takahiro and {Ishii}, Miki and
         {Mayama}, Satoshi and {Morino}, Jun-ichi and {Yamashita}, Takuya and
         {Pyo}, Tae-Soo and {Nishikawa}, Takayuki and {Usuda}, Tomonori and
         {Murakawa}, Koji and {Suto}, Hiroshi and {Oya}, Shin and
         {Takato}, Naruhisa and {Ando}, Hiroyasu and {Miyama}, Shoken M. and
         {Kobayashi}, Naoto and {Kaifu}, Norio},
        title = "{A Young Brown Dwarf Companion to DH Tauri}",
      journal = {ApJ},
         year = 2005,
        month = feb,
       volume = {620},
       number = {2},
        pages = {984-993},
          doi = {10.1086/427086},
archivePrefix = {arXiv},
       eprint = {astro-ph/0411177},
 primaryClass = {astro-ph},
       adsurl = {https://ui.adsabs.harvard.edu/abs/2005ApJ...620..984I}
}

@ARTICLE{Kasper2021,
       author = {{Kasper}, M. and {Cerpa Urra}, N. and {Pathak}, P. and {Bonse}, M. and {Nousiainen}, J. and {Engler}, B. and {Heritier}, C.~T. and {Kammerer}, J. and {Leveratto}, S. and {Rajani}, C. and {Bristow}, P. and {Le Louarn}, M. and {Madec}, P. -Y. and {Str{\"o}bele}, S. and {Verinaud}, C. and {Glauser}, A. and {Quanz}, S.~P. and {Helin}, T. and {Keller}, C. and {Snik}, F. and {Boccaletti}, A. and {Chauvin}, G. and {Mouillet}, D. and {Kulcs{\'a}r}, C. and {Raynaud}, H. -F.},
        title = "{PCS {\textemdash} A Roadmap for Exoearth Imaging with the ELT}",
      journal = {The Messenger},
         year = 2021,
        month = mar,
       volume = {182},
        pages = {38-43},
          doi = {10.18727/0722-6691/5221},
archivePrefix = {arXiv},
       eprint = {2103.11196},
 primaryClass = {astro-ph.IM},
       adsurl = {https://ui.adsabs.harvard.edu/abs/2021Msngr.182...38K}
}

@ARTICLE{Keppler2018,
       author = {{Keppler}, M. and {Benisty}, M. and {M{\"u}ller}, A. and {Henning}, Th. and
         {van Boekel}, R. and {Cantalloube}, F. and {Ginski}, C. and
         {van Holstein}, R.~G. and {Maire}, A. -L. and {Pohl}, A. and {Samland
        }, M. and {Avenhaus}, H. and {Baudino}, J. -L. and {Boccaletti}, A. and
         {de Boer}, J. and {Bonnefoy}, M. and {Chauvin}, G. and {Desidera}, S. and
         {Langlois}, M. and {Lazzoni}, C. and {Marleau}, G. -D. and
         {Mordasini}, C. and {Pawellek}, N. and {Stolker}, T. and {Vigan}, A. and
         {Zurlo}, A. and {Birnstiel}, T. and {Brandner}, W. and {Feldt}, M. and
         {Flock}, M. and {Girard}, J. and {Gratton}, R. and {Hagelberg}, J. and
         {Isella}, A. and {Janson}, M. and {Juhasz}, A. and {Kemmer}, J. and
         {Kral}, Q. and {Lagrange}, A. -M. and {Launhardt}, R. and {Matter}, A. and
         {M{\'e}nard}, F. and {Milli}, J. and {Molli{\`e}re}, P. and
         {Olofsson}, J. and {P{\'e}rez}, L. and {Pinilla}, P. and {Pinte}, C. and
         {Quanz}, S.~P. and {Schmidt}, T. and {Udry}, S. and {Wahhaj}, Z. and
         {Williams}, J.~P. and {Buenzli}, E. and {Cudel}, M. and {Dominik}, C. and
         {Galicher}, R. and {Kasper}, M. and {Lannier}, J. and {Mesa}, D. and
         {Mouillet}, D. and {Peretti}, S. and {Perrot}, C. and {Salter}, G. and
         {Sissa}, E. and {Wildi}, F. and {Abe}, L. and {Antichi}, J. and
         {Augereau}, J. -C. and {Baruffolo}, A. and {Baudoz}, P. and
         {Bazzon}, A. and {Beuzit}, J. -L. and {Blanchard}, P. and
         {Brems}, S.~S. and {Buey}, T. and {De Caprio}, V. and {Carbillet}, M. and
         {Carle}, M. and {Cascone}, E. and {Cheetham}, A. and {Claudi}, R. and
         {Costille}, A. and {Delboulb{\'e}}, A. and {Dohlen}, K. and
         {Fantinel}, D. and {Feautrier}, P. and {Fusco}, T. and {Giro}, E. and
         {Gluck}, L. and {Gry}, C. and {Hubin}, N. and {Hugot}, E. and
         {Jaquet}, M. and {Le Mignant}, D. and {Llored}, M. and {Madec}, F. and
         {Magnard}, Y. and {Martinez}, P. and {Maurel}, D. and {Meyer}, M. and
         {M{\"o}ller-Nilsson}, O. and {Moulin}, T. and {Mugnier}, L. and
         {Orign{\'e}}, A. and {Pavlov}, A. and {Perret}, D. and {Petit}, C. and
         {Pragt}, J. and {Puget}, P. and {Rabou}, P. and {Ramos}, J. and
         {Rigal}, F. and {Rochat}, S. and {Roelfsema}, R. and {Rousset}, G. and
         {Roux}, A. and {Salasnich}, B. and {Sauvage}, J. -F. and {Sevin}, A. and
         {Soenke}, C. and {Stadler}, E. and {Suarez}, M. and {Turatto}, M. and
         {Weber}, L.},
        title = "{Discovery of a planetary-mass companion within the gap of the transition disk around PDS 70}",
      journal = {A\&A},
         year = "2018",
        month = "Sep",
       volume = {617},
          eid = {A44},
        pages = {A44},
          doi = {10.1051/0004-6361/201832957},
archivePrefix = {arXiv},
       eprint = {1806.11568},
 primaryClass = {astro-ph.EP},
       adsurl = {https://ui.adsabs.harvard.edu/abs/2018A&A...617A..44K}
}

@ARTICLE{Kipping2022,
       author = {{Kipping}, David and {Bryson}, Steve and {Burke}, Chris and {Christiansen}, Jessie and {Hardegree-Ullman}, Kevin and {Quarles}, Billy and {Hansen}, Brad and {Szul{\'a}gyi}, Judit and {Teachey}, Alex},
        title = "{An exomoon survey of 70 cool giant exoplanets and the new candidate Kepler-1708 b-i}",
      journal = {Nature Astronomy},
         year = 2022,
        month = jan,
       volume = {6},
        pages = {367-380},
          doi = {10.1038/s41550-021-01539-1},
archivePrefix = {arXiv},
       eprint = {2201.04643},
 primaryClass = {astro-ph.EP},
       adsurl = {https://ui.adsabs.harvard.edu/abs/2022NatAs...6..367K}
}

@ARTICLE{Kouwenhoven2005,
       author = {{Kouwenhoven}, M.~B.~N. and {Brown}, A.~G.~A. and {Zinnecker}, H. and {Kaper}, L. and {Portegies Zwart}, S.~F.},
        title = "{The primordial binary population.  I. A near-infrared adaptive optics search for close visual companions to A star members of Scorpius OB2}",
      journal = {\aap},
         year = 2005,
        month = jan,
       volume = {430},
        pages = {137-154},
          doi = {10.1051/0004-6361:20048124},
archivePrefix = {arXiv},
       eprint = {astro-ph/0410106},
 primaryClass = {astro-ph},
       adsurl = {https://ui.adsabs.harvard.edu/abs/2005A&A...430..137K}
}

@ARTICLE{Lachapelle2015,
       author = {{Lachapelle}, Fran{\c{c}}ois-Ren{\'e} and {Lafreni{\`e}re}, David and
         {Gagn{\'e}}, Jonathan and {Jayawardhana}, Ray and {Janson}, Markus and
         {Helling}, Christiane and {Witte}, Soeren},
        title = "{Characterization of Low-mass, Wide-separation Substellar Companions to Stars in Upper Scorpius: Near-infrared Photometry and Spectroscopy}",
      journal = {\apj},
         year = 2015,
        month = mar,
       volume = {802},
       number = {1},
          eid = {61},
        pages = {61},
          doi = {10.1088/0004-637X/802/1/61},
archivePrefix = {arXiv},
       eprint = {1503.07586},
 primaryClass = {astro-ph.SR},
       adsurl = {https://ui.adsabs.harvard.edu/abs/2015ApJ...802...61L}
}

@ARTICLE{Lafreniere2009,
       author = {{Lafreni{\`e}re}, David and {Marois}, Christian and {Doyon}, Ren{\'e} and {Barman}, Travis},
        title = "{HST/NICMOS Detection of HR 8799 b in 1998}",
      journal = {\apjl},
         year = 2009,
        month = apr,
       volume = {694},
       number = {2},
        pages = {L148-L152},
          doi = {10.1088/0004-637X/694/2/L148},
archivePrefix = {arXiv},
       eprint = {0902.3247},
 primaryClass = {astro-ph.EP},
       adsurl = {https://ui.adsabs.harvard.edu/abs/2009ApJ...694L.148L}
}

@ARTICLE{Lafreniere2010,
       author = {{Lafreni{\`e}re}, David and {Jayawardhana}, Ray and {van Kerkwijk}, Marten H.},
        title = "{The Directly Imaged Planet Around the Young Solar Analog 1RXS J160929.1 - 210524: Confirmation of Common Proper Motion, Temperature, and Mass}",
      journal = {\apj},
         year = 2010,
        month = aug,
       volume = {719},
       number = {1},
        pages = {497-504},
          doi = {10.1088/0004-637X/719/1/497},
archivePrefix = {arXiv},
       eprint = {1006.3070},
 primaryClass = {astro-ph.EP},
       adsurl = {https://ui.adsabs.harvard.edu/abs/2010ApJ...719..497L}
}

@ARTICLE{Lafreniere2011,
       author = {{Lafreni{\`e}re}, David and {Jayawardhana}, Ray and {Janson}, Markus and {Helling}, Christiane and {Witte}, Soeren and {Hauschildt}, Peter},
        title = "{Discovery of an \raisebox{-0.5ex}\textasciitilde23 M $_{Jup}$ Brown Dwarf Orbiting \raisebox{-0.5ex}\textasciitilde700 AU from the Massive Star HIP 78530 in Upper Scorpius}",
      journal = {\apj},
         year = 2011,
        month = mar,
       volume = {730},
       number = {1},
          eid = {42},
        pages = {42},
          doi = {10.1088/0004-637X/730/1/42},
archivePrefix = {arXiv},
       eprint = {1101.4666},
 primaryClass = {astro-ph.SR},
       adsurl = {https://ui.adsabs.harvard.edu/abs/2011ApJ...730...42L}
}

@article{Lagrange2010,
	Adsurl = {http://adsabs.harvard.edu/abs/2010Sci...329...57L},
	Archiveprefix = {arXiv},
	Author = {{Lagrange}, A.-M. and {Bonnefoy}, M. and {Chauvin}, G. and {Apai}, D. and {Ehrenreich}, D. and {Boccaletti}, A. and {Gratadour}, D. and {Rouan}, D. and {Mouillet}, D. and {Lacour}, S. and {Kasper}, M.},
	Doi = {10.1126/science.1187187},
	Eprint = {1006.3314},
	Journal = {Science},
	Month = jul,
	Pages = {57},
	Primaryclass = {astro-ph.EP},
	Title = {{A Giant Planet Imaged in the Disk of the Young Star {$\beta$} Pictoris}},
	Volume = 329,
	Year = 2010}

@ARTICLE{Lagrange2025,
       author = {{Lagrange}, A. -M. and {Wilkinson}, C. and {M{\^a}lin}, M. and {Boccaletti}, A. and {Perrot}, C. and {Matr{\`a}}, L. and {Combes}, F. and {Beust}, H. and {Rouan}, D. and {Chomez}, A. and {Milli}, J. and {Charnay}, B. and {Mazevet}, S. and {Flasseur}, O. and {Olofsson}, J. and {Bayo}, A. and {Kral}, Q. and {Carter}, A. and {Crotts}, K.~A. and {Delorme}, P. and {Chauvin}, G. and {Thebault}, P. and {Rubini}, P. and {Kiefer}, F. and {Radcliffe}, A. and {Mazoyer}, J. and {Bodrito}, T. and {Stasevic}, S. and {Langlois}, M.},
        title = "{Evidence for a sub-Jovian planet in the young TWA 7 disk}",
      journal = {\nat},
         year = 2025,
        month = jun,
       volume = {642},
       number = {8069},
        pages = {905-908},
          doi = {10.1038/s41586-025-09150-4},
archivePrefix = {arXiv},
       eprint = {2502.15081},
 primaryClass = {astro-ph.EP},
       adsurl = {https://ui.adsabs.harvard.edu/abs/2025Natur.642..905L}
}

@ARTICLE{Lazzoni2020a,
       author = {{Lazzoni}, C. and {Zurlo}, A. and {Desidera}, S. and {Mesa}, D. and {Fontanive}, C. and {Bonavita}, M. and {Ertel}, S. and {Rice}, K. and {Vigan}, A. and {Boccaletti}, A. and {Bonnefoy}, M. and {Chauvin}, G. and {Delorme}, P. and {Gratton}, R. and {Houll{\'e}}, M. and {Maire}, A.~L. and {Meyer}, M. and {Rickman}, E. and {Spalding}, E.~A. and {Asensio-Torres}, R. and {Langlois}, M. and {M{\"u}ller}, A. and {Baudino}, J. -L. and {Beuzit}, J. -L. and {Biller}, B. and {Brandner}, W. and {Buenzli}, E. and {Cantalloube}, F. and {Cheetham}, A. and {Cudel}, M. and {Feldt}, M. and {Galicher}, R. and {Janson}, M. and {Hagelberg}, J. and {Henning}, T. and {Kasper}, M. and {Keppler}, M. and {Lagrange}, A. -M. and {Lannier}, J. and {LeCoroller}, H. and {Mouillet}, D. and {Peretti}, S. and {Perrot}, C. and {Salter}, G. and {Samland}, M. and {Schmidt}, T. and {Sissa}, E. and {Wildi}, F.},
        title = "{The search for disks or planetary objects around directly imaged companions: a candidate around DH Tauri B}",
      journal = {\aap},
         year = 2020,
        month = sep,
       volume = {641},
          eid = {A131},
        pages = {A131},
          doi = {10.1051/0004-6361/201937290},
archivePrefix = {arXiv},
       eprint = {2007.10097},
 primaryClass = {astro-ph.EP},
       adsurl = {https://ui.adsabs.harvard.edu/abs/2020A&A...641A.131L}
}

@ARTICLE{Lazzoni2024,
       author = {{Lazzoni}, C. and {Rice}, K. and {Zurlo}, A. and {Hinkley}, S. and {Desidera}, S.},
        title = "{Binary planet formation through tides}",
      journal = {\mnras},
         year = 2024,
        month = jan,
       volume = {527},
       number = {2},
        pages = {3837-3846},
          doi = {10.1093/mnras/stad3443},
archivePrefix = {arXiv},
       eprint = {2311.01618},
 primaryClass = {astro-ph.EP},
       adsurl = {https://ui.adsabs.harvard.edu/abs/2024MNRAS.527.3837L}
}

@ARTICLE{MacGregor2017,
       author = {{MacGregor}, Meredith A. and {Wilner}, David J. and {Czekala}, Ian and
         {Andrews}, Sean M. and {Dai}, Y. Sophia and {Herczeg}, Gregory J. and
         {Kratter}, Kaitlin M. and {Kraus}, Adam L. and {Ricci}, Luca and
         {Testi}, Leonardo},
        title = "{ALMA Measurements of Circumstellar Material in the GQ Lup System}",
      journal = {ApJ},
         year = "2017",
        month = "Jan",
       volume = {835},
       number = {1},
          eid = {17},
        pages = {17},
          doi = {10.3847/1538-4357/835/1/17},
archivePrefix = {arXiv},
       eprint = {1611.06229},
 primaryClass = {astro-ph.SR},
       adsurl = {https://ui.adsabs.harvard.edu/abs/2017ApJ...835...17M}
}

@ARTICLE{Macintosh2014,
       author = {{Macintosh}, Bruce and {Graham}, James R. and {Ingraham}, Patrick and
         {Konopacky}, Quinn and {Marois}, Christian and {Perrin}, Marshall and
         {Poyneer}, Lisa and {Bauman}, Brian and {Barman}, Travis and
         {Burrows}, Adam S. and {Cardwell}, Andrew and {Chilcote}, Jeffrey and
         {De Rosa}, Robert J. and {Dillon}, Daren and {Doyon}, Rene and
         {Dunn}, Jennifer and {Erikson}, Darren and {Fitzgerald}, Michael P. and
         {Gavel}, Donald and {Goodsell}, Stephen and {Hartung}, Markus and
         {Hibon}, Pascale and {Kalas}, Paul and {Larkin}, James and
         {Maire}, Jerome and {Marchis}, Franck and {Marley}, Mark S. and
         {McBride}, James and {Millar-Blanchaer}, Max and {Morzinski}, Katie and
         {Norton}, Andrew and {Oppenheimer}, B.~R. and {Palmer}, David and
         {Patience}, Jennifer and {Pueyo}, Laurent and {Rantakyro}, Fredrik and
         {Sadakuni}, Naru and {Saddlemyer}, Leslie and {Savransky}, Dmitry and
         {Serio}, Andrew and {Soummer}, Remi and {Sivaramakrishnan}, Anand and
         {Song}, Inseok and {Thomas}, Sandrine and {Wallace}, J. Kent and
         {Wiktorowicz}, Sloane and {Wolff}, Schuyler},
        title = "{First light of the Gemini Planet Imager}",
      journal = {Proceedings of the National Academy of Science},
         year = "2014",
        month = "Sep",
       volume = {111},
       number = {35},
        pages = {12661-12666},
          doi = {10.1073/pnas.1304215111},
archivePrefix = {arXiv},
       eprint = {1403.7520},
 primaryClass = {astro-ph.EP},
       adsurl = {https://ui.adsabs.harvard.edu/abs/2014PNAS..11112661M}
}

@INPROCEEDINGS{Maire2016b,
       author = {{Maire}, Anne-Lise and {Langlois}, Maud and {Dohlen}, Kjetil and
         {Lagrange}, Anne-Marie and {Gratton}, Raffaele and
         {Chauvin}, Ga{\"e}l. and {Desidera}, Silvano and {Girard}, Julien H. and
         {Milli}, Julien and {Vigan}, Arthur and {Zins}, Gerard and
         {Delorme}, Philippe and {Beuzit}, Jean-Luc and {Claudi}, Riccardo U. and
         {Feldt}, Markus and {Mouillet}, David and {Puget}, Pascal and
         {Turatto}, Massimo and {Wildi}, Fran{\c{c}}ois},
        title = "{SPHERE IRDIS and IFS astrometric strategy and calibration}",
    booktitle = {Proc. SPIE},
         year = "2016",
       series = {Society of Photo-Optical Instrumentation Engineers (SPIE) Conference Series},
       volume = {9908},
        month = "Aug",
          eid = {990834},
        pages = {990834},
          doi = {10.1117/12.2233013},
archivePrefix = {arXiv},
       eprint = {1609.06681},
 primaryClass = {astro-ph.IM},
       adsurl = {https://ui.adsabs.harvard.edu/abs/2016SPIE.9908E..34M}
}

@ARTICLE{Maire2021a,
       author = {{Maire}, A. -L. and {Chauvin}, G. and {Vigan}, A. and {Gratton}, R. and {Langlois}, M. and {Girard}, J.~H. and {Kenworthy}, M.~A. and {Pott}, J. -U. and {Henning}, T. and {Kervella}, P. and {Lacour}, S. and {Rickman}, E.~L. and {Boccaletti}, A. and {Delorme}, P. and {Meyer}, M.~R. and {Nowak}, M. and {Quanz}, S.~P. and {Zurlo}, A.},
        title = "{High-precision Astrometric Studies in Direct Imaging with SPHERE}",
      journal = {The Messenger},
         year = 2021,
        month = jun,
       volume = {183},
        pages = {7-12},
          doi = {10.18727/0722-6691/5228},
archivePrefix = {arXiv},
       eprint = {2103.13700},
 primaryClass = {astro-ph.IM},
       adsurl = {https://ui.adsabs.harvard.edu/abs/2021Msngr.183....7M}
}

@INPROCEEDINGS{Males2018,
       author = {{Males}, Jared R. and {Close}, Laird M. and {Miller}, Kelsey and {Schatz}, Lauren and {Doelman}, David and {Lumbres}, Jennifer and {Snik}, Frans and {Rodack}, Alex and {Knight}, Justin and {Van Gorkom}, Kyle and {Long}, Joseph D. and {Hedglen}, Alex and {Kautz}, Maggie and {Jovanovic}, Nemanja and {Morzinski}, Katie and {Guyon}, Olivier and {Douglas}, Ewan and {Follette}, Katherine B. and {Lozi}, Julien and {Bohlman}, Chris and {Durney}, Olivier and {Gasho}, Victor and {Hinz}, Phil and {Ireland}, Michael and {Jean}, Madison and {Keller}, Christoph and {Kenworthy}, Matt and {Mazin}, Ben and {Noenickx}, Jamison and {Alfred}, Dan and {Perez}, Kevin and {Sanchez}, Anna and {Sauve}, Corwynn and {Weinberger}, Alycia and {Conrad}, Al},
        title = "{MagAO-X: project status and first laboratory results}",
    booktitle = {Adaptive Optics Systems VI},
         year = 2018,
       editor = {{Close}, Laird M. and {Schreiber}, Laura and {Schmidt}, Dirk},
       series = {Society of Photo-Optical Instrumentation Engineers (SPIE) Conference Series},
       volume = {10703},
        month = jul,
          eid = {1070309},
        pages = {1070309},
          doi = {10.1117/12.2312992},
archivePrefix = {arXiv},
       eprint = {1807.04315},
 primaryClass = {astro-ph.IM},
       adsurl = {https://ui.adsabs.harvard.edu/abs/2018SPIE10703E..09M}
}

@INPROCEEDINGS{Males2024,
       author = {{Males}, Jared R. and {Close}, Laird M. and {Haffert}, Sebastiaan Y. and {Kautz}, Maggie Y. and {Kueny}, Jay and {Long}, Joseph D. and {McEwen}, Eden and {Swimmer}, Noah and {Bailey}, John I. and {Foster}, Warren and {Mazin}, Benjamin A. and {Pearce}, Logan and {Liberman}, Joshua and {Twitchell}, Katie and {Weinberger}, Alycia J. and {Guyon}, Olivier and {Hedglen}, Alexander D. and {McLeod}, Avalon and {Roberts}, Roz and {Van Gorkom}, Kyle and {Li}, Jialin and {Doty}, Isabella and {Gasho}, Victor},
        title = "{MagAO-X: commissioning results and status of ongoing upgrades}",
    booktitle = {Adaptive Optics Systems IX},
         year = 2024,
       editor = {{Jackson}, Kathryn J. and {Schmidt}, Dirk and {Vernet}, Elise},
       series = {Society of Photo-Optical Instrumentation Engineers (SPIE) Conference Series},
       volume = {13097},
        month = aug,
          eid = {1309709},
        pages = {1309709},
          doi = {10.1117/12.3019464},
       adsurl = {https://ui.adsabs.harvard.edu/abs/2024SPIE13097E..09M}
}

@article{Marois2010a,
	Adsurl = {http://adsabs.harvard.edu/abs/2010Natur.468.1080M},
	Archiveprefix = {arXiv},
	Author = {{Marois}, C. and {Zuckerman}, B. and {Konopacky}, Q.~M. and {Macintosh}, B. and {Barman}, T.},
	Doi = {10.1038/nature09684},
	Eprint = {1011.4918},
	Journal = {\nat},
	Month = dec,
	Pages = {1080-1083},
	Primaryclass = {astro-ph.EP},
	Title = {{Images of a fourth planet orbiting HR 8799}},
	Volume = 468,
	Year = 2010}

@ARTICLE{Martinez2022,
       author = {{Martinez}, Raquel A. and {Kraus}, Adam L.},
        title = "{A Mid-infrared Study of Directly Imaged Planetary-mass Companions Using Archival Spitzer/IRAC Images}",
      journal = {\aj},
         year = 2022,
        month = jan,
       volume = {163},
       number = {1},
          eid = {36},
        pages = {36},
          doi = {10.3847/1538-3881/ac3745},
archivePrefix = {arXiv},
       eprint = {2111.03087},
 primaryClass = {astro-ph.EP},
       adsurl = {https://ui.adsabs.harvard.edu/abs/2022AJ....163...36M}
}

@ARTICLE{Mawet2014,
       author = {{Mawet}, D. and {Milli}, J. and {Wahhaj}, Z. and {Pelat}, D. and
         {Absil}, O. and {Delacroix}, C. and {Boccaletti}, A. and {Kasper}, M. and
         {Kenworthy}, M. and {Marois}, C. and {Mennesson}, B. and {Pueyo}, L.},
        title = "{Fundamental Limitations of High Contrast Imaging Set by Small Sample Statistics}",
      journal = {\apj},
         year = 2014,
        month = sep,
       volume = {792},
       number = {2},
          eid = {97},
        pages = {97},
          doi = {10.1088/0004-637X/792/2/97},
archivePrefix = {arXiv},
       eprint = {1407.2247},
 primaryClass = {astro-ph.IM},
       adsurl = {https://ui.adsabs.harvard.edu/abs/2014ApJ...792...97M}
}

@ARTICLE{Mesa2019,
       author = {{Mesa}, D. and {Keppler}, M. and {Cantalloube}, F. and {Rodet}, L. and
         {Charnay}, B. and {Gratton}, R. and {Langlois}, M. and
         {Boccaletti}, A. and {Bonnefoy}, M. and {Vigan}, A. and {Flasseur}, O. and
         {Bae}, J. and {Benisty}, M. and {Chauvin}, G. and {de Boer}, J. and
         {Desidera}, S. and {Henning}, T. and {Lagrange}, A. -M. and
         {Meyer}, M. and {Milli}, J. and {M{\"u}ller}, A. and {Pairet}, B. and
         {Zurlo}, A. and {Antoniucci}, S. and {Baudino}, J. -L. and
         {Brown Sevilla}, S. and {Cascone}, E. and {Cheetham}, A. and
         {Claudi}, R.~U. and {Delorme}, P. and {D'Orazi}, V. and {Feldt}, M. and
         {Hagelberg}, J. and {Janson}, M. and {Kral}, Q. and {Lagadec}, E. and
         {Lazzoni}, C. and {Ligi}, R. and {Maire}, A. -L. and {Martinez}, P. and
         {Menard}, F. and {Meunier}, N. and {Perrot}, C. and {Petrus}, S. and
         {Pinte}, C. and {Rickman}, E.~L. and {Rochat}, S. and {Rouan}, D. and
         {Samland}, M. and {Sauvage}, J. -F. and {Schmidt}, T. and {Udry}, S. and
         {Weber}, L. and {Wildi}, F.},
        title = "{VLT/SPHERE exploration of the young multiplanetary system PDS70}",
      journal = {\aap},
         year = 2019,
        month = dec,
       volume = {632},
          eid = {A25},
        pages = {A25},
          doi = {10.1051/0004-6361/201936764},
archivePrefix = {arXiv},
       eprint = {1910.11169},
 primaryClass = {astro-ph.EP},
       adsurl = {https://ui.adsabs.harvard.edu/abs/2019A&A...632A..25M}
}

@ARTICLE{Mesa2023,
       author = {{Mesa}, D. and {Gratton}, R. and {Kervella}, P. and {Bonavita}, M. and {Desidera}, S. and {D'Orazi}, V. and {Marino}, S. and {Zurlo}, A. and {Rigliaco}, E.},
        title = "{AF Lep b: The lowest-mass planet detected by coupling astrometric and direct imaging data}",
      journal = {\aap},
         year = 2023,
        month = apr,
       volume = {672},
          eid = {A93},
        pages = {A93},
          doi = {10.1051/0004-6361/202345865},
archivePrefix = {arXiv},
       eprint = {2302.06213},
 primaryClass = {astro-ph.EP},
       adsurl = {https://ui.adsabs.harvard.edu/abs/2023A&A...672A..93M}
}

@INPROCEEDINGS{Milli2018,
       author = {{Milli}, J. and {Kasper}, M. and {Bourget}, P. and {Pannetier}, C. and
         {Mouillet}, D. and {Sauvage}, J. -F. and {Reyes}, C. and {Fusco}, T. and
         {Cantalloube}, F. and {Tristam}, K. and {Wahhaj}, Z. and
         {Beuzit}, J. -L. and {Girard}, J.~H. and {Mawet}, D. and {Telle}, A. and
         {Vigan}, A. and {N'Diaye}, M.},
        title = "{Low wind effect on VLT/SPHERE: impact, mitigation strategy, and results}",
    booktitle = {Proc. SPIE},
         year = "2018",
       series = {Society of Photo-Optical Instrumentation Engineers (SPIE) Conference Series},
       volume = {10703},
        month = "Jul",
          eid = {107032A},
        pages = {107032A},
          doi = {10.1117/12.2311499},
archivePrefix = {arXiv},
       eprint = {1806.05370},
 primaryClass = {astro-ph.IM},
       adsurl = {https://ui.adsabs.harvard.edu/abs/2018SPIE10703E..2AM}
}

@ARTICLE{Neuhauser2003,
       author = {{Neuh{\"a}user}, R. and {Guenther}, E.~W. and {Alves}, J. and {Hu{\'e}lamo}, N. and {Ott}, Th. and {Eckart}, A.},
        title = "{An infrared imaging search for low-mass companions to members of the young nearby {\ensuremath{\beta}} Pic and Tucana/Horologium associations}",
      journal = {Astronomische Nachrichten},
         year = 2003,
        month = nov,
       volume = {324},
       number = {6},
        pages = {535-542},
          doi = {10.1002/asna.200310167},
       adsurl = {https://ui.adsabs.harvard.edu/abs/2003AN....324..535N}
}

@ARTICLE{Neuhauser2005b,
       author = {{Neuh{\"a}user}, R. and {Guenther}, E.~W. and {Wuchterl}, G. and {Mugrauer}, M. and {Bedalov}, A. and {Hauschildt}, P.~H.},
        title = "{Evidence for a co-moving sub-stellar companion of GQ Lup}",
      journal = {\aap},
         year = 2005,
        month = may,
       volume = {435},
       number = {1},
        pages = {L13-L16},
          doi = {10.1051/0004-6361:200500104},
archivePrefix = {arXiv},
       eprint = {astro-ph/0503691},
 primaryClass = {astro-ph},
       adsurl = {https://ui.adsabs.harvard.edu/abs/2005A&A...435L..13N}
}

@ARTICLE{Neuhauser2008,
       author = {{Neuh{\"a}user}, R. and {Mugrauer}, M. and {Seifahrt}, A. and {Schmidt}, T.~O.~B. and {Vogt}, N.},
        title = "{Astrometric and photometric monitoring of GQ Lupi and its sub-stellar companion}",
      journal = {\aap},
         year = 2008,
        month = jun,
       volume = {484},
       number = {1},
        pages = {281-291},
          doi = {10.1051/0004-6361:20078493},
archivePrefix = {arXiv},
       eprint = {0801.2287},
 primaryClass = {astro-ph},
       adsurl = {https://ui.adsabs.harvard.edu/abs/2008A&A...484..281N}
}

@ARTICLE{Neuhauser2011,
       author = {{Neuh{\"a}user}, R. and {Ginski}, C. and {Schmidt}, T.~O.~B. and {Mugrauer}, M.},
        title = "{Further deep imaging of HR 7329 A ({\ensuremath{\eta}} Tel A) and its brown dwarf companion B}",
      journal = {\mnras},
         year = 2011,
        month = sep,
       volume = {416},
       number = {2},
        pages = {1430-1435},
          doi = {10.1111/j.1365-2966.2011.19139.x},
archivePrefix = {arXiv},
       eprint = {1106.1388},
 primaryClass = {astro-ph.SR},
       adsurl = {https://ui.adsabs.harvard.edu/abs/2011MNRAS.416.1430N}
}

@INPROCEEDINGS{Nielsen2020,
       author = {{Nielsen}, E.~L. and {De Rosa}, R. and {Macintosh}, B. and {Wang}, J. and
         {Ruffio}, J. and {Chiang}, E. and {Marley}, M. and {Saumon}, D. and
         {Savransky}, D. and {Gemini Planet Imager Exoplanet Survey Team}},
        title = "{The Gemini Planet Imager Exoplanet Survey: Giant Planet and Brown Dwarf Demographics from 10-100 AU}",
    booktitle = {American Astronomical Society Meeting Abstracts \#235},
         year = 2020,
       series = {American Astronomical Society Meeting Abstracts},
       volume = {235},
        month = jan,
          eid = {280.02},
        pages = {280.02},
       adsurl = {https://ui.adsabs.harvard.edu/abs/2020AAS...23528002N}
}

@ARTICLE{Nogueira2024,
       author = {{Nogueira}, P.~H. and {Lazzoni}, C. and {Zurlo}, A. and {Bhowmik}, T. and {Donoso-Oliva}, C. and {Desidera}, S. and {Milli}, J. and {P{\'e}rez}, S. and {Delorme}, P. and {Fernadez}, A. and {Langlois}, M. and {Petrus}, S. and {Cabrera-Vives}, G. and {Chauvin}, G.},
        title = "{Astrometric and photometric characterization of {\ensuremath{\eta}} Tel B combining two decades of observations}",
      journal = {\aap},
         year = 2024,
        month = jul,
       volume = {687},
          eid = {A301},
        pages = {A301},
          doi = {10.1051/0004-6361/202449222},
archivePrefix = {arXiv},
       eprint = {2405.04723},
 primaryClass = {astro-ph.SR},
       adsurl = {https://ui.adsabs.harvard.edu/abs/2024A&A...687A.301N}
}

@ARTICLE{Palma-Bifani2023,
       author = {{Palma-Bifani}, P. and {Chauvin}, G. and {Bonnefoy}, M. and {Rojo}, P.~M. and {Petrus}, S. and {Rodet}, L. and {Langlois}, M. and {Allard}, F. and {Charnay}, B. and {Desgrange}, C. and {Homeier}, D. and {Lagrange}, A. -M. and {Beuzit}, J. -L. and {Baudoz}, P. and {Boccaletti}, A. and {Chomez}, A. and {Delorme}, P. and {Desidera}, S. and {Feldt}, M. and {Ginski}, C. and {Gratton}, R. and {Maire}, A. -L. and {Meyer}, M. and {Samland}, M. and {Snellen}, I. and {Vigan}, A. and {Zhang}, Y.},
        title = "{Peering into the young planetary system AB Pic. Atmosphere, orbit, obliquity, and second planetary candidate}",
      journal = {\aap},
         year = 2023,
        month = feb,
       volume = {670},
          eid = {A90},
        pages = {A90},
          doi = {10.1051/0004-6361/202244294},
archivePrefix = {arXiv},
       eprint = {2211.01474},
 primaryClass = {astro-ph.EP},
       adsurl = {https://ui.adsabs.harvard.edu/abs/2023A&A...670A..90P}
}

@ARTICLE{Patience2012,
       author = {{Patience}, J. and {King}, R.~R. and {De Rosa}, R.~J. and {Vigan}, A. and {Witte}, S. and {Rice}, E. and {Helling}, Ch. and {Hauschildt}, P.},
        title = "{Spectroscopy across the brown dwarf/planetary mass boundary. I. Near-infrared JHK spectra}",
      journal = {\aap},
         year = 2012,
        month = apr,
       volume = {540},
          eid = {A85},
        pages = {A85},
          doi = {10.1051/0004-6361/201118058},
archivePrefix = {arXiv},
       eprint = {1201.3921},
 primaryClass = {astro-ph.EP},
       adsurl = {https://ui.adsabs.harvard.edu/abs/2012A&A...540A..85P}
}

@ARTICLE{Perez2019,
       author = {{P{\'e}rez}, Sebasti{\'a}n and {Marino}, Sebasti{\'a}n and
         {Casassus}, Simon and {Baruteau}, Cl{\'e}ment and {Zurlo}, Alice and
         {Flores}, Christian and {Chauvin}, Gael},
        title = "{Upper limits on protolunar disc masses using ALMA observations of directly imaged exoplanets}",
      journal = {MNRAS},
         year = "2019",
        month = "Sep",
       volume = {488},
       number = {1},
        pages = {1005-1011},
          doi = {10.1093/mnras/stz1775},
archivePrefix = {arXiv},
       eprint = {1906.11774},
 primaryClass = {astro-ph.EP},
       adsurl = {https://ui.adsabs.harvard.edu/abs/2019MNRAS.488.1005P}
}

@ARTICLE{Phillips2020,
       author = {{Phillips}, M.~W. and {Tremblin}, P. and {Baraffe}, I. and {Chabrier}, G. and {Allard}, N.~F. and {Spiegelman}, F. and {Goyal}, J.~M. and {Drummond}, B. and {H{\'e}brard}, E.},
        title = "{A new set of atmosphere and evolution models for cool T-Y brown dwarfs and giant exoplanets}",
      journal = {\aap},
         year = 2020,
        month = may,
       volume = {637},
          eid = {A38},
        pages = {A38},
          doi = {10.1051/0004-6361/201937381},
archivePrefix = {arXiv},
       eprint = {2003.13717},
 primaryClass = {astro-ph.SR},
       adsurl = {https://ui.adsabs.harvard.edu/abs/2020A&A...637A..38P}
}

@ARTICLE{Potter2002,
       author = {{Potter}, D. and {Mart{\'\i}n}, E.~L. and {Cushing}, M.~C. and
         {Baudoz}, P. and {Brandner}, W. and {Guyon}, O. and {Neuh{\"a}user}, R.},
        title = "{Hokupa'a-Gemini Discovery of Two Ultracool Companions to the Young Star HD 130948}",
      journal = {ApJ},
         year = "2002",
        month = "Mar",
       volume = {567},
       number = {2},
        pages = {L133-L136},
          doi = {10.1086/339999},
archivePrefix = {arXiv},
       eprint = {astro-ph/0201431},
 primaryClass = {astro-ph},
       adsurl = {https://ui.adsabs.harvard.edu/abs/2002ApJ...567L.133P}
}

@ARTICLE{Schmidt2008,
       author = {{Schmidt}, T.~O.~B. and {Neuh{\"a}user}, R. and {Seifahrt}, A. and
         {Vogt}, N. and {Bedalov}, A. and {Helling}, Ch. and {Witte}, S. and
         {Hauschildt}, P.~H.},
        title = "{Direct evidence of a sub-stellar companion around CT Chamaeleontis}",
      journal = {\aap},
         year = 2008,
        month = nov,
       volume = {491},
       number = {1},
        pages = {311-320},
          doi = {10.1051/0004-6361:20078840},
archivePrefix = {arXiv},
       eprint = {0809.2812},
 primaryClass = {astro-ph},
       adsurl = {https://ui.adsabs.harvard.edu/abs/2008A&A...491..311S}
}

@ARTICLE{Schwarz2016,
       author = {{Schwarz}, Henriette and {Ginski}, Christian and {de Kok}, Remco J. and
         {Snellen}, Ignas A.~G. and {Brogi}, Matteo and {Birkby}, Jayne L.},
        title = "{The slow spin of the young substellar companion GQ Lupi b and its orbital configuration}",
      journal = {A\&A},
         year = "2016",
        month = "Sep",
       volume = {593},
          eid = {A74},
        pages = {A74},
          doi = {10.1051/0004-6361/201628908},
archivePrefix = {arXiv},
       eprint = {1607.00012},
 primaryClass = {astro-ph.EP},
       adsurl = {https://ui.adsabs.harvard.edu/abs/2016A&A...593A..74S}
}

@ARTICLE{Seifahrt2007,
       author = {{Seifahrt}, A. and {Neuh{\"a}user}, R. and {Hauschildt}, P.~H.},
        title = "{Near-infrared integral-field spectroscopy of the companion to GQ Lupi}",
      journal = {A\&A},
         year = "2007",
        month = "Feb",
       volume = {463},
       number = {1},
        pages = {309-313},
          doi = {10.1051/0004-6361:20066463},
archivePrefix = {arXiv},
       eprint = {astro-ph/0612250},
 primaryClass = {astro-ph},
       adsurl = {https://ui.adsabs.harvard.edu/abs/2007A&A...463..309S}
}

@ARTICLE{Sheehan2019,
       author = {{Sheehan}, Patrick D. and {Wu}, Ya-Lin and {Eisner}, Josh A. and {Tobin}, John J.},
        title = "{High-precision Dynamical Masses of Pre-main-sequence Stars with ALMA and Gaia}",
      journal = {\apj},
         year = 2019,
        month = apr,
       volume = {874},
       number = {2},
          eid = {136},
        pages = {136},
          doi = {10.3847/1538-4357/ab09f9},
archivePrefix = {arXiv},
       eprint = {1903.00032},
 primaryClass = {astro-ph.SR},
       adsurl = {https://ui.adsabs.harvard.edu/abs/2019ApJ...874..136S}
}

@ARTICLE{Soummer2012,
       author = {{Soummer}, R{\'e}mi and {Pueyo}, Laurent and {Larkin}, James},
        title = "{Detection and Characterization of Exoplanets and Disks Using Projections on Karhunen-Lo{\`e}ve Eigenimages}",
      journal = {ApJ},
         year = "2012",
        month = "Aug",
       volume = {755},
       number = {2},
          eid = {L28},
        pages = {L28},
          doi = {10.1088/2041-8205/755/2/L28},
archivePrefix = {arXiv},
       eprint = {1207.4197},
 primaryClass = {astro-ph.IM},
       adsurl = {https://ui.adsabs.harvard.edu/abs/2012ApJ...755L..28S}
}

@ARTICLE{Stolker2021,
       author = {{Stolker}, Tomas and {Haffert}, Sebastiaan Y. and {Kesseli}, Aurora Y. and {van Holstein}, Rob G. and {Aoyama}, Yuhiko and {Brinchmann}, Jarle and {Cugno}, Gabriele and {Girard}, Julien H. and {Marleau}, Gabriel-Dominique and {Meyer}, Michael R. and {Milli}, Julien and {Quanz}, Sascha P. and {Snellen}, Ignas A.~G. and {Todorov}, Kamen O.},
        title = "{Characterizing the Protolunar Disk of the Accreting Companion GQ Lupi B}",
      journal = {\aj},
         year = 2021,
        month = dec,
       volume = {162},
       number = {6},
          eid = {286},
        pages = {286},
          doi = {10.3847/1538-3881/ac2c7f},
archivePrefix = {arXiv},
       eprint = {2110.04307},
 primaryClass = {astro-ph.EP},
       adsurl = {https://ui.adsabs.harvard.edu/abs/2021AJ....162..286S}
}

@ARTICLE{Teachey2018,
       author = {{Teachey}, Alex and {Kipping}, David M.},
        title = "{Evidence for a large exomoon orbiting Kepler-1625b}",
      journal = {Science Advances},
         year = "2018",
        month = "Oct",
       volume = {4},
       number = {10},
          eid = {eaav1784},
        pages = {eaav1784},
          doi = {10.1126/sciadv.aav1784},
archivePrefix = {arXiv},
       eprint = {1810.02362},
 primaryClass = {astro-ph.EP},
       adsurl = {https://ui.adsabs.harvard.edu/abs/2018SciA....4.1784T}
}

@ARTICLE{Torres2006,
   author = {{Torres}, C.~A.~O. and {Quast}, G.~R. and {da Silva}, L. and 
	{de La Reza}, R. and {Melo}, C.~H.~F. and {Sterzik}, M.},
    title = "{Search for associations containing young stars (SACY). I. Sample and searching method}",
  journal = {\aap},
   eprint = {arXiv:astro-ph/0609258},
     year = 2006,
    month = dec,
   volume = 460,
    pages = {695-708},
      doi = {10.1051/0004-6361:20065602},
   adsurl = {http://adsabs.harvard.edu/abs/2006A%26A...460..695T}
}

@ARTICLE{VanHolstein2021,
       author = {{van Holstein}, R.~G. and {Stolker}, T. and {Jensen-Clem}, R. and {Ginski}, C. and {Milli}, J. and {de Boer}, J. and {Girard}, J.~H. and {Wahhaj}, Z. and {Bohn}, A.~J. and {Millar-Blanchaer}, M.~A. and {Benisty}, M. and {Bonnefoy}, M. and {Chauvin}, G. and {Dominik}, C. and {Hinkley}, S. and {Keller}, C.~U. and {Keppler}, M. and {Langlois}, M. and {Marino}, S. and {M{\'e}nard}, F. and {Perrot}, C. and {Schmidt}, T.~O.~B. and {Vigan}, A. and {Zurlo}, A. and {Snik}, F.},
        title = "{A survey of the linear polarization of directly imaged exoplanets and brown dwarf companions with SPHERE-IRDIS. First polarimetric detections revealing disks around DH Tau B and GSC 6214-210 B}",
      journal = {\aap},
         year = 2021,
        month = mar,
       volume = {647},
          eid = {A21},
        pages = {A21},
          doi = {10.1051/0004-6361/202039290},
archivePrefix = {arXiv},
       eprint = {2101.04033},
 primaryClass = {astro-ph.EP},
       adsurl = {https://ui.adsabs.harvard.edu/abs/2021A&A...647A..21V}
}

@article{VanLeeuwen2007,
	Adsurl = {http://adsabs.harvard.edu/abs/2007A\%26A...474..653V},
	Archiveprefix = {arXiv},
	Author = {{Van Leeuwen}, F.},
	Doi = {10.1051/0004-6361:20078357},
	Eprint = {0708.1752},
	Journal = {A\&A},
	Month = nov,
	Pages = {653-664},
	Title = {{Validation of the new Hipparcos reduction}},
	Volume = 474,
	Year = 2007}

@ARTICLE{Vigan2017,
   author = {{Vigan}, A. and {Bonavita}, M. and {Biller}, B. and {Forgan}, D. and 
	{Rice}, K. and {Chauvin}, G. and {Desidera}, S. and {Meunier}, J.-C. and 
	{Delorme}, P. and {Schlieder}, J.~E. and {Bonnefoy}, M. and 
	{Carson}, J. and {Covino}, E. and {Hagelberg}, J. and {Henning}, T. and 
	{Janson}, M. and {Lagrange}, A.-M. and {Quanz}, S.~P. and {Zurlo}, A. and 
	{Beuzit}, J.-L. and {Boccaletti}, A. and {Buenzli}, E. and {Feldt}, M. and 
	{Girard}, J.~H.~V. and {Gratton}, R. and {Kasper}, M. and {Le Coroller}, H. and 
	{Mesa}, D. and {Messina}, S. and {Meyer}, M. and {Montagnier}, G. and 
	{Mordasini}, C. and {Mouillet}, D. and {Moutou}, C. and {Reggiani}, M. and 
	{Segransan}, D. and {Thalmann}, C.},
    title = "{The VLT/NaCo large program to probe the occurrence of exoplanets and brown dwarfs at wide orbits. IV. Gravitational instability rarely forms wide, giant planets}",
  journal = {A\&A},
archivePrefix = "arXiv",
   eprint = {1703.05322},
 primaryClass = "astro-ph.EP",
     year = 2017,
    month = jun,
   volume = 603,
      eid = {A3},
    pages = {A3},
      doi = {10.1051/0004-6361/201630133},
   adsurl = {http://adsabs.harvard.edu/abs/2017A\%26A...603A...3V}
}

@ARTICLE{Wahhaj2011,
       author = {{Wahhaj}, Zahed and {Liu}, Michael C. and {Biller}, Beth A. and {Clarke}, Fraser and {Nielsen}, Eric L. and {Close}, Laird M. and {Hayward}, Thomas L. and {Mamajek}, Eric E. and {Cushing}, Michael and {Dupuy}, Trent and {Tecza}, Matthias and {Thatte}, Niranjan and {Chun}, Mark and {Ftaclas}, Christ and {Hartung}, Markus and {Reid}, I. Neill and {Shkolnik}, Evgenya L. and {Alencar}, Silvia H.~P. and {Artymowicz}, Pawel and {Boss}, Alan and {de Gouveia Dal Pino}, Elisabethe and {Gregorio-Hetem}, Jane and {Ida}, Shigeru and {Kuchner}, Marc and {Lin}, Douglas N.~C. and {Toomey}, Douglas W.},
        title = "{The Gemini NICI Planet-finding Campaign: Discovery of a Substellar L Dwarf Companion to the Nearby Young M Dwarf CD-35 2722}",
      journal = {\apj},
         year = 2011,
        month = mar,
       volume = {729},
       number = {2},
          eid = {139},
        pages = {139},
          doi = {10.1088/0004-637X/729/2/139},
archivePrefix = {arXiv},
       eprint = {1101.2893},
 primaryClass = {astro-ph.SR},
       adsurl = {https://ui.adsabs.harvard.edu/abs/2011ApJ...729..139W}
}

@ARTICLE{Wahhaj2021,
       author = {{Wahhaj}, Z. and {Milli}, J. and {Romero}, C. and {Cieza}, L. and {Zurlo}, A. and {Vigan}, A. and {Pe{\~n}a}, E. and {Valdes}, G. and {Cantalloube}, F. and {Girard}, J. and {Pantoja}, B.},
        title = "{A search for a fifth planet around HR 8799 using the star-hopping RDI technique at VLT/SPHERE}",
      journal = {\aap},
         year = 2021,
        month = apr,
       volume = {648},
          eid = {A26},
        pages = {A26},
          doi = {10.1051/0004-6361/202038794},
archivePrefix = {arXiv},
       eprint = {2101.08268},
 primaryClass = {astro-ph.EP},
       adsurl = {https://ui.adsabs.harvard.edu/abs/2021A&A...648A..26W}
}

@ARTICLE{Weible2025,
       author = {{Weible}, Gabriel and {Wagner}, Kevin and {Stone}, Jordan and {Ertel}, Steve and {Apai}, D{\'a}niel and {Kratter}, Kaitlin and {Leisenring}, Jarron},
        title = "{Orbital and Atmospheric Modeling of H II 1348B: An Eccentric Young Substellar Companion in the Pleiades}",
      journal = {\aj},
         year = 2025,
        month = apr,
       volume = {169},
       number = {4},
          eid = {197},
        pages = {197},
          doi = {10.3847/1538-3881/adadf6},
archivePrefix = {arXiv},
       eprint = {2501.16429},
 primaryClass = {astro-ph.EP},
       adsurl = {https://ui.adsabs.harvard.edu/abs/2025AJ....169..197W}
}

@ARTICLE{Wolff2017,
       author = {{Wolff}, Schuyler G. and {M{\'e}nard}, Fran{\c{c}}ois and
         {Caceres}, Claudio and {Lef{\`e}vre}, Charlene and {Bonnefoy}, Mickael and
         {C{\'a}novas}, H{\'e}ctor and {Maret}, S{\'e}bastien and
         {Pinte}, Christophe and {Schreiber}, Matthias R. and
         {van der Plas}, Gerrit},
        title = "{An Upper Limit on the Mass of the Circumplanetary Disk for DH Tau b}",
      journal = {AJ},
         year = "2017",
        month = "Jul",
       volume = {154},
       number = {1},
          eid = {26},
        pages = {26},
          doi = {10.3847/1538-3881/aa74cd},
archivePrefix = {arXiv},
       eprint = {1705.08470},
 primaryClass = {astro-ph.EP},
       adsurl = {https://ui.adsabs.harvard.edu/abs/2017AJ....154...26W}
}

@ARTICLE{Wu2015a,
       author = {{Wu}, Ya-Lin and {Close}, Laird M. and {Males}, Jared R. and
         {Barman}, Travis S. and {Morzinski}, Katie M. and
         {Follette}, Katherine B. and {Bailey}, Vanessa and
         {Rodigas}, Timothy J. and {Hinz}, Philip and {Puglisi}, Alfio and
         {Xompero}, Marco and {Briguglio}, Runa},
        title = "{New Extinction and Mass Estimates from Optical Photometry of the Very Low Mass Brown Dwarf Companion CT Chamaeleontis B with the Magellan AO System}",
      journal = {\apj},
         year = 2015,
        month = mar,
       volume = {801},
       number = {1},
          eid = {4},
        pages = {4},
          doi = {10.1088/0004-637X/801/1/4},
archivePrefix = {arXiv},
       eprint = {1501.01396},
 primaryClass = {astro-ph.SR},
       adsurl = {https://ui.adsabs.harvard.edu/abs/2015ApJ...801....4W}
}

@ARTICLE{Wu2015b,
       author = {{Wu}, Ya-Lin and {Close}, Laird M. and {Males}, Jared R. and {Barman}, Travis S. and {Morzinski}, Katie M. and {Follette}, Katherine B. and {Bailey}, Vanessa P. and {Rodigas}, Timothy J. and {Hinz}, Philip and {Puglisi}, Alfio and {Xompero}, Marco and {Briguglio}, Runa},
        title = "{New Extinction and Mass Estimates of the Low-mass Companion 1RXS 1609 B with the Magellan AO System: Evidence of an Inclined Dust Disk}",
      journal = {\apjl},
         year = 2015,
        month = jul,
       volume = {807},
       number = {1},
          eid = {L13},
        pages = {L13},
          doi = {10.1088/2041-8205/807/1/L13},
archivePrefix = {arXiv},
       eprint = {1506.05816},
 primaryClass = {astro-ph.EP},
       adsurl = {https://ui.adsabs.harvard.edu/abs/2015ApJ...807L..13W}
}

@ARTICLE{Wu2017a,
       author = {{Wu}, Ya-Lin and {Close}, Laird M. and {Eisner}, Josh A. and
         {Sheehan}, Patrick D.},
        title = "{An Explanation of the Very Low Radio Flux of Young Planet-mass Companions}",
      journal = {AJ},
         year = 2017,
        month = dec,
       volume = {154},
       number = {6},
          eid = {234},
        pages = {234},
          doi = {10.3847/1538-3881/aa93db},
archivePrefix = {arXiv},
       eprint = {1710.07489},
 primaryClass = {astro-ph.EP},
       adsurl = {https://ui.adsabs.harvard.edu/abs/2017AJ....154..234W}
}

@ARTICLE{Wu2017b,
       author = {{Wu}, Ya-Lin and {Sheehan}, Patrick D. and {Males}, Jared R. and
         {Close}, Laird M. and {Morzinski}, Katie M. and {Teske}, Johanna K. and
         {Haug-Baltzell}, Asher and {Merchant}, Nirav and {Lyons}, Eric},
        title = "{An ALMA and MagAO Study of the Substellar Companion GQ Lup B*}",
      journal = {ApJ},
         year = "2017",
        month = "Feb",
       volume = {836},
       number = {2},
          eid = {223},
        pages = {223},
          doi = {10.3847/1538-4357/aa5b96},
archivePrefix = {arXiv},
       eprint = {1701.07541},
 primaryClass = {astro-ph.SR},
       adsurl = {https://ui.adsabs.harvard.edu/abs/2017ApJ...836..223W}
}

@ARTICLE{Wu2020,
       author = {{Wu}, Ya-Lin and {Bowler}, Brendan P. and {Sheehan}, Patrick D. and
         {Andrews}, Sean M. and {Herczeg}, Gregory J. and {Kraus}, Adam L. and
         {Ricci}, Luca and {Wilner}, David J. and {Zhu}, Zhaohuan},
        title = "{ALMA 0.88 mm Survey of Disks around Planetary-mass Companions}",
      journal = {\aj},
         year = 2020,
        month = may,
       volume = {159},
       number = {5},
          eid = {229},
        pages = {229},
          doi = {10.3847/1538-3881/ab818c},
archivePrefix = {arXiv},
       eprint = {2003.08658},
 primaryClass = {astro-ph.EP},
       adsurl = {https://ui.adsabs.harvard.edu/abs/2020AJ....159..229W}
}

@ARTICLE{Zhou2014,
       author = {{Zhou}, Yifan and {Herczeg}, Gregory J. and {Kraus}, Adam L. and
         {Metchev}, Stanimir and {Cruz}, Kelle L.},
        title = "{Accretion onto Planetary Mass Companions of Low-mass Young Stars}",
      journal = {ApJ},
         year = 2014,
        month = mar,
       volume = {783},
       number = {1},
          eid = {L17},
        pages = {L17},
          doi = {10.1088/2041-8205/783/1/L17},
archivePrefix = {arXiv},
       eprint = {1401.6545},
 primaryClass = {astro-ph.SR},
       adsurl = {https://ui.adsabs.harvard.edu/abs/2014ApJ...783L..17Z}
}

@ARTICLE{Zurlo2023,
       author = {{Zurlo}, Alice and {Gratton}, Raffaele and {P{\'e}rez}, Sebasti{\'a}n and {Cieza}, Lucas},
        title = "{Observations of planet forming disks in multiple stellar systems}",
      journal = {European Physical Journal Plus},
         year = 2023,
        month = may,
       volume = {138},
       number = {5},
          eid = {411},
        pages = {411},
          doi = {10.1140/epjp/s13360-023-04041-x},
archivePrefix = {arXiv},
       eprint = {2304.14450},
 primaryClass = {astro-ph.SR},
       adsurl = {https://ui.adsabs.harvard.edu/abs/2023EPJP..138..411Z}
}

@ARTICLE{Zurlo2024,
       author = {{Zurlo}, Alice},
        title = "{Direct imaging of exoplanets}",
      journal = {arXiv e-prints},
         year = 2024,
        month = apr,
          eid = {arXiv:2404.05797},
        pages = {arXiv:2404.05797},
          doi = {10.48550/arXiv.2404.05797},
archivePrefix = {arXiv},
       eprint = {2404.05797},
 primaryClass = {astro-ph.EP},
       adsurl = {https://ui.adsabs.harvard.edu/abs/2024arXiv240405797Z}
}

@ARTICLE{Blunt2020,
       author = {{Blunt}, S. and {Wang}, J. J. and {Angelo}, I. and {Ngo}, H. and {Cody}, D. and {De Rosa}, R. J. and {Graham}, J. R. and {Hirsch}, L. and {Nagpal}, V. and {Nielsen}, E. L. and {Pearce}, L. and {Rice}, M. and {Tejada}, R.},
        title = "{orbitize!: A Comprehensive Orbit-fitting Software Package for the High-contrast Imaging Community}",
      journal = {\aj},
         year = 2020,
        month = march,
       volume = {159},
       number = {3},
       eid = {89},
        pages = {8},
          doi = {10.3847/1538-3881/ab6663},
       adsurl = {https://ui.adsabs.harvard.edu/abs/2020AJ....159...89B/abstract}
}

@ARTICLE{Vousden2016,
       author = {{Vousden}, W. D. and {Farr}, W. M. and {Mandel}, I.},
        title = "{Dynamic temperature selection for parallel tempering in Markov chain Monte Carlo simulations}",
      journal = {\mnras},
         year = 2016,
        month = jan,
       volume = {455},
       number = {2},
        pages = {1919-1937},
          doi = {10.1093/mnras/stv2422},
       adsurl = {https://ui.adsabs.harvard.edu/abs/2016MNRAS.455.1919V/abstract}
}

@ARTICLE{Blunt2024,
       author = {{Blunt}, S. and {Wang}, J. and {Hirsch}, L and {Tejada}, R. and {Nagpal}, V. and {Surti}, T. and {Covarrubias}, S. and {McKenna}, T. and {Chávez}, R. and {Llop-Sayson}, J. and {Arora}, M. and {Chavez}, A. and {Cody}, D. and {Choudhary}, S. and {Smith}, A. and {Balmer}, W. and {Stolker}, T. and {Gallamore}, H. and {Ó}, C. and {Nielsen}, E. and {De Rosa}, R. J.},
        title = "{orbitize! v3: Orbit fitting for the High-contrast Imaging Community}",
      journal = {Journal of Open Source Software},
         year = 2024,
        month = sep,
       volume = {9},
       number = {101},
        id = {6756},
          doi = {10.21105/joss.06756},
       adsurl = {https://ui.adsabs.harvard.edu/abs/2024JOSS....9.6756B/abstract}
}

@ARTICLE{Lindegren2021,
       author = {{Lindegren}, L. and {Klioner}, S.~A. and {Hern{\'a}ndez}, J. and {Bombrun}, A. and {Ramos-Lerate}, M. and {Steidelm{\"u}ller}, H. and {Bastian}, U. and {Biermann}, M. and {de Torres}, A. and {Gerlach}, E. and {Geyer}, R. and {Hilger}, T. and {Hobbs}, D. and {Lammers}, U. and {McMillan}, P.~J. and {Stephenson}, C.~A. and {Casta{\~n}eda}, J. and {Davidson}, M. and {Fabricius}, C. and {Gracia-Abril}, G. and {Portell}, J. and {Rowell}, N. and {Teyssier}, D. and {Torra}, F. and {Bartolom{\'e}}, S. and {Clotet}, M. and {Garralda}, N. and {Gonz{\'a}lez-Vidal}, J.~J. and {Torra}, J. and {Abbas}, U. and {Altmann}, M. and {Anglada Varela}, E. and {Balaguer-N{\'u}{\~n}ez}, L. and {Balog}, Z. and {Barache}, C. and {Becciani}, U. and {Bernet}, M. and {Bertone}, S. and {Bianchi}, L. and {Bouquillon}, S. and {Brown}, A.~G.~A. and {Bucciarelli}, B. and {Busonero}, D. and {Butkevich}, A.~G. and {Buzzi}, R. and {Cancelliere}, R. and {Carlucci}, T. and {Charlot}, P. and {Cioni}, M.-R.~L. and {Crosta}, M. and {Crowley}, C. and {del Peloso}, E.~F. and {del Pozo}, E. and {Drimmel}, R. and {Esquej}, P. and {Fienga}, A. and {Fraile}, E. and {Gai}, M. and {Garcia-Reinaldos}, M. and {Guerra}, R. and {Hambly}, N.~C. and {Hauser}, M. and {Jan{\ss}en}, K. and {Jordan}, S. and {Kostrzewa-Rutkowska}, Z. and {Lattanzi}, M.~G. and {Liao}, S. and {Licata}, E. and {Lister}, T.~A. and {L{\"o}ffler}, W. and {Marchant}, J.~M. and {Masip}, A. and {Mignard}, F. and {Mints}, A. and {Molina}, D. and {Mora}, A. and {Morbidelli}, R. and {Murphy}, C.~P. and {Pagani}, C. and {Panuzzo}, P. and {Pe{\~n}alosa Esteller}, X. and {Poggio}, E. and {Re Fiorentin}, P. and {Riva}, A. and {Sagrist{\`a} Sell{\'e}s}, A. and {Sanchez Gimenez}, V. and {Sarasso}, M. and {Sciacca}, E. and {Siddiqui}, H.~I. and {Smart}, R.~L. and {Souami}, D. and {Spagna}, A. and {Steele}, I.~A. and {Taris}, F. and {Utrilla}, E. and {van Reeven}, W. and {Vecchiato}, A.},
        title = "{Gaia Early Data Release 3. The astrometric solution}",
      journal = {\aap},
         year = 2021,
        month = may,
       volume = {649},
          eid = {A2},
        pages = {A2},
          doi = {10.1051/0004-6361/202039709},
archivePrefix = {arXiv},
       eprint = {2012.03380},
 primaryClass = {astro-ph.IM},
       adsurl = {https://ui.adsabs.harvard.edu/abs/2021A&A...649A...2L}
}

\begin{appendix}

\section{Azimuthal effects on contrast curves}
\label{appA}

\begin{figure*}
    \centering
    \includegraphics[width=0.8\linewidth]{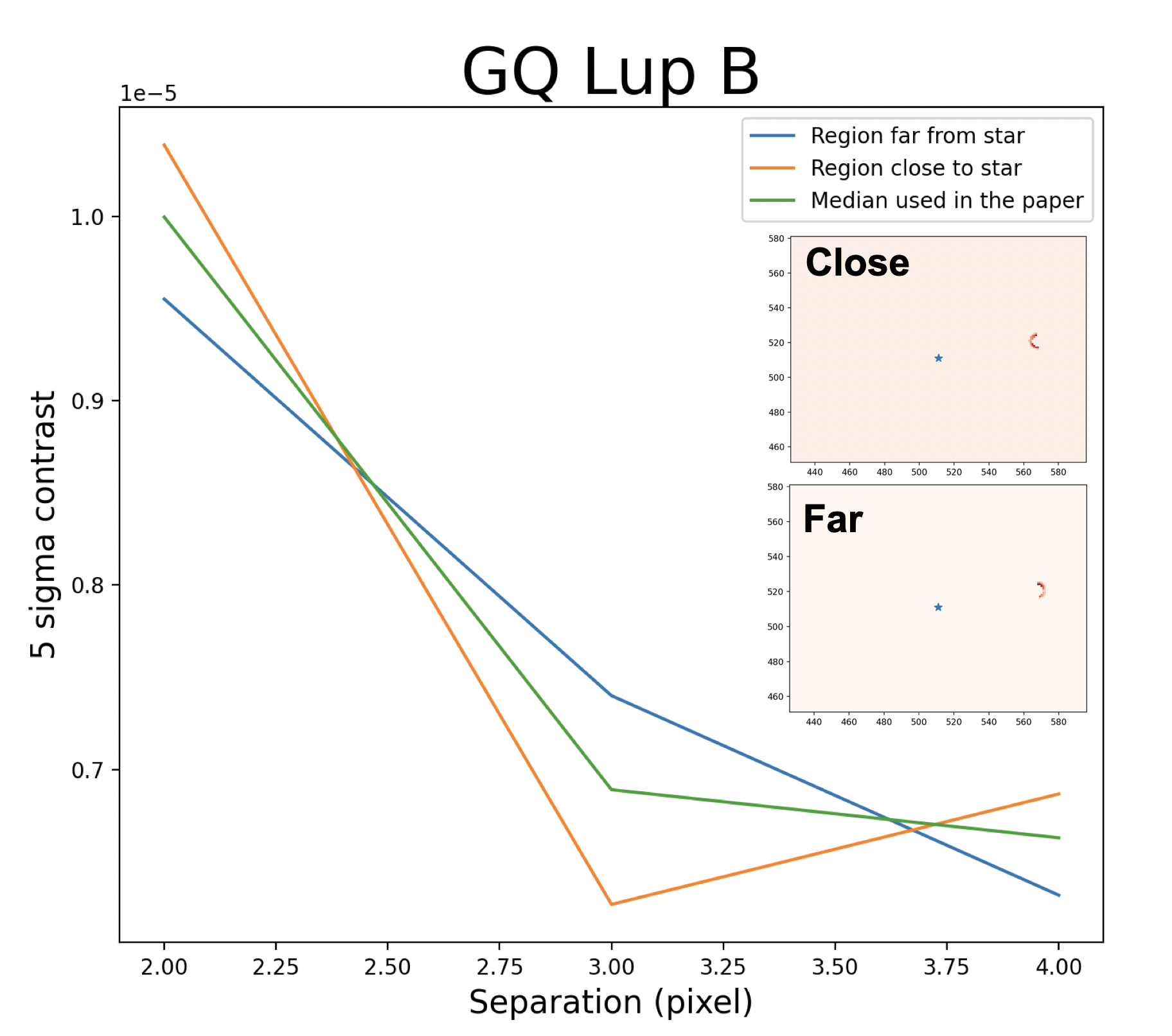}
    \caption{{GQ Lup B contrast curves derived from the semi-annuli located on the star-facing side (orange) and on the opposite side (blue) are shown and compared with the nominal contrast curve obtained using full annuli (green), as adopted in the main analysis. Insets illustrate the geometry of the semi-annuli close to (top) and far from (bottom) the host star used to compute the contrasts.}}
    \label{cc_compGQ}
\end{figure*}

\begin{figure*}[t]
    \centering

    \begin{subfigure}[t]{0.49\textwidth}
        \includegraphics[width=\linewidth]{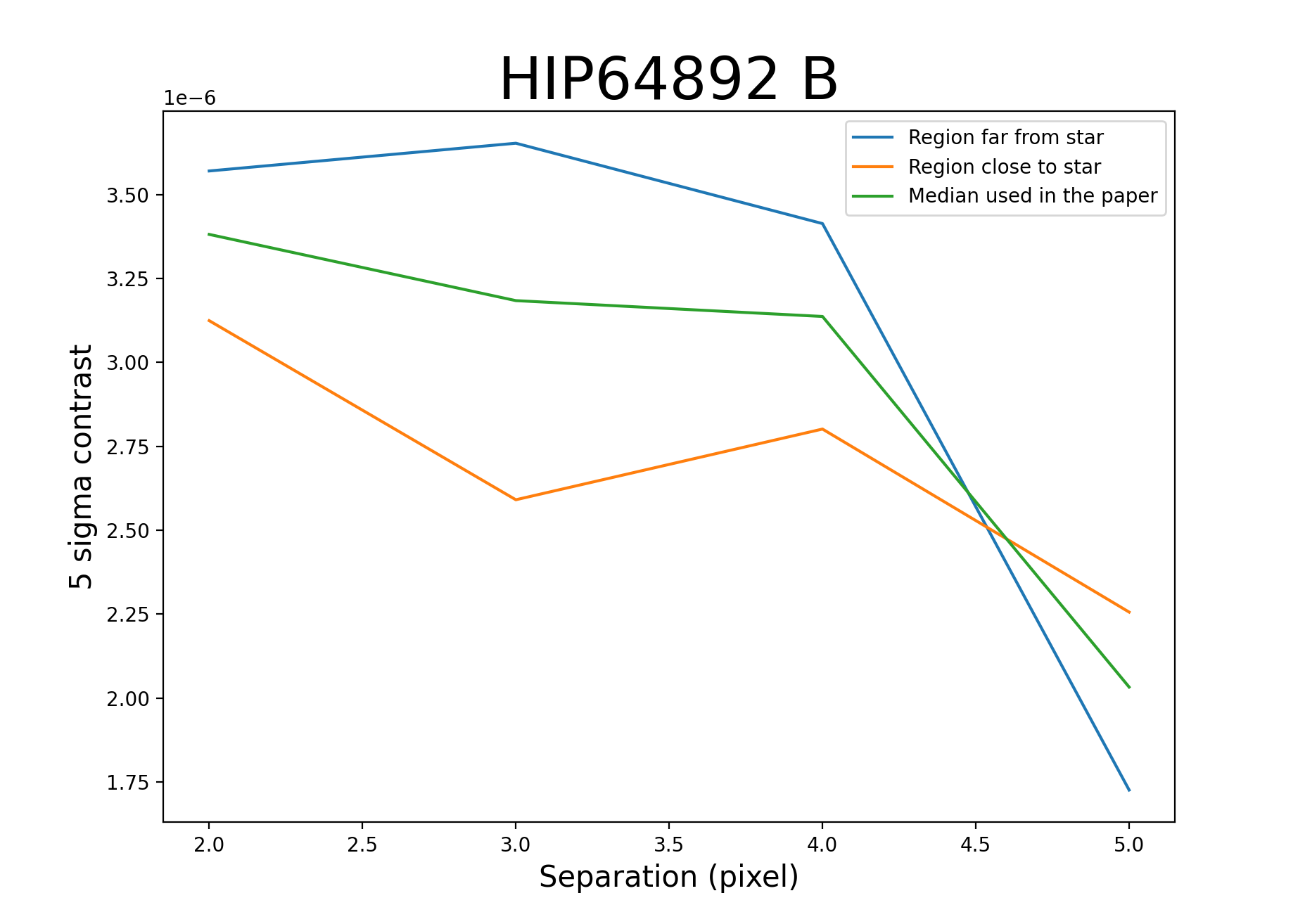}
    \end{subfigure}
    \hfill
    \begin{subfigure}[t]{0.49\textwidth}
        \includegraphics[width=\linewidth]{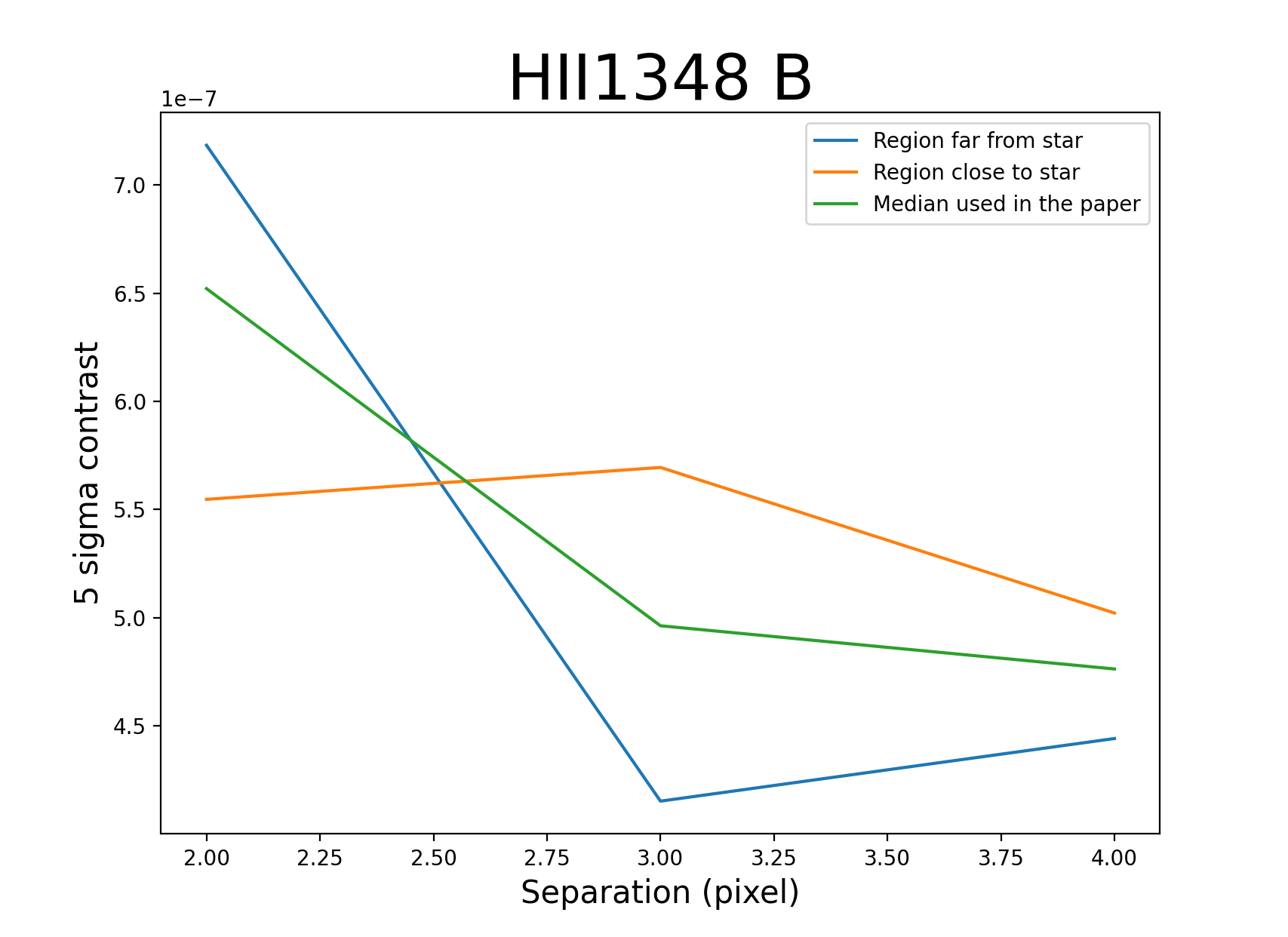}
    \end{subfigure}

    \caption{{HIP 64892 B (left) and HII 1348 B (right) contrast curves derived from the semi-annuli located on the star-facing side (orange) and on the opposite side (blue) are shown and compared with the nominal contrast curve obtained using full annuli (green), as adopted in the main analysis.}}
    \label{cc_compHIPHII}
\end{figure*}

\section{Comparison of contrast curves with one and multiple PSFs}
\begin{figure*}[t]
    \centering

    \begin{subfigure}[t]{0.49\textwidth}
        \includegraphics[width=\linewidth]{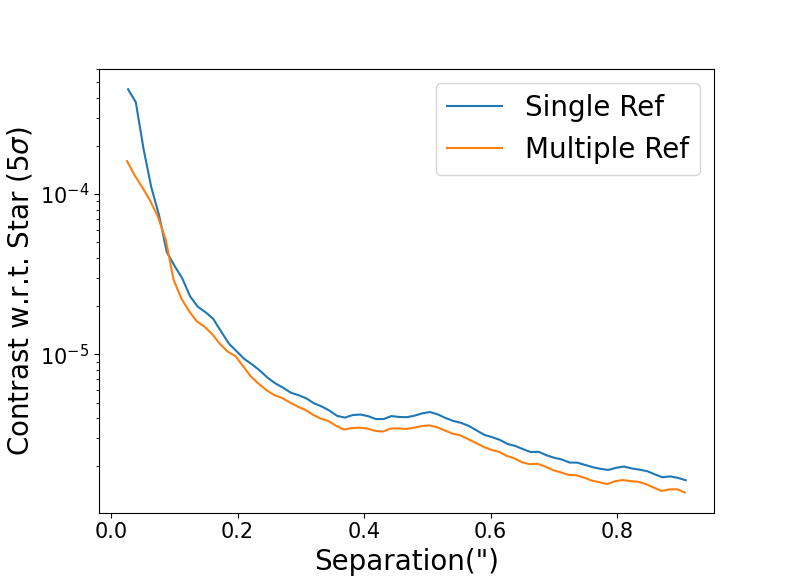}
        \caption{TYC 7084-794-1 B}
        \label{fig:panel1}
    \end{subfigure}
    \hfill
    \begin{subfigure}[t]{0.49\textwidth}
        \includegraphics[width=\linewidth]{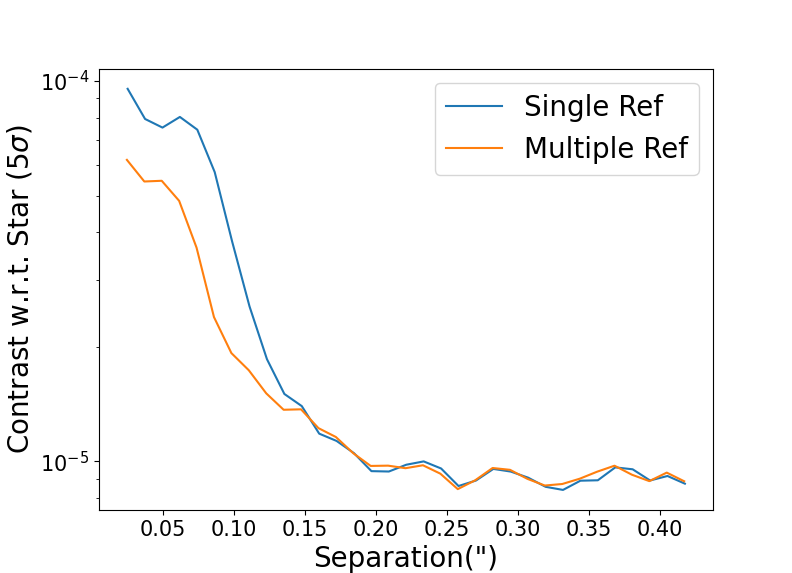}
        \caption{CT Cha B}
        \label{fig:panel2}
    \end{subfigure}
    \hfill
    \begin{subfigure}[t]{0.49\textwidth}
        \includegraphics[width=\linewidth]{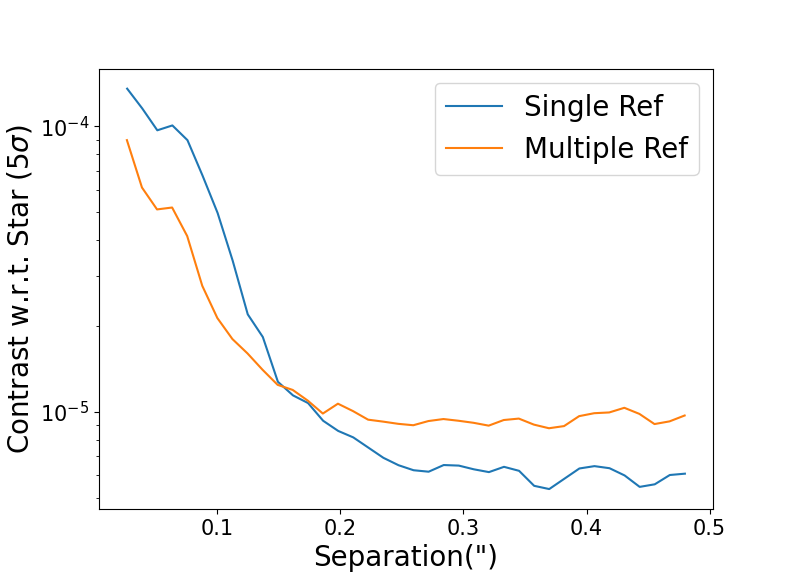}
        \caption{DH Tau B (2nd epoch)}
        \label{fig:panel3}
    \end{subfigure}
    \hfill
    \begin{subfigure}[t]{0.49\textwidth}
        \includegraphics[width=\linewidth]{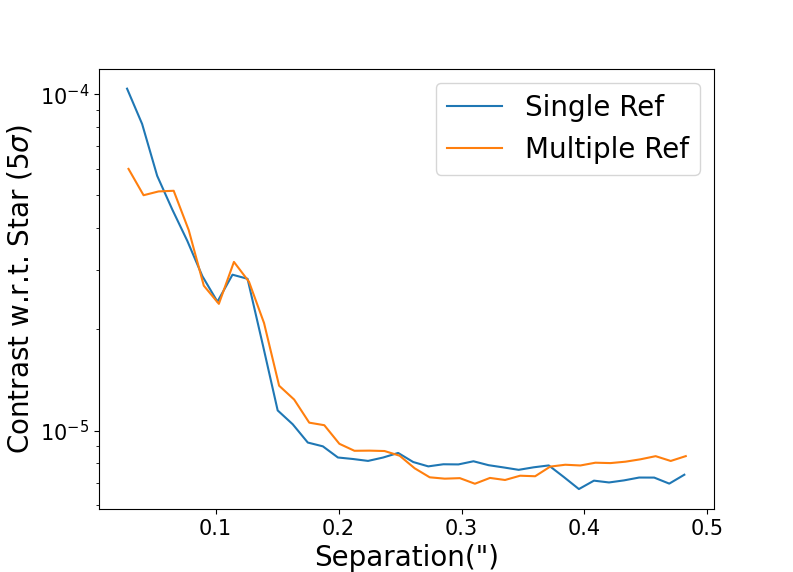}
        \caption{DH Tau B (1st epoch)}
        \label{fig:panel4}
    \end{subfigure}

    \caption{Representative contrast curves obtained with two subtraction strategies: a single PSF model (blue) and a PSF library (orange). Four cases are shown: uniform gain across all separations (TYC 7084-794-1 B), improvement at small separations (CT Cha b), inner-separation gain but degradation at large separations (DH Tau b, second epoch), and alternating performance (DH Tau b, first epoch).}
    \label{cc_SHVSsing}
\end{figure*}

\section{Test for the mock ADI}

\begin{figure}
    \centering
    \includegraphics[width=\columnwidth]{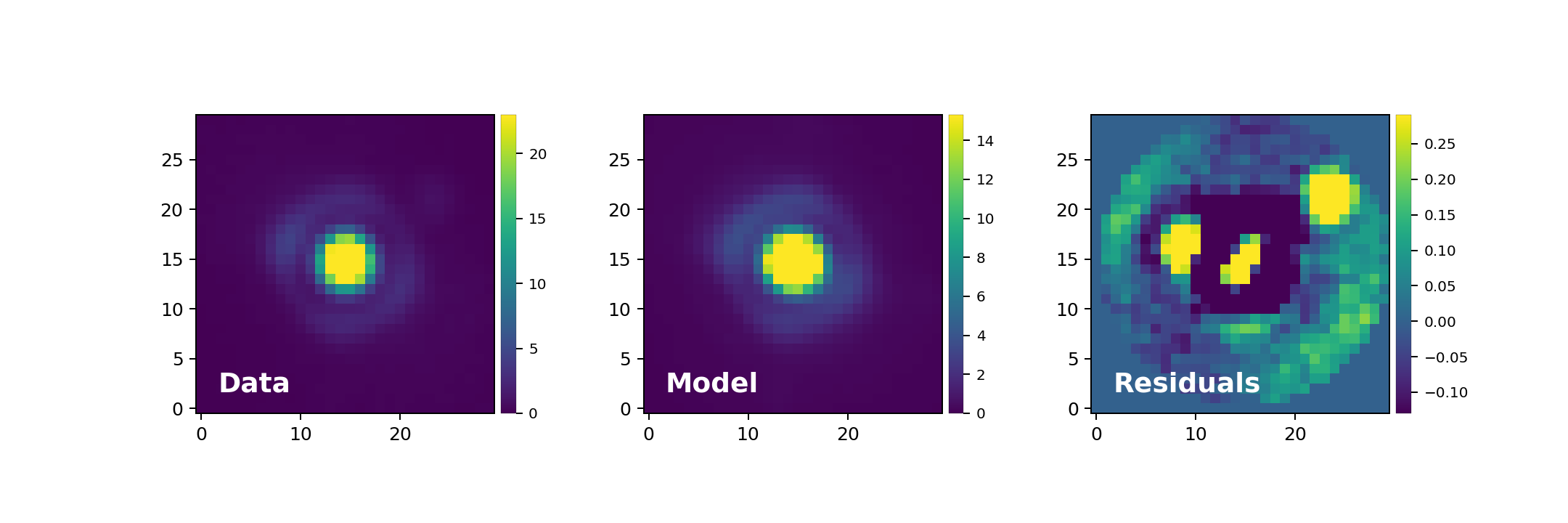}
    \includegraphics[width=\columnwidth]{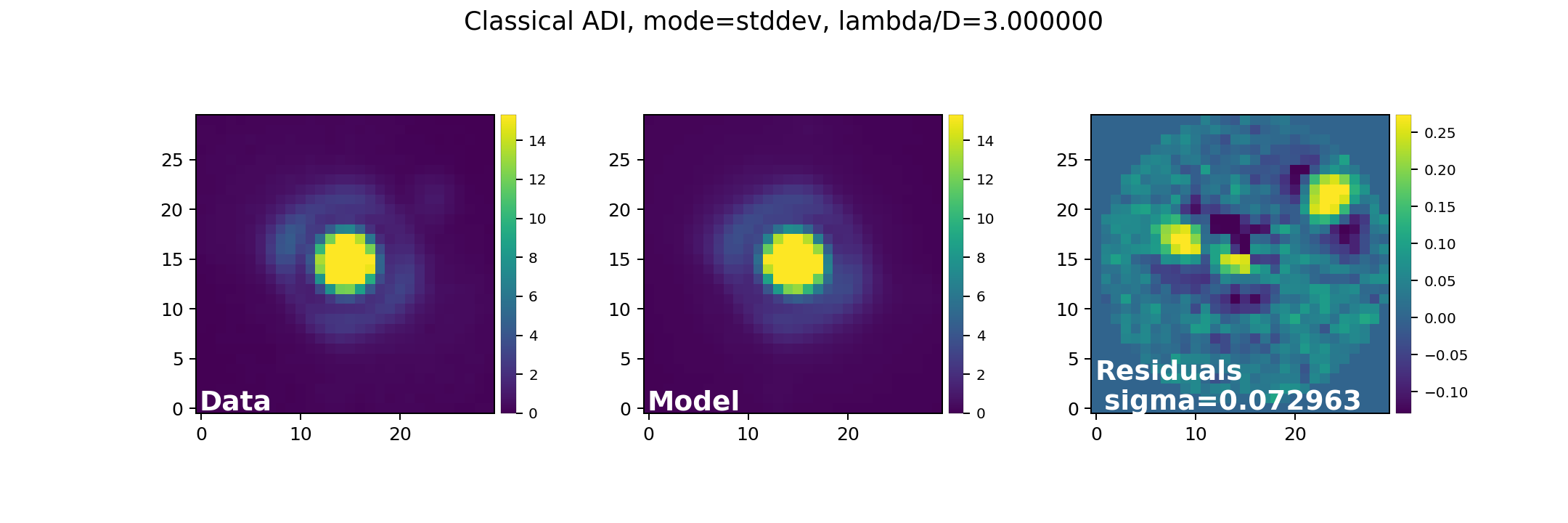}
    \caption{{Example of the performance of the mock ADI for a crowded field around the star HIP\,87836. {\it Top panel}: classical PSF modelling applied in this paper. {\it Bottom panel:} mock ADI model as explained in Sec.~\ref{data_red}. The wings of self-subtraction are evident in the case of the mock ADI, making it not suitable for very faint and close companions.  } }
    \label{mock}
\end{figure}

\begin{figure}
    \centering
    \includegraphics[width=\columnwidth]{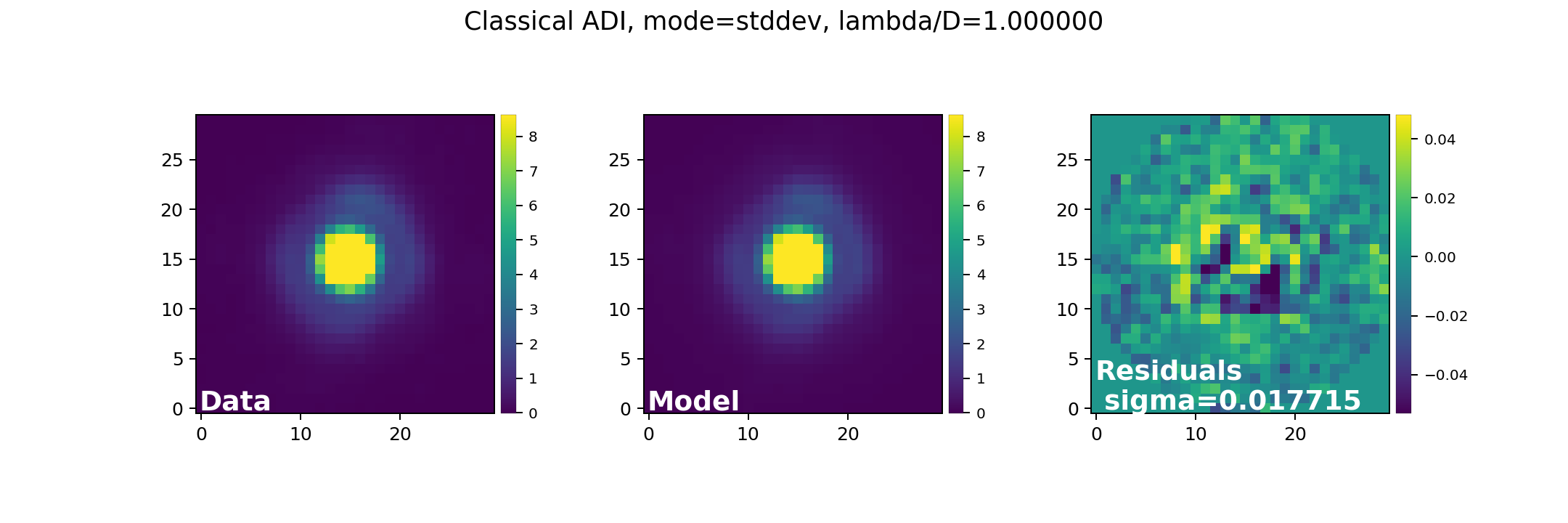}
    \caption{{ Result of the mock ADI in a synthetic cube reproducing the dataset of DH\,Tau b and DH\,Tau bb. The synthetic satellite is not visible in the residuals.   } }
    \label{mockDHTau}
\end{figure}

\section{Astrometry for background sources}

\begin{table*}
\centering
\caption{Astrometry for the background sources identified in the images.}
\begin{tabular}{lccc}
\hline\hline
  Name  &  & Separation & Pos Angle \\
        &  & (mas)          &  ($^{\circ}$) \\  
\hline \\
TYC 7084-794-1   &          & 5491    &   284.89\\[0.75ex]
CT Cha           &          & 2124    &    70.30\\[0.75ex]
GQ Lup           & source 1 & 1023    &   148.20\\[0.75ex]
                 & source 2 & 6468    &   320.40\\[0.75ex]
DH Tau           &          & 2873    &   348.69\\[0.75ex]
HIP 64892        &          & 5799    &   200.28\\[0.75ex]
RX J1609.5-2105  & source 1 & 6176    &   44.27 \\[0.75ex]
                 & source 2 & 4430    &   349.15\\[0.75ex]
                 & source 3 & 4818    &   271.75\\[0.75ex]
                 & source 4 & 3840    &   223.40\\[0.75ex]

\hline
\end{tabular}
\label{bgastro}
\end{table*}

\end{appendix}
\end{document}